\input epsf
\documentclass[12pt]{article}
\usepackage{latexsym,amsmath,amssymb,graphicx}
\renewcommand{\theequation}{\thesection.\arabic{equation}}
\renewcommand{\thefootnote}{\fnsymbol{footnote}}

\usepackage{amsmath}
\usepackage{amsfonts}

\numberwithin{equation}{section}

\def\doubleset#1#2{\bgroup%
\def\doit#1#2{%
\setbox\dblsetbox=\hbox{$\cstyle #1$}%
\raise#2\ht\dblsetbox\copy\dblsetbox%
\hskip-\wd\dblsetbox%
\raise-#2\ht\dblsetbox\box\dblsetbox}%
\mathchoice%
{\def\cstyle{\displaystyle}\doit#1#2}%
{\def\cstyle{\textstyle}\doit#1#2}%
{\def\cstyle{\scriptstyle}\doit#1#2}%
{\def\cstyle{\scriptscriptstyle}\doit#1#2}\egroup}
\def\underarrow#1{\vbox{\ialign{##\crcr$\hfil\displaystyle
 {#1}\hfil$\crcr\noalign{\kern1pt\nointerlineskip}$\longrightarrow$\crcr}}}

\newbox\dblsetbox

\newlength{\extraspace}
\newlength{\extraspaces}
\newcommand{\be}{\begin{equation}
\addtolength{\abovedisplayskip}{\extraspaces}
\addtolength{\belowdisplayskip}{\extraspaces}
\addtolength{\abovedisplayshortskip}{\extraspace}
\addtolength{\belowdisplayshortskip}{\extraspace}}
\newcommand{\ee}{\end{equation}}

\newcommand{\ba}{\begin{eqnarray}
\addtolength{\abovedisplayskip}{\extraspaces}
\addtolength{\belowdisplayskip}{\extraspaces}
\addtolength{\abovedisplayshortskip}{\extraspace}
\addtolength{\belowdisplayshortskip}{\extraspace}}
\newcommand{\ea}{\end{eqnarray}}

\newcommand{\bd}{\begin{displaymath}
\addtolength{\abovedisplayskip}{\extraspaces}
\addtolength{\belowdisplayskip}{\extraspaces}
\addtolength{\abovedisplayshortskip}{\extraspace}
\addtolength{\belowdisplayshortskip}{\extraspace}}
\newcommand{\ed}{\end{displaymath}}

\newcounter{saveeqn}

\newcommand{\newsection}[1]{
\vspace{12mm} \pagebreak[3] \addtocounter{section}{1}
\setcounter{equation}{0} \setcounter{subsection}{0}
\noindent{\bf \thesection. #1} \nopagebreak
\medskip
\nopagebreak}

\newcommand{\newsubsection}[1]{
\vspace{0.8cm} \pagebreak[3] \addtocounter{subsection}{1}
\noindent{\it \thesubsection. #1} \nopagebreak \vspace{2mm}
\nopagebreak}

\begin{document}
\addtolength{\baselineskip}{1.5mm}

\thispagestyle{empty}
\begin{flushright}
hep-th/   \\
\end{flushright}
\vbox{} \vspace{1.5cm}

\begin{center}
\centerline{\LARGE{Gauging Spacetime Symmetries On The
Worldsheet}}
\bigskip
\centerline{\LARGE{And The Geometric Langlands Program}}

\vspace{1.8cm}

{Meng-Chwan~Tan \footnote{On leave of absence from the National
University of Singapore.}}
\\[2mm]
{\it School of Natural Sciences, Institute for Advanced Study\\
Princeton, New Jersey 08540, USA} \\[1mm]
e-mail: tan@ias.edu\\
\end{center}

\vspace{1.8 cm}

\centerline{\bf Abstract}\smallskip \noindent

We study the two-dimensional twisted $(0,2)$ sigma-model on
various smooth complex flag manifolds $G/B$, and explore its
relevance to the geometric Langlands program. We find that an
equivalence - at the level of the holomorphic chiral algebra -
between a bosonic string on $G/B$ and a $B$-gauged version of
itself on $G$, will imply an isomorphism of classical $\cal
W$-algebras and a level relation which underlie a geometric
Langlands correspondence for $G=SL(N, \bf{C})$. This furnishes an
alternative physical interpretation of the geometric Langlands
correspondence for $G=SL(N, \bf{C})$, to that demonstrated earlier
by Kapustin and Witten via an electric-magnetic duality of
four-dimensional gauge theory. Likewise, the Hecke operators and
Hecke eigensheaves will have an alternative physical
interpretation in terms of the correlation functions of local
operators in the holomorphic chiral algebra of a
{\it{quasi-topological}} sigma-model {\it{without}} boundaries. A
forthcoming paper will investigate the interpretation of a
``quantum'' geometric Langlands correspondence for $G=SL(N,
\bf{C})$ in a similar setting, albeit with fluxes of the
sigma-model moduli which induce a ``quantum'' deformation of the
relevant classical algebras turned on.

\newpage

\renewcommand{\thefootnote}{\arabic{footnote}}
\setcounter{footnote}{0}

\newsection{Introduction}

The geometric Langlands correspondence has recently been given an
elegant physical interpretation by Kapustin and Witten in their
seminal paper \cite{KW} - by considering a certain twisted ${\cal
N}=4$ supersymmetric Yang-Mills theory in four-dimensions
compactified on a complex Riemann surface $C$, the geometric
Langlands correspondence associated to a holomorphic $G$-bundle on
$C$ can be shown to arise naturally from an electric-magnetic
duality in four-dimensions. To be more specific, it can be shown
\cite{KW} that one can, for example, relate various mathematical
objects and concepts of the correspondence such as Hecke
eigensheaves and the action of the Hecke operator, to the
boundaries and the 't Hooft line operator of the underlying
four-dimensional quantum gauge theory. Through a four-dimensional
electric-magnetic duality, or a mirror symmetry of the resulting
two-dimensional topological sigma-model at low-energies, one can
then map the relevant objects on either side of the correspondence
to their corresponding partners on the other side, thus furnishing
a purely physical interpretation of the geometric Langlands
conjecture.

The work of Kapustin and Witten centres around a gauge-theoretic
interpretation of the geometric Langlands correspondence. However,
it does not shed any light on the utility of two-dimensional
$\it{axiomatic}$ conformal field theory in the geometric Langlands
program, which, incidentally, is ubiquitous in the mathematical
literature on the subject \cite{2a, 2b, 2c, 2d, Frenkel}. This
seems rather puzzling. Afterall, the various axiomatic definitions
of a conformal field theory that fill the mathematical literature,
are based on established physical concepts, and it is therefore
natural to expect that in any physical interpretation of the
geometric Langlands correspondence, a two-dimensional conformal
field theory of some sort will be involved. It will certainly be
illuminating for the geometric Langlands program as a whole, if
one can deduce the conformal field-theoretic approach developed in
the mathematical literature, from the gauge-theoretic approach of
Kapustin and Witten, or vice-versa.

Note that the gauge-theoretic approach to the program necessarily
involves a certain two-dimensional quantum field theory in its
formulation, a generalised topological sigma-model to be exact.
This strongly suggests that perhaps a good starting point towards
elucidating the connection between the conformal field-theoretic
and gauge-theoretic approaches, would be to explore other physical
models in two-dimensions which will enable us to make direct
contact with the central results of the correspondence derived
from the axiomatic conformal field-theoretic approach. The work in
this paper represents our modest attempt towards this aim.

The key ingredients in the conformal field-theoretic approach to
the geometric Langlands correspondence are, affine Lie algebras at
the critical level without stress tensors \cite{book}, and $\cal
W$-algebras (defined by a Drinfeld-Sokolov or DS reduction
procedure) associated to the affine versions of the Langlands dual
of the Lie algebras \cite{book, dikii}. The duality between
$\it{classical}$ $\cal W$-algebras which underlies the conformal
field-theoretic approach to the correspondence, is just an
isomorphism between the $\it{Poisson}$ algebra generated by the
centre ${\frak z}(\widehat {\frak g})$ of the $\textrm{\it
completed universal enveloping algebra}$ of the affine Lie algebra
$\widehat {\frak g}$ at the critical level, where $\frak g$ is the
simple Lie algebra of the group $G$, and the classical $\cal
W$-algebra associated to the affine Lie algebra $^L {\widehat
{\frak g}}$ in the limit of large level $k'$ - ${\cal
W}_{\infty}(^L\widehat {\frak g})$, where $^L{\frak g}$ is the
simple Lie algebra of the Langlands dual group $^LG$; in other
words, a geometric Langlands correspondence for $G$ simply
originates from an isomorphism ${\frak z}(\widehat {\frak g})
\simeq {\cal W}_{\infty}(^L\widehat {\frak g})$ of $\it{Poisson}$
algebras \cite{Frenkel, Langlands-Drinfeld}. This statement is
accompanied by a relation $(k+h^{\vee})r^{\vee}= (k' +
{^Lh}^{\vee})^{-1}$ between the generic levels $k$ and $k'$ of
$\widehat {\frak g}$ and $^L{\widehat {\frak g}}$ respectively
(where $r^{\vee}$ is the lacing number of $\frak g$, and
$h^{\vee}$ and ${^Lh}^{\vee}$ are the dual Coxeter numbers of
$\frak g$ and $^L{\frak g}$ respectively), which defines a
``quantum" generalisation of the above isomorphism of classical
$\cal W$-algebras \cite{Frenkel, Langlands-Drinfeld}, whereby the
$k=-h^{\vee}$ and $k' = \infty$ limits  just correspond to the
classical isomorphism mentioned herein that we shall be discussing
in this paper.

A strong hint that one should be considering for our purpose a
two-dimensional twisted $(0,2)$ sigma-model on a flag manifold,
stems from our recent understanding of the role sheaves of
``Chiral Differential Operators" (or CDO's) play in the
description of its holomorphic chiral algebra \cite{CDO}, and from
the fact that global sections of CDO's on a flag manifold furnish
a module of an affine Lie algebra at the critical level \cite{CDO,
MSV}. On the other hand, since Toda field theories lead to
free-field realisations of the $\cal W$-algebras defined by the DS
reduction scheme mentioned above (see Sect. 6 of \cite{review},
and the references therein), and since the Toda theory can be
obtained as a gauge-invariant content of a certain gauged WZW
theory \cite{review 23, review 24},  it ought to be true that by
relating a relevant aspect of the sigma-model on a flag manifold
to a gauged WZW model, one should be able to uncover a physical
manifestation of the isomorphism of (classical) $\cal W$-algebras.
Indeed, we shall show that an equivalence - at the level of the
$\it{holomorphic}$ chiral algebra - between a bosonic string on a
smooth flag manifold $G/B$ and a $B$-gauged WZW model on $G$,
where $G=SL(N, \bf{C})$, will necessarily imply an isomorphism of
classical $\cal W$-algebras and the relation
$(k+h^{\vee})r^{\vee}= (k' + {^Lh}^{\vee})^{-1}$ which underlie a
geometric Langlands correspondence for $G=SL(N, \bf{C})$. Since a
string on a group manifold $G$ can be expressed as a WZW model on
$G$ \cite{Polchinski 2}, it would mean that an equivalence, at the
level of the holomorphic chiral algebra, between a bosonic string
on a smooth coset manifold $G/B$ and a $B$-gauged version of
itself on $G$ - a statement which stems from the ubiquitous notion
that one can always physically interpret a geometrical symmetry of
the target space as a gauge symmetry in the worldsheet theory -
will imply a geometric Langlands correspondence for $G=SL(N,
\bf{C})$. This furnishes an alternative physical interpretation of
the geometric Langlands correspondence for $G=SL(N, \bf{C})$ to
that of an electric-magnetic duality of four-dimensional gauge
theory. Likewise, the Hecke operators and Hecke eigensheaves will
also lend themselves to different physical interpretations -
instead of line operators and branes in a two-dimensional
topological sigma-model, they are, in our case, associated to the
correlation functions of local operators that span the holomorphic
chiral algebra of a {\it{quasi-topological}} sigma-model
{\it{without}} boundaries. Our results therefore open up a new way
of looking at the correspondence, thus providing the prospect of
novel mathematical and physical insights for the program in
general.

\smallskip\noindent{\it A Brief Summary and Plan of the Paper}

We shall now give a brief summary and plan of the paper.

In $\S$2, we shall show that an equivalence - at the level of the
holomorphic chiral algebra - between a bosonic string on $G/B$ and
a $B$-gauged version of itself on $G$, will necessarily imply a
geometric Langlands correspondence for $G=SL(N, \bf{C})$, where
$N=2,3$. We begin by considering the twisted $(0,2)$ sigma-model
on a complex flag manifold defined by the coset space $SL(N)/ B$
where $N=2,3$, and $B$ is a Borel subgroup containing upper
triangular matrices of $SL(N)$.\footnote{Here and henceforth, in
writing $SL(N)$, we really mean $SL(N,\bf{C})$.} We then explain
how a subspace of the global sections of the sheaf of CDO's,
describing the holomorphic chiral subalgebra of the sigma-model on
$SL(N)/ B$, will furnish a module of an affine $SL(N)$ algebra at
the critical level. This in turn will allow us to show, using the
results in appendix A, that the $\it{classical}$ centre of the
completed universal enveloping algebra of the affine $SL(N)$
algebra at the critical level - ${\frak z}(\widehat {\frak
{sl}}_N)$, is spanned by the Laurent modes of certain local fields
of spins 2 and 3 in the classical $\it{holomorphic}$ chiral
algebra of the $\it{purely}$ $\it{bosonic}$ sector of the
sigma-model on $SL(N)/B$. Next, we turn to a dual description of
this classical, holomorphic chiral algebra in the purely bosonic
sector (or bosonic string part) of the sigma-model  on $SL(N)/B$ -
the classical, $\it{holomorphic}$ BRST-cohomology (or chiral
algebra) of a $B$-gauged WZW model on $SL(N)$. One can then show
that an equivalence between these classical, holomorphic chiral
algebras will necessarily imply an isomorphism ${\frak z}(\widehat
{\frak g}) \simeq {\cal W}_{\infty}(^L\widehat {\frak g})$ of
Poisson algebras and the level relation $(k+h^{\vee})r^{\vee}= (k'
+ {^Lh}^{\vee})^{-1}$ that underlie a geometric Langlands
correspondence for $G=SL(N)$, where $N=2,3$.

In $\S$3, we will generalise our arguments in $\S$2 to arbitrary
$N$. To this end,  we will first discuss the twisted sigma-model
on any complex flag manifold $SL(N)/B$, and the global sections of
the sheaf of CDO's on $SL(N)/B$ associated to the chiral algebra
of the purely bosonic sector of the sigma-model that will furnish
a module of an affine $SL(N)$ algebra at the critical level. We
will then proceed to discuss the construction of higher-spin
analogs of the Segal-Sugawara tensor from the affine $SL(N)$
algebra, and show that these fields of higher spins which are in
the $\it{classical}$ holomorphic chiral algebra of the
$\it{purely}$ $\it{bosonic}$ sector of the sigma-model on
$SL(N)/B$, will have Laurent modes that span the classical centre
$\frak{z}({\widehat{\frak{sl}}_N})$ of the completed universal
enveloping algebra of the affine $SL(N)$ algebra at critical
level.   Next, we will outline the mathematical Drinfeld-Sokolov
reduction procedure in \cite{book} of defining ${\cal
W}_{k'}(\widehat {\frak g})$, a $\cal W$-algebra associated to
${\widehat {\frak g}}$ at level $k'$, via a Hecke algebra.
Thereafter, we will show that the $\it{holomorphic}$ sector of the
BRST-cohomology of the $B$-gauged WZW model on $SL(N)$ physically
realises, in all generality, this particular Hecke algebra, i.e.,
the holomorphic BRST-cohomology of the $B$-gauged WZW model on
$SL(N)$ always consist of local operators which generate a ${\cal
W}_{k'}(\widehat {\frak {sl}}_N)$ OPE algebra. Hence, its
classical, holomorphic BRST-cohomology will always consist of
local fields with Laurent modes that generate a classical ${\cal
W}_{\infty}(\widehat {\frak {sl}}_N)$-algebra. By specialising our
analysis (in the classical limit) to $N=2,3$, we will make contact
with the results in $\S$2. One can now extend the arguments in
$\S$2 for $N=2,3$ to $\it{any}$ $N$; since $\frak g = \frak{sl}_N
= {^L\frak g}$, an equivalence - at the level of the classical
holomorphic chiral algebra - between the purely bosonic sector (or
bosonic string part) of the sigma-model on $SL(N)/B$ and the
$B$-gauged WZW model on $SL(N)$, will necessarily imply an
isomorphism $\frak z(\widehat{\frak g}) \simeq {\cal
W}_{\infty}(^L\widehat {\frak g})$ of Poisson algebras and the
level relation $(k+h^{\vee})r^{\vee}= (k' + {^Lh}^{\vee})^{-1}$
which underlie a geometric Langlands correspondence for $G=SL(N)$.
That is, an equivalence - at the level of the holomorphic chiral
algebra - between a bosonic string on $G/B$ and a $B$-gauged
version of itself on $G$, will necessarily imply a geometric
Langlands correspondence for any $G=SL(N, \bf{C})$.

In $\S$4, we shall derive, via the isomorphism of classical $\cal
W$-algebras discussed in $\S$3, a correspondence between flat
holomorphic $^LG$-bundles on the worldsheet $\Sigma$ and Hecke
eigensheaves on the moduli space $\textrm{Bun}_G$ of holomorphic
$G$-bundles on $\Sigma$, where $G=SL(N)$. Lastly, we shall
physically interpret the Hecke eigensheaves and Hecke operators of
the geometric Langlands program in terms of the correlation
functions of purely bosonic local operators in the holomorphic
chiral algebra of the twisted $(0,2)$ sigma-model on the complex
flag manifold $SL(N)/B$.

In appendix A, we will review the two-dimensional twisted $(0,2)$
sigma-model  considered in \cite{CDO}, and explain its relation to
the theory of CDO's. In particular, we will describe how the
relevant physical features of the sigma-model and its holomorphic
chiral algebra, can be interpreted in terms of the sheaf of CDO's
and its Cech-cohomology.

\smallskip\noindent{\it Relation to the Gauge-Theoretic Approach}

Though we have not made any explicit connections to the
gauge-theoretic approach of Kapustin and Witten yet, we hope to be
able to address this important issue in a later publication,
perhaps with the insights gained in this paper.

\smallskip\noindent{\it A ``Quantum'' Geometric Langlands Correspondence}

A forthcoming paper will investigate the interpretation of a
``quantum'' geometric Langlands correspondence for $G=SL(N)$ in a
similar physical context, albeit with fluxes of the sigma-model
moduli turned on, such that the level of the affine $SL(N)$
algebra with a module furnished by the global sections of the
sheaf CDO's on $X= SL(N)/B$, can be deformed away from the
critical value, whereby a ``quantum'' deformation of our present
setup can be defined.

\newsection{An Equivalence of Classical Holomorphic Chiral Algebras and the Geometric Langlands Correspondence for $G=SL(2)$ and $SL(3)$}

In this section, we shall study explicit examples of the twisted
sigma-model on the complex flag manifolds $SL(2)/B$ and $SL(3)/B$,
and the corresponding sheaves of CDO's that describe its
holomorphic chiral algebra. We shall also study the holomorphic
BRST-cohomology of a $B$-gauged WZW model on $SL(2)$ and $SL(3)$.
We will then show that an equivalence - at the level of the
holomorphic chiral algebra - between a bosonic string on $SL(N)/B$
and a $B$-gauged version of itself on $SL(N)$, where $N=2,3$, will
imply an isomorphism of classical $\cal W$-algebras and the
relation $(k+h^{\vee})r^{\vee}= (k' + h^{\vee})^{-1}$ which
underlie a geometric Langlands correspondence for $G=SL(2)$ and
$SL(3)$ respectively.

\newsubsection{The Twisted Sigma-Model on $SL(2)/B$ and its Classical Holomorphic Chiral Algebra}
Let us take $X= SL(2)/B$. In other words, since $SL(2)/B \cong
\mathbb{CP}^1$, we will be exploring and analysing the chiral
algebra $\cal A$ of operators in the twisted $(0,2)$ model on
$\mathbb {CP}^1$. To this end, we will work locally on the
worldsheet $\Sigma$, choosing a local complex parameter $z$.

Now, $\mathbb{CP}^1$ can be regarded as the complex $\gamma$-plane
plus a point at infinity.  Thus, we can cover it by two open sets,
$U_1$ and $U_2$, where $U_1$ is the complex $\gamma$-plane, and
$U_2$ is the complex $\tilde \gamma$-plane, where
$\tilde\gamma=1/\gamma$.

Since $U_1$ is isomorphic to $\mathbb{C}$, the sheaf of CDO's in
$U_1$ can be described by a single free $\beta\gamma$ system with
action \be I={1\over 2\pi}\int|d^2z| \ \beta \partial_{\bar z}
\gamma. \label{actionU1} \ee

Here  $\beta$ and $\gamma$,  are fields of dimension $(1,0)$ and
$(0,0)$ respectively. They obey the usual free-field OPE's; there
are no singularities in the operator products $\beta(z)\cdot
\beta(z')$ and $\gamma(z)\cdot\gamma(z')$, while \be
\beta(z)\gamma(z')  \sim  -{1\over z-z'}. \ee

Similarly, the sheaf of CDO's in $U_2$ is described by a single
free $\tilde\beta\tilde\gamma$ system with action \be I= {1\over
2\pi}\int|d^2z| \ \tilde \beta \partial_{\bar z} \tilde\gamma,
\label{actionU2} \ee where the fields $\tilde \beta$, and $\tilde
\gamma$ obey the same OPE's as $\beta$ and $\gamma$. In other
words, the non-trivial OPE's are given by \be \tilde \beta(z)
\tilde \gamma(z')  \sim  -{1\over z-z'}. \ee

In order to describe a globally-defined sheaf of CDO's, one will
need to glue the free conformal field theories with actions
(\ref{actionU1}) and (\ref{actionU2}) in the overlap region $U_1
\cap U_2$. To do so,  one must use the admissible automorphisms of
the free conformal field theories defined in
(\ref{autoCDOgamma})-(\ref{autoCDObeta}) to glue the free-fields
together. In the case of $X = \mathbb {CP}^1$, the automorphisms
will be given by
\begin{eqnarray}
\label{autoCP1gamma}
{\tilde \gamma} & = & {1 \over \gamma},\\
\label{autoCP1beta} {\tilde \beta} & = &   - \gamma^2 \beta + 2
\partial_z \gamma.
\end{eqnarray}
As there is no obstruction to this gluing in the twisted
sigma-model on any flag manifold $SL(N)/B$ \cite{MSV}, a sheaf of
CDO's can be globally-defined on the $\mathbb {CP}^1$
target-space.

\bigskip\noindent{\it Global Sections of the Sheaf of CDO's on $X=SL(2)/B$}

Recall that for a general manifold $X$ of complex dimension $n$,
the chiral algebra $\cal A$ will be given by ${\cal A} =
\bigoplus_{g_R = 0}^{g_R = n} H^{g_R}( X, {\widehat {\cal
O}^{ch}_X})$ as a vector space. Since $\mathbb {CP}^1$ has complex
dimension 1, we will have, for $X=\mathbb {CP}^1$, the relation
${\cal A} = \bigoplus_{g_R = 0}^{g_R = 1} H^{g_R}( \mathbb {CP}^1,
{\widehat {\cal O}^{ch}_{\mathbb P^1}})$. Thus, in order to
understand the chiral algebra of the twisted sigma-model, one
needs only to study the global sections of the sheaf ${\widehat
{\cal O}^{ch}_{\mathbb P^1}}$, and its first Cech cohomology $H^1(
\mathbb {CP}^1, {\widehat {\cal O}^{ch}_{\mathbb P^1}})$. However,
for our purpose, it would suffice to study just the purely bosonic
sector of $\cal A$ - from our $\overline Q_+$-Cech cohomology
dictionary, this translates to studying the global sections
$H^0({\mathbb {CP}^1}, {\widehat {\cal O}^{ch}_{\mathbb P^1}})$
only.

At dimension 0, the space of global sections $H^0( \mathbb {CP}^1,
{\widehat {\cal O}}^{ch}_{\mathbb P^1; 0})$ must be spanned by
functions of arbitrary degree in $\gamma$. Since all regular,
holomorphic functions on a compact Riemann surface such as
$\mathbb {CP}^1$ must be constants, we find that the space of
global sections at dimension 0, given by $H^0( \mathbb {CP}^1,
{\widehat {\cal O}}^{ch}_{\mathbb P^1; 0})$, is one-dimensional
and generated by 1.

Let us now ascertain the space $H^0( \mathbb {CP}^1, {\widehat
{\cal O}}^{ch}_{\mathbb P^1; 1})$ of global sections of dimension
1. In order to get a global section of ${\widehat {\cal
O}}^{ch}_{\mathbb P^1}$ of dimension 1, we can act on a global
section of ${{\widehat {\cal O}}^{ch}_{\mathbb P^1}}$ of dimension
0 with the partial derivative $\partial_z$. Since $\partial_z 1 =
0$, this prescription will not apply here.

One could also consider operators of the form $f(\gamma)
\partial_z \gamma$, where $f(\gamma)$ is a holomorphic function of
$\gamma$. However, there are no such global sections either - such
an operator, by virtue of the way it transforms purely
geometrically under (\ref{autoCP1gamma}), would correspond to a
section of $\Omega^1(\mathbb {CP}^1)$, the sheaf of holomorphic
differential forms $f(\gamma) d\gamma$ on $\mathbb {CP}^1$, and
from the classical result $H^0(\mathbb {CP}^1, \Omega^1(\mathbb
{CP}^1)) = 0$, which continues to hold in the quantum theory, we
see that $f(\gamma) \partial_z \gamma$ cannot be a dimension  1
global section of ${{\widehat {\cal O}}^{ch}_{\mathbb P^1}}$.

Other possibilities include operators which are linear in $\beta$.
In fact, from the automorphism relation of (\ref{autoCP1beta}), we
find an immediate example as the LHS, $\tilde \beta$, is by
definition regular in $U_2$, while the RHS, being polynomial in
$\gamma$, $\partial_z \gamma$ and $\beta$, is manifestly regular
in $U_1$. Their being equal means that they represent a dimension
1 global section of ${\widehat {\cal O}}^{ch}_{\mathbb {P}^1}$
that we will call $J_+$: \be J_+  = - \gamma^2 \beta + 2
\partial_z \gamma = {\tilde \beta}. \label{J_+} \ee The
construction is completely symmetric between $U_1$ and $U_2$, with
$\gamma\leftrightarrow \tilde\gamma$,
$\beta\leftrightarrow\tilde\beta$, so a reciprocal formula gives
another dimension 1 global section $J_-$: \be J_- =\beta =
-\tilde\gamma^2\tilde \beta + 2\partial_z \tilde\gamma.
\label{J_-}\ee Hence,  $J_+$ and $J_-$ give us two dimension 1
global sections of the sheaf ${\widehat{\cal O}}^{ch}_{\mathbb
P^1}$. Since these are global sections of a sheaf of chiral vertex
operators, we can construct more of them from their OPE's. There
are no singularities in the $J_+ \cdot J_+$ or $J_-\cdot J_-$
operator products, but \be J_+ J_- \sim {2J_3\over z-z'} - {2
\over{(z-z')^2}}, \ee where $J_3$ is another global section of
dimension 1 given by \be J_3 = -\gamma\beta. \label{J_3} \ee (Note
that normal-ordering is again understood for all operators above
and below).

Notice that since $\{J_+, J_-, J_3 \}$ are $\psi^{\bar
i}$-independent, they are purely bosonic operators that belong in
$H^0(\mathbb {CP}^1, {\widehat {\cal O}}^{ch}_{\mathbb{P}^1 ;
1})$. One can verify that they satisfy the following closed OPE
algebra:
\begin{eqnarray}
{J}_3 (z) {J}_+ (z') & \sim & {{+{ J}_+ (z')} \over z-z'},  \\
{J}_3 (z) {J}_- (z') & \sim & {{-{ J}_- (z')} \over z-z'}, \\
{J}_3 (z) {J}_3 (z') & \sim & - {1\over (z-z')^2}, \\
J_+ (z) J_- (z') & \sim & {2J_3\over z-z'} - {2 \over{(z-z')^2}}.
\end{eqnarray}
From the above OPE algebra, we learn that the $J$'s furnish a
module of an affine algebra of $SL(2)$ at level $-2$, which here,
as noted in \cite{MSV}, appears in the Wakimoto free-field
representation. Indeed, these chiral vertex operators are
holomorphic in $z$, which means that one can expand them in a
Laurent series that allows an affinisation of the $SL(2)$ algebra
generated by their resulting zero modes. Thus, the space of global
sections of ${\widehat{\cal O}}^{ch}_{\mathbb {P}^1}$ furnishes a
module of an affine algebra of $SL(2)$ at level
$-2$.\footnote{Note that one can consistently introduce
appropriate fluxes to deform the level away from $-2$ - recall
from our discussion in Appendix A that the $E_{ij}= \partial_i
B_j$ term in (\ref{autoCDObeta}) is related to the fluxes that
correspond to the moduli of the chiral algebra, and since this
term will determine the level $k$ of the affine $SL(2)$ algebra
via the term $-k \partial_z\gamma$ of $\tilde \beta$, (which is
set to $k=-2$ in the current undeformed case), turning on the
relevant fluxes will deform $k$ away from $-2$. Henceforth,
whenever we consider $k\neq -2$, we really mean turning on fluxes
in this manner.} case The space of these operators obeys all the
physical axioms of a chiral algebra except for reparameterisation
invariance on the $z$-plane or worldsheet $\Sigma$. We will
substantiate this last statement momentarily by showing that the
holomorphic stress tensor fails to exist in the $\overline
Q_+$-cohomology at the quantum level. As we shall see shortly,
this observation will be crucial to our results in this section.

\bigskip\noindent{\it The Segal-Sugawara Tensor and the Classical Holomorphic Chiral Algebra}

Recall from section 2.6 and our $\overline Q_+$-Cech cohomology
dictionary, that there will be a $\psi^{\bar i}$-independent
stress tensor operator ${T}(z)$ in the quantum $\overline
Q_+$-cohomology of the underlying twisted sigma-model on $\mathbb
{CP}^1$, if and only if the corresponding ${\widehat T}(z)$
operator of the free $\beta\gamma$ system belongs in $H^0(\mathbb
{CP}^1, {{\widehat {\cal O}}^{ch}_{\mathbb P^1}})$ - the space of
global sections of ${\widehat {\cal O}}^{ch}_{\mathbb P^1}$. Let's
look at this more closely.

Now, note that  for $X= \mathbb {CP}^1$, we have
\begin{eqnarray}
{\widehat T}(z) & = &  -: \beta \partial_z \gamma: (z).
\end{eqnarray}
where the above operators are defined and regular in $U_1$.
Similarly, we also have
\begin{eqnarray}
\label{A1} {\widetilde{\widehat {{T}}}(z)} & = &  -: \tilde\beta
\partial_z \tilde\gamma: (z).
\end{eqnarray}
where the above operators are defined and regular in $U_2$. By
substituting the automorphism relations
(\ref{autoCP1gamma})-(\ref{autoCP1beta}) into (\ref{A1}), a small
computation shows that in $U_1 \cap U_2$, we have \be
{\widetilde{\widehat T}(z)} - {\widehat T}(z) =
\partial_z({{\partial_z \gamma}\over{\gamma}}).
\label{B1} \ee where an operator that is a global section of
${{\widehat {\cal O}}^{ch}_{\mathbb P^1}}$ must agree in $U_1 \cap
U_2$.

The only way to consistently modify $\widehat T$ and $\widetilde
{\widehat T}$ so as to agree on $U_1\cap U_2$, is to shift them by
a multiple of the term ${(\partial^2_z \gamma) / \gamma}$ and
$(\partial_z \gamma)^2 / {\gamma^2}$. However, any linear sum of
these two terms has a pole at both $\gamma=0$ and
$\tilde\gamma=0$. Thus, it cannot be used to redefine $\widehat T$
or $\widetilde {\widehat T} $ (which has to be regular in $U_1$ or
$U_2$ respectively). Therefore, we conclude that ${\widehat T}(z)$
does not belong in $H^0(\mathbb {CP}^1, {\widehat {\cal
O}}^{ch}_{\mathbb P^1})$. This means that $T(z)$ does not exist in
the $\overline Q_+$-cohomology of the underlying twisted
sigma-model on $\mathbb {CP}^1$ at the quantum level.

This last statement is in perfect agreement with the physical
picture presented in section 2.3, which states that since
$c_1(\mathbb {CP}^1) \neq 0$, there are now one-loop corrections
to the action of $\overline Q_+$, such that the $T(z)$ is no
longer annihilated by $\overline Q_+$. This just corresponds to
the mathematical fact that the sheaf ${\widehat {\cal
O}}^{ch}_{X}$ of CDO on $X$ has a structure of a conformal vertex
algebra if and only if the conformal anomaly measured by $c_1(X)$
vanishes.  Note also that (\ref{B1}) is a counterpart in Cech
cohomology to the sigma-model relation \be [\overline Q_+, T_{zz}]
= \partial_z(R_{i \bar j} \partial_z \phi^i \psi^{\bar j}).
\label{counterpart} \ee Since $\phi^i$ corresponds to $\gamma^i$,
we see from (\ref{counterpart}) that the sigma-model operator
$R_{i \bar j} \psi^{\bar j}$ must correspond to $1 / \gamma$.
Hence, we have an interpretation of the one-loop beta function
(which is proportional to $R_{i \bar j}$) in terms of holomorphic
data. This has been emphasised in \cite{CDO} as a novel way to
view the one-loop beta function from a purely mathematical
viewpoint.

The absence of $T(z)$ in the quantum holomorphic chiral algebra of
the twisted sigma-model on $\mathbb {CP}^1$, can also be observed
from a different but crucial viewpoint. To this end, note that for
any affine algebra $\widehat {\frak g}$ at level $k \neq
-h^{\vee}$, where $h^{\vee}$ is the dual Coxeter number of the Lie
algebra $\frak g$, one can construct the corresponding stress
tensor out of the currents of $\widehat {\frak g}$ via a
Segal-Sugawara construction \cite{Ketov}. In the present case of
an affine $SL(2)$ algebra, the stress tensor can be constructed as
\be T(z) = {{: (J_+ J_- + J_3^2)(z) :} \over {k+2}}, \label{SS def
T(z)} \ee where because ${\frak g} = {{\frak {sl}}_2}$, $h^{\vee}
=2$. As required, for every $k \neq {-2}$, the modes of the
Laurent expansion of $T(z)$ will span a Virasoro algebra. In
particular, $T(z)$ will generate holomorphic reparametrisations of
the coordinates on the worldsheet $\Sigma$. Notice that this
definition of $T(z)$ in (\ref{SS def T(z)}) is ill-defined when
$k=-2$. Nevertheless,  one can always associate $T(z)$ with an
operator $S(z)$ that is well-defined at any finite level, such
that \be S(z) =  (k+2) T(z) \label{S(z)} \ee is known as the
Segal-Sugawara tensor. It is given by \be S(z) = {: (J_+ J_- +
J_3^2)(z) :}. \label{s(z)} \ee From (\ref{S(z)}), we see that
$S(z)$ generates, in its OPE's with other field operators, $(k+2)$
times the transformations usually generated by the stress tensor
$T(z)$. Therefore, at the level $k= -2$, $S(z)$ generates no
transformations at all - its OPE's with all other field operators
are trivial. This is equivalent to saying that the holomorphic
stress tensor does not exist at all, since $S(z)$, which is the
only well-defined operator at this level that could possibly
generate the transformation of fields under an arbitrary
holomorphic reparametrisation of the worldsheet coordinates on
$\Sigma$, acts by zero.

Note that $T(z)$ will fail to exist in the chiral algebra and
therefore $S(z)$ will act by zero, only at the quantum level,
i.e., $T(z)$ and $S(z)$ still exist as local fields of spin two in
the $\overline Q_+$-cohomology of the sigma-model at the classical
level. To substantiate this statement, first recall from section
2.3 that $[\overline Q _+, T(z)] = 0$ classically in the absence
of quantum corrections to the action of $\overline Q_+$. Next,
note that the integer $2$ in the factor $(k+2)$ of the expression
$S(z)$ in (\ref{S(z)}), is due to a shift by $h^{\vee}=2$ in the
level $k$ because of quantum renormalisation effects \cite{Fuchs},
i.e., the classical expression of $S(z)$ for a general level $k$
can actually be written as \be S(z) = k T(z), \label{classical
S(z)} \ee and therefore, one will also have $[\overline Q_+, S(z)]
= 0$ in the classical theory. Moreover, since in our case, we
actually have $S(z) = -2 T(z)$ in the classical theory, it will
also be true that under quantum corrections of the action of
$\overline Q_+$, we will have \be [\overline Q_+, S_{zz}] = -2
\partial_z(R_{i \bar j} \partial_z \phi^i \psi^{\bar j}).
\label{corrections} \ee This corresponds, in the Cech cohomology
picture, to the expression \be {\widetilde{\widehat S}(z)} -
{\widehat S}(z) = -2
\partial_z({{\partial_z \gamma}\over{\gamma}}),
\ee which means that ${\widehat S}(z)$, the Cech cohomology
counterpart to the $S(z)$ operator, fails to be in $H^0(\mathbb
{CP}^1, {\widehat {\cal O}}^{ch}_{\mathbb P^1})$. This is again
consistent with the fact that $S(z)$ does not belong in the
quantum chiral algebra of the sigma-model, but rather, $S(z)$
belongs in its $\it{classical}$ chiral algebra. In other words,
one can always represent $S(z)$ by a classical $c$-number. This
point will be important when we discuss how one can define Hecke
eigensheaves that will correspond to flat $^LG$-bundles on a
Riemann surface $\Sigma$ in our physical interpretation of the
geometric Langlands correspondence for $G=SL(2)$.

The fact that $S(z)$ fails to correspond to any element in
$H^0(\mathbb {CP}^1, {\widehat {\cal O}}^{ch}_{\mathbb P^1})$
means that it will act trivially in any OPE with other field
operators. This in turn implies that its Laurent modes will
commute with the Laurent modes of any other existing operator; in
particular, the Laurent modes of $S(z)$ will commute with the
Laurent modes of the currents $J_+(z)$, $J_-(z)$ and $J_3(z)$ - in
other words, the Laurent modes of $S(z)$ will generate the centre
${\frak z}(\widehat{\frak{sl}}_2)$ of the completed universal
enveloping algebra of the affine $SL(2)$ algebra
$\widehat{\frak{sl}}_2$ at the critical level $k=-2$ (spanned by
the Laurent modes of $J_+(z)$, $J_-(z)$ and $J_3(z)$ in the
quantum chiral algebra of the twisted sigma-model on
$SL(2)/B$).\footnote{Recall that $S(z)$ is constructed out of the
currents of the affine $SL(2)$ algebra by using the invariant
tensors of the corresponding Lie algebra usually employed to
define higher-order Casimir invariants. Consequently, its Laurent
modes will span not the centre of the affine algebra, but rather
the centre of the completed universal enveloping algebra of the
affine algebra.} Last but not least, notice that $S(z)$ is also
$\psi^{\bar j}$-independent and must therefore be purely bosonic
in nature. In other words, $S(z)$ exist only in the
$\it{classical}$ holomorphic chiral algebra of the $\it{purely}$
$\it{bosonic}$ (or $\psi^{\bar j}$-independent) sector of the
twisted sigma-model.

Note that since $S(z)$ is a classical field, ${\frak
z}(\widehat{\frak{sl}}_2)$, which is generated by its Laurent
modes, must also be classical in nature. This statement can be
further substantiated as follows. Firstly, note that since $S(z)$
is holomorphic in $z$ and is of conformal weight two, one can
expand it in terms of a Laurent expansion as \be S(z) = \sum_{n
\in {\mathbb Z}} {\hat S}_n z^{-n-2}.\label{S(z) expansion}\ee Let
us begin with the general case of $k \neq -h^{\vee}$ for any
affine algebra $\widehat {\frak g}$, whereby a quantum definition
of $S(z)$ exists, so that the ${\hat S}_n$ modes of its Laurent
expansion can be related to the $J^a_n$ modes of the currents of
$\widehat{\frak g}$ through the quantum commutator relations
\begin{eqnarray}
[{\hat S}_n, J^a_m] & = & -(k + h^{\vee}) m J^a_{n+m},\\
{[{\hat S}_n, {\hat S}_m ]} & = & (k+ h^{\vee}) \left( (n-m) {\hat
S}_{n+m} + {k \over 12} \ \textrm{dim}\ {\frak g}\ (n^3-n) \
\delta_{n, -m}\right), \label{S_n}
\end{eqnarray}
where $a = 1, 2, \dots, \textrm{dim}{\frak g}$. If we now let $k=-
h^{\vee}$ and $\frak g = {\frak {sl}_2}$, we will have $[{\hat
S}_n, J^a_m] = [{\hat S}_n, {\hat S}_m] =0$. Hence, one can define
simultaneous eigenstates of the ${\hat S}_n$ and $J^a_n$ mode
operators. In particular, one would be able to properly define a
general state ${\Psi} = {\hat S}_{-l} {\hat S}_{-q} \dots {\hat
S}_{-p} |0, \alpha \rangle$, where $| 0, \alpha \rangle$ is a
vacuum state associated to a representation of $\frak {sl}_2$
labelled by $\alpha$, such that $J^a_0 | 0, \alpha \rangle =
\alpha^a | 0, \alpha \rangle$. However, note that any $\Psi$ will
correspond to a null-state, i.e., $\Psi$ decouples from the real,
physical Hilbert space of quantum states spanned by the
representations of $\frak {sl}_2$ \cite{lindstrom}. This means
that the ${\hat S}_{m} $'s which generate ${\frak
z}(\widehat{\frak{sl}}_2)$ cannot exist as quantum mode operators.
Hence, ${\frak z}(\widehat{\frak{sl}}_2)$ must be a classical
algebra.

\bigskip\noindent{\it A Classical Virasoro Algebra}

Since we now understand that $S(z)$ must be a holomorphic
classical field at $k =-2$, let us rewrite, for interpretive
clarity, the Laurent expansion of $S(z)$ as \be S(z) = \sum_{n \in
{\mathbb Z}} S_n z^{-n-2}, \label{S(z) classical}\ee so as to
differentiate the classical modes of expansion $S_n$ from their
quantum counterpart ${\hat S}_n$ in (\ref{S(z) expansion}). Unlike
the ${\hat S}_n$'s which obey the quantum commutator relations in
(\ref{S_n}) for an arbitrary level $k \neq -2$, the $S_n$'s, being
the modes of a Laurent expansion of a classical field, will
instead obey Poisson bracket relations that define a certain
classical algebra at $k=-2$.

Based on our arguments thus far, we see that the quantum version
of $S(z)$ as expressed in (\ref{S(z) expansion}), must reduce to
its classical counterpart as expressed in (\ref{S(z) classical}),
when $k \to -2$. In other words, one can see that by taking $(k+2)
\to 0$, we are going to the classical limit of this operator. This
is analogous to taking the limit ${\hbar} \to 0$ in any quantum
mechanical theory so as to obtain its classical counterpart. In
fact, by identifying $(k+ h^{\vee})$ or in this case $(k+2)$ with
$i \hbar$, and by noting that one must make the replacement from
Possion brackets to commutators via $\{S_n, S_m \}_{P.B.}
\rightarrow {1\over {i \hbar}} [ {\hat S}_n, {\hat S}_m ]$  in
quantising the $S_n$'s into operators, we can ascertain the
classical algebra generated by the $S_n$'s from (\ref{S_n}) as \be
\{S_n, S_m\}_{P.B.} =  (n-m) {S}_{n+m} - {6 \over 12}\ (n^3-n) \
\delta_{n, -m}. \label{poisson brackets of S_n} \ee Since we have
the classical relation $S(z) \sim T(z)$, it means that we can
interpret the $S_n$ modes as the Virasoro modes of the Laurent
expansion of the classical stress tensor field $T(z)$. In other
words, the $S_n$'s span a classical Virasoro algebra with central
charge $-6$ as given by (\ref{poisson brackets of S_n}). This is
sometimes denoted as the Virasoro Poisson algebra $Sym'(vir_{-6})$
in the mathematical literature \cite{Frenkel}. Hence, we have the
identification ${\frak z}(\widehat{\frak{sl}}_2) \simeq
Sym'(vir_{-6})$

\newsubsection{A Gauged WZW Model and the Geometric Langlands Correspondence for $G=SL(2)$}

Let us now seek a dual description of the above classical,
holomorphic chiral algebra of the twisted sigma-model on $SL(2)/B$
spanned by $S(z)$. To this end, let us first generalise the action
of the twisted sigma-model by making the replacement $g_{i \bar j}
\to g_{i \bar j} + b_{i \bar j}$ in $V$ of $S_{\textrm twist}$ in
(\ref{Stwist}), where $b_{i \bar j}$ is a $(1,1)$-form on the
target space $X$ associated to a B-field. This just adds to
$S_{\textrm{twist}}$ a $\it{cohomologically}$-$\it{trivial}$
$\overline Q_+$-exact term $\{\overline Q_+, - b_{i \bar j}
\psi^i_{\bar z}
\partial_z \phi^{\bar j} \}$, and does nothing to change our above
discussions about the classical chiral algebra of the sigma-model.
This generalised action can be explicitly written as
\begin{eqnarray}
S_{\mathrm gen}& = & \int_{\Sigma} |d^2z| \  (g_{i{\bar j}} + b_{i
\bar j})(
\partial_z \phi^{\bar j} \partial_{\bar z}\phi^i) + g_{i \bar j}
\psi_{\bar z}^i {D}_z \psi^{\bar j} + b_{i \bar j} \psi_{\bar z}^i
{\partial}_z \psi^{\bar j} + b_{{i \bar l},{\bar j}} \psi^i_{\bar
z}\partial _z \phi^{\bar l} \psi^{\bar j}. \label{actiongen}
\end{eqnarray}

Now recall that $S(z)$ exists in the classical holomorphic chiral
algebra of the $\psi^{\bar j}$-independent, purely bosonic sector
of the twisted sigma-model on $SL(2)/B$. This means that in order
for one to ascertain the dual description of $S(z)$, it suffices
to confine oneself to the study of the holomorphic chiral algebra
of the $\psi^{\bar j}$-independent sector of the twisted
sigma-model on $SL(2)/B$. The purely bosonic, $\psi^{\bar
j}$-independent specialisation of $S_{\textrm{equiv}}$, which
describes this particular sector of interest, can be written as
\be S_{\textrm{bosonic}} = \int_{\Sigma} |d^2z| \ (g_{i\bar j} +
b_{i \bar j}) \partial_{\bar z} \phi^i \partial_z \phi^{\bar
j}.\label{Sbosonic explicit} \ee Notice that
$S_{\textrm{bosonic}}$ just describes a free bosonic string which
propagates in an $SL(2)/B$ target-space. Hence, one can actually
describe the holomorphic chiral algebra associated to the
$\psi^{\bar j}$-independent sector of the twisted sigma-model on
$SL(2)/B$ in terms of the $\it{holomorphic}$ BRST-cohomology (or
chiral algebra) of a $B$-gauged WZW model on
$SL(2)$.\footnote{Note that a non-linear sigma-model on any
homogenous coset space such as $G/H$, will be described by an
asymmetrically $H$-gauged WZW model on $G$ associated with the
action $g \to gh^{-1}$, where $g \in G$ and $h \in H$. However,
note that the BRST-cohomology of an asymmetrically $H$-gauged WZW
model on $G$ coincides exactly with the $\it{holomorphic}$ (i.e.,
purely left-moving) sector of the BRST-cohomology of a
$\it{symmetrically}$ $H$-gauged WZW model on $G$ that is genuinely
gauge-invariant, and that which we are thus considering in this
paper. In other words, at the level of the holomorphic chiral
algebra, a physically equivalent description of the $\psi^{\bar
j}$-independent, non-supersymmetric sector of the twisted
sigma-model on $SL(2)/B$, will be given by a $B$-gauged WZW model
on $SL(2)$ that is gauge-invariant on the worldsheet. This
argument applies for any $G= SL(N)$ and $H=B$ as well.} In other
words, $S(z)$ should correspond to an observable in the classical
holomorphic BRST-cohomology of the $B$-gauged WZW model on
$SL(2)$.

Note that what would be relevant to all our later discussions is
the classical, holomorphic chiral algebra of the $\psi^{\bar
j}$-independent, non-supersymmetric sector of the twisted
sigma-model on $X= SL(N)/B$, for any $N \geq 3$. Note also that
the above arguments would apply for all $X = SL(N)/B$. As such,
let us now proceed to describe the $B$-gauged WZW model on any
$SL(N)$ in greater detail.\footnote{It may be disconcerting to
some readers that the Borel subgroup of $SL(N, \bf{C})$ which we
are gauging the $SL(N, \bf{C})$ WZW model by, is non-compact in
general. Apart from citing several well-known examples in the
physical literature \cite{review 23, review 24, sourdis,O&B, ref
for gauged WZW} that have done likewise to consider non-compact
WZW models gauged to non-compact (sometimes Borel) subgroups, one
can also argue that our model is actually equivalent - within our
context - to a physically consistent model which gauges a
$\it{compact}$ subgroup instead. Firstly, note that for a complex
flag manifold $SL(N, \bf {C})$, we have the relation ${SL(N,
\bf{C})} /B = SU(N) / C(T)$, where $C(T)$ is the centralizer of
the torus of $SU(N)$ spanned by purely diagonal matrices in
$SU(N)$ \cite{intro to lie groups} - in other words, $C(T)$ is an
anomaly-free, $\it{compact}$ diagonal subgroup in the context of a
$C(T)$-gauged WZW model on $SU(N)$. Secondly, note that the OPE
algebras of the affine algebras $\widehat{\frak{su}}_N$ and
$\widehat{\frak{sl}}_N$ are the same. Together with the previous
footnote, these two points imply that the $B$-gauged WZW model on
$SL(N, \bf{C})$ and the $C(T)$-gauged WZW model on $SU(N)$ (which
can be physically consistently defined, and whose gauge group is
also compact), are equivalent at the level of their holomorphic
BRST-cohomologies. However, since one of our main aims is to
relate the gauged WZW model to the algebraic DS-reduction scheme
in $\S$3, we want to consider the B-gauged WZW model on
$SL(N,\bf{C})$. Last but not least, note that we will ultimately
be interested in the $\it{classical}$ spectrum of the gauged WZW
model only, whereby the compactness or non-compactness of the
gauge group will be irrelevant.}

\bigskip\noindent{\it The $B$-Gauged WZW Model on $SL(N)$}

 First, note that the action of a general WZW model can be written as
 \be S_{\textrm{WZ}}(g) = { k ' \over {4 \pi}} \int_{\Sigma} d^2z \  \textrm{Tr}
(\partial_{z} g^{-1} \partial_{\bar z} g)  + { i k ' \over {24
\pi}} \int_{B; \partial B = \Sigma} d^3 x \ \textrm{Tr} (g^{-1} d
g)^3, \label{WZW action}
 \ee
where $k'$ is the level, and $g$ is a worldsheet scalar field
valued in any simple, maximally non-compact, connected Lie group
$G$ (such as $SL(N, \bf{C})$ which we are considering in this
paper), that is also periodic along one of the worldsheet
directions with period $2 \pi$. The trace $\textrm{Tr}$ is the
usual matrix trace in the defining representation of $G$.

A gauged version of (\ref{WZW action}) can be written as
\begin{eqnarray}
S_{\textrm{gauged}} (g, A_z, A_{\bar z}) & = & S_{\textrm{WZ}} (g)  + {k' \over {2\pi}}\int_{\Sigma} d^2z \  \textrm{Tr}  [ A_z (\partial_{\bar z} g g^{-1} + {\bar M}) -  A_{\bar z}(g^{-1} \partial_z g + {M})   \nonumber \\
&& \hspace{3.0cm}  + A_z g A_{\bar z} g^{-1} - A_z A_{\bar z}],
\label{gauged WZW action}
\end{eqnarray}
where the worldsheet one-form gauge field $A= A_z dz + A_{\bar z}
d{\bar z}$ is valued in $\frak h$, the Lie algebra of a subgroup
$H$ of $G$. Notice that $S_{\textrm{gauged}} (g, A_z, A_{\bar z})$
differs slightly from the standard form of a gauged WZW model
commonly found in the physical literature - additional $\bar M$
and $M$ constant matrices have been incorporated in the
$\partial_{\bar z} g g^{-1}$ and $g^{-1} \partial_z g $ terms of
the standard action, so that one can later use them to derive the
correct form of the holomorphic stress tensor without reference to
a coset formalism. Setting $\bar M$ and $M$ to the zero matrices
simply takes us back to the standard action for the gauged WZW
model. As required, $S_{\textrm{gauged}} (g, A_z, A_{\bar z})$ is
invariant under the standard (chiral) local gauge transformations
\be g \to hgh^{-1}; \ \ \ A_z \to
\partial_z h \cdot h^{-1} + h A_z h^{-1}; \ \ \ A_{\bar z} \to
\partial_{\bar z} h \cdot h^{-1} + h A_{\bar z} h^{-1},
\label{gauge tx} \ee where $h = e^{\lambda (z, \bar z)} \in H$ for
any $\lambda (z, \bar z) \in {\frak h}$.\footnote{A similar model
has been considered in \cite{ref for gauged WZW}. However, the
action in that context is instead invariant under a
$\it{non}$-$\it{chiral}$ local gauge transformation. Moreover, it
does not contain the $A_zA_{\bar z}$ term present in a standard
gauged WZW model.} The invariance of (\ref{gauged WZW action})
under the gauge transformations in (\ref{gauge tx}) can be
verified as follows. Firstly, note that the $\bar
M(M)$-independent terms make up the usual Lagrangian for the
standard gauged WZW action, which is certainly invariant under the
gauge transformations of (\ref{gauge tx}). Next, note that under
an infinitesimal gauge transformation $h \simeq 1+ \lambda$, the
terms $\textrm{Tr} (A_{z} \ \bar M)$ and $\textrm{Tr} (A_{\bar z}
\ M)$ change as
\begin{eqnarray} \label{variation Tr 1}\delta \textrm{Tr} (A_{z}
\ \bar M) & = & \textrm{Tr} (\partial_{z} \lambda \
\bar M) - \textrm{Tr} (\bar M \ [\lambda, A_z]),  \\
\delta \textrm{Tr} (A_{\bar z} \ M) & = & \textrm{Tr}
(\partial_{\bar z} \lambda \ M) - \textrm{Tr} (M \ [\lambda,
A_{\bar z}]). \label{variation Tr} \end{eqnarray} Since we will be
considering the case where $H$ is the Borel subgroup of $G$ and
therefore, $\lambda$ and $A$ will be valued in the Lie algebra of
a maximally $\it{solvable}$ (Borel) subgroup of $G$, the second
term on the R.H.S. of (\ref{variation Tr 1}) and (\ref{variation
Tr}) will be zero \cite{ref for gauged WZW}. What  remains are
total divergence terms that will vanish upon integration on
$\Sigma$ because it is a worldsheet with no boundaries. Therefore,
unless $H$ is a Borel subgroup of $G$ (or any other solvable
subgroup of $G$), one cannot incorporate $\bar M$ and $M$ in the
action and still maintain the requisite gauge invariance. This
explains why generalisations of gauged WZW models with these
constant matrices $\bar M$ and $M$ have not appeared much in the
physical literature. Nevertheless, this generalisation can be
considered in our case. As we shall see shortly, this
generalisation will allow us to obtain the correct form of the
holomorphic stress tensor of the $B$-gauged WZW model on $SL(N)$
without any explicit reference to a coset formalism.

The classical equations of motion that follow from the field
variations in (\ref{gauge tx}) are
\begin{eqnarray}
\label{1}
\delta A_z & : & D_{\bar z} g g^{-1}|_H = -M_+, \\
\label{2}
\delta A_{\bar z} & : & g^{-1} D_z g |_H = -M-, \\
\label{3}
\delta g & : & D_{\bar z} (g^{-1} D_z g ) = F_{z \bar z}, \\
\label{4} \delta g & : & D_{z} (D_{\bar z} g g^{-1}) = F_{\bar z
z},
\end{eqnarray}
where $F_{z \bar z} = \partial_z A_{\bar z} - \partial_{\bar z}
A_z + [A_z, A_{\bar z}]$ and $F_{\bar z z} = \partial_{\bar z} A_z
- \partial_{z} A_{\bar z} + [A_{\bar z}, A_{z}]$ are the
non-vanishing components of the field strength, and the covariant
derivatives are given by $D_z =
\partial_z + [A_z, \ ]$ and $D_{\bar z} = \partial_{\bar z} +
[A_{\bar z}, \ ]$. By imposing the condition of (\ref{2}) in
(\ref{3}), and by imposing the condition of (\ref{1}) in
(\ref{4}), since $M_{\pm}$ are constant matrices, we find that we
have the zero curvature condition $F_{z \bar z} = F_{\bar z z} =
0$ as expected of a non-dynamically gauged WZW model. This means
that $A_z$ and $A_{\bar z}$ are trivial on-shell. One is then free
to use the gauge invariance to set $A_z$ and/or $A_{\bar z}$ to a
constant such as zero. In setting $A_z = A_{\bar z} =0$ in
(\ref{3}) and (\ref{4}), noting that $F_{z \bar z}= F_{\bar z z} =
0$, we have the relations \be
\partial_{\bar z} (g^{-1} \partial_z g ) = 0 \qquad \textrm{and} \qquad \partial_{z} (\partial_{\bar z} g g^{-1}) =
0. \label{con} \ee In other words, we have a $\frak g$-valued,
holomorphic conserved current $J(z) = g^{-1} \partial_z g$, and a
$\frak g$-valued antiholomorphic conserved current $\bar J(\bar z)
= \partial_{\bar z} g g^{-1}$, both of which are dimension one and
generate affine symmetries on $\Sigma$. The action in (\ref{gauged
WZW action}) can thus be written as \begin{eqnarray} \label{gauged
WZW action simplified}
S_{\textrm{gauged}} (g, A_z, A_{\bar z}) & = & S_{\textrm{WZ}} (g)  + {k' \over {2\pi}}\int_{\Sigma} d^2z \  \textrm{Tr}  [ A_z ({\bar J}(\bar z) + \bar M) -  A_{\bar z}(J(z) + M)   \nonumber \\
&& \hspace{3.0cm}  + A_z g A_{\bar z} g^{-1} - A_z A_{\bar z}],
\end{eqnarray}

For our case where $H$ is a Borel subgroup $B$ of $G$, one can
further simplify (\ref{gauged WZW action simplified}) as follows.
Firstly, since $G$ is a connected simple group, it will have a
simple Lie algebra $\frak g$. As such, $\frak g$ will have a
Cartan decomposition ${\frak g} = {\frak n}_- \oplus {\frak c}
\oplus {\frak n}_+$, where $\frak c$ is the Cartan subalgebra, and
${\frak n}_{\pm}$ are the nilpotent subalgebras of the the upper
and lower triangular matrices of $G$. The Borel subalgebras will
then be given by ${\frak b}_{\pm} = {\frak c} \oplus {\frak
n}_{\pm}$, and they correspond to the Borel subgroups $B_{\pm}$.
For the complex flag manifolds that we will be considering in this
paper, $B_+$ will be the Borel subgroup of interest. $B$ will
henceforth mean $B_+$ in all of our proceeding discussions. With
respect to this decomposition of the Lie algebra of $G$, we can
write $ J(z) = \sum_{a=1}^{\textrm{dim}{{\frak n}_-}} J^a_- (z)
t^{-}_a + \sum_{a=1}^{\textrm{dim}{{\frak c}}} J^a_c(z) t^{c}_a +
\sum_{a=1}^{\textrm{dim}{{\frak n}_+}} J^a_+(z) t^{+}_a$, and $
{\bar J}(\bar z) = \sum_{a=1}^{\textrm{dim}{{\frak n}_-}} {\bar
J}^a_- (\bar z) t^{-}_a + \sum_{a=1}^{\textrm{dim}{{\frak c}}}
{\bar J}^a_c(\bar z) t^{c}_a + \sum_{a=1}^{\textrm{dim}{{\frak
n}_+}} {\bar J}^a_+(\bar z) t^{+}_a$, where $t^{-}_a \in {\frak
n}_-$, $t^{c}_a \in {\frak c}$, and $t^{+}_a \in {\frak n}_+$. One
can also write $M = \sum_{a=1}^{\textrm{dim}{{\frak n}_-}} M^a_-
t^{-}_a + \sum_{a=1}^{\textrm{dim}{{\frak c}}} M^a_c t^{c}_a  +
\sum_{a=1}^{\textrm{dim}{{\frak n}_+}} M^a_+ t^{+}_a$, and $\bar M
= \sum_{a=1}^{\textrm{dim}{{\frak n}_-}} {\bar M}^a_- t^{-}_a +
\sum_{a=1}^{\textrm{dim}{{\frak c}}} {\bar M}^a_c t^{c}_a  +
\sum_{a=1}^{\textrm{dim}{{\frak n}_+}} {\bar M}^a_+ t^{+}_a$,
where $M^a_{\pm ; c}$(${\bar M}^a_{\pm ; c}$) are arbitrary number
constants. Next, note that $H=B$, and $B \simeq N_+$, where $N_+ =
[B,B]$ is the subgroup of $G$ generated by its Lie algebra ${\frak
n}_+$ of strictly upper triangular matrices which are traceless,
i.e., for $t,t' \in {\frak n}_+$, we have $\textrm{Tr}_{L} (tt') -
\textrm{Tr}_{R}(t't) = 0$, where the trace $\textrm{Tr}_{L}$ and
$\textrm{Tr}_{R}$ are taken over some $L$ and $R$ representation
of $G$ respectively. In other words, $N_+$ is the non-anomalous
subgroup to be gauged, and we can write $A_{z} = \sum_{a
=1}^{\textrm{dim}{\frak n}_+} {\tilde A}_{z}^a t^+_a$, and
$A_{\bar z} =   \sum_{a =1}^{\textrm{dim}{\frak n}_+} {\tilde
A}_{\bar z}^a t^+_a$. Next, note that since $\textrm{Tr}
(t^{\alpha}_a t^{\beta}_b) = \delta_{a,b}\delta^{\alpha, \beta}$,
the trace of the second term on the R.H.S. of (\ref{gauged WZW
action simplified}) will be non-vanishing only  for components of
$J(z)$($\bar J(\bar z)$) and $M$($\bar M$) that are associated to
their expansion in ${\frak n}_+$. Let us denote ${J}^+(z) =
\sum_{a=1}^{\textrm{dim}{\frak n}_+} {J}^a_+ (z) t^{+}_a$ and
${M}^+ = \sum_{a=1}^{\textrm{dim} {{\frak n}_+}} {M}^a_+ t^{+}_a$.
Let us also denote ${\bar J}^+(\bar z) =
\sum_{a=1}^{\textrm{dim}{\frak n}_+} {\bar J}^a_+ (\bar z)
t^{+}_a$ and ${\bar M}^+ = \sum_{a=1}^{\textrm{dim} {{\frak n}_+}}
{\bar M}^a_+ t^{+}_a$. Then, one can write the action of a
$B$-gauged WZW model on $G = SL(N)$ as
\begin{eqnarray} S_{\textrm{B-gauged}} (g, A_{z}, A_{\bar z}, J^+,
{\bar J}^+)& = & S_{\textrm{WZ}} (g) - {k' \over {2\pi}}
\int_{\Sigma} d^2z \ \textrm{Tr} [ A_{\bar z}( J^+(z) + {M}^+)-
A_{z} ( {\bar J}^+(\bar z) + {\bar M}^+)\nonumber \\ &&
\hspace{4.0cm} - A_z g A_{\bar z} g^{-1} + A_z A_{\bar z}].
\label{B-gauged WZW action SL(N)}
\end{eqnarray}

\smallskip\noindent{\it The $B$-Gauged WZW Model on $SL(2)$}

Now that we have derived the action $S_{\textrm{B-gauged}} (g,
A_z, A_{\bar z}, J^+, {\bar J}^+)$ of a $B$-gauged WZW model on
$\it{any}$ $SL(N)$, we will proceed to specialise to the case
where $G=SL(2)$. In this case, $\textrm{dim}\ {\frak n}_- =
\textrm{dim} \ {\frak c} = \textrm{dim} \ {\frak n}_+ = 1$, and so
$J^+(z) = J^1_+(z) t^+_1$, $\bar J^+(z) = \bar J^1_+(z) t^+_1$,
$M^+ = M^1_+ t^{+}_1$, $\bar M^+ = \bar M^1_+ t^{+}_1$, $A_{z}=
{\tilde A}^1_{z} (z) t^+_1$ and $A_{\bar z}= {\tilde A}^1_{\bar z}
t^+_1$. The gauged WZW action is then given by \begin{eqnarray}
S_{{SL(2)}} (g, A_z, A_{\bar z}, J^+, \bar J^+) & =&
S_{\textrm{WZ}} (g) - {k' \over {2\pi}} \int_{\Sigma} d^2z \
{\tilde A}^1_{\bar z} ( J^1_+(z) + {M}^1_+) - {\tilde A}^1_{z}(
\bar J^1_+(z) + {\bar M}^1_+) \nonumber \\
&& \hspace{3.5cm} - {\tilde A}^1_z g {\tilde A}^1_{\bar z} g^{-1}
+ {\tilde A}^1_z {\tilde A}^1_{\bar z}. \label{B-gauged WZW action
general}
\end{eqnarray}

Due to the $B$-gauge invariance of the theory, we must divide the
measure in any path integral computation by the volume of the
$B$-gauge symmetry. That is, the partition function has to take
the form \be Z_{SL(2)} = \int_{\Sigma} { {[g^{-1}dg, d{\tilde
A}^1_{z}, d{\tilde A}^1_{\bar z}]} \over {(\textrm{gauge
volume})}} \ \textrm{exp} \left(i S_{SL(2)}(g, A_z, A_{\bar z},
J^+, \bar J^+) \right). \ee One must now fix this gauge invariance
to eliminate the non-unique degrees of freedom. One can do this by
employing the BRST formalism which requires the introduction of
Faddev-Popov ghost fields.

In order to obtain the $\it{holomorphic}$ BRST transformations of
the fields, one simply replaces the position-dependent
infinitesimal gauge parameter $\epsilon$ of $h= B= \textrm{exp}(-
\epsilon t^+_1)$ in the corresponding $\textrm{\it left-sector}$
of the gauge transformations in (\ref{gauge tx}) with the ghost
field $c$, which then gives us \be\delta_{\textrm{BRST}}(g) = - c
t^+_1 g, \quad \delta_{\textrm{BRST}}({\tilde A}^1_{\bar z}) = -
\partial_{\bar z} c, \quad \delta_{\textrm{BRST}}(\textrm{others}) =0. \label{BRST tx}\ee The ghost field $c$ and its
anti-ghost partner $b$ will transform as \be
\delta_{\textrm{BRST}} (c) = 0, \quad \delta_{\textrm{BRST}}(b) =
{\tilde B}, \quad \delta_{\textrm{BRST}} {(\tilde B)} = 0. \ee In
the above, $\tilde B$ is the Nakanishi-Lautrup auxiliary field
that is the holomorphic BRST transform of $b$. It also serves as a
Lagrange multiplier to impose the gauge-fixing condition.

In order to obtain the $\it{antiholomorphic}$ BRST transformations
of the fields, one employs the same recipe to the corresponding
$\textrm{\it right-sector}$ of the gauge transformations in
(\ref{gauge tx}) with the infinitesimal position-dependent gauge
parameter now replaced by the ghost field $\bar c$, which then
gives us \be \bar \delta_{\textrm{BRST}}(g) = \bar c t^+_1 g,
\quad \bar \delta_{\textrm{BRST}}({\tilde A}^1_{z}) = -
\partial_{z} \bar c, \quad \bar\delta_{\textrm{BRST}}(\textrm{others}) =0.\label{BRST tx 1}\ee The ghost field $\bar c$ and its
anti-ghost partner $\bar b$ will transform as \be
\bar\delta_{\textrm{BRST}} (\bar c) = 0, \quad
\bar\delta_{\textrm{BRST}}(\bar b) = {\tilde {\bar B}}, \quad
\bar\delta_{\textrm{BRST}} {(\tilde {\bar B})} = 0. \ee In the
above, $\tilde {\bar B}$ is the Nakanishi-Lautrup auxiliary field
that is the antiholomorphic BRST transform of $\bar b$. It also
serves as a Lagrange multiplier to impose the gauge-fixing
condition.

Since the BRST transformations in (\ref{BRST tx}) and (\ref{BRST
tx 1}) are just infinitesimal versions of the gauge
transformations in (\ref{gauge tx}), $S_{{SL(2)}} (g, A_z, A_{\bar
z}, J^+, \bar J^+)$ will be invariant under them. An important
point to note at this juncture is that in addition to
$(\delta_{\textrm{BRST}} + \bar\delta_{\textrm{BRST}}) \cdot
(\delta_{\textrm{BRST}} + \bar\delta_{\textrm{BRST}}) = 0$, the
holomorphic and antiholomorphic BRST-variations are also
separately nilpotent, i.e., $\delta^2_{\textrm{BRST}} = 0$ and
$\bar\delta^2_{\textrm{BRST}}=0$. Moreover,
$\delta_{\textrm{BRST}}\cdot \bar\delta_{\textrm{BRST}}= - \bar
\delta_{\textrm{BRST}} \cdot \delta_{\textrm{BRST}}$. In fact, one
can easily inspect this from the field variations themselves. This
means that the BRST-cohomology of the $B$-gauged WZW model on
$SL(2)$ can be decomposed into $\it{independent}$ holomorphic and
antiholomorphic sectors that are just complex conjugate of each
other, and that it can be computed via a spectral sequence,
whereby the first two complexes will be furnished by its
holomorphic and antiholomorphic BRST-cohomologies respectively.
Since we will only be interested in the holomorphic chiral algebra
of the $B$-gauged WZW model on $SL(2)$ (which as mentioned, is
just identical to its antiholomorphic chiral algebra by a complex
conjugation), we shall henceforth focus on the $\it{holomorphic}$
BRST-cohomology of the $B$-gauged WZW model on $SL(2)$ (as well as
for all other cases of $SL(N)$ in this paper, since this
observation of a polarisation of the BRST-cohomology will be true
of any $B$-gauged WZW model on $SL(N)$ as we will see.)

By the usual recipe of the BRST formalism, one can fix the gauge
by adding to the BRST-invariant action $S_{{SL(2)}} (g, A_z,
A_{\bar z}, J^+, \bar J^+)$, a BRST-exact term. Since the BRST
transformation by $(\delta_{\textrm{BRST}} + \bar
\delta_{\textrm{BRST}})$ is nilpotent, the new total action will
still be BRST-invariant as required. The choice of the BRST-exact
operator will then define the gauge-fixing condition. A consistent
choice of the BRST-exact operator will give us the requisite
action for the ghost and anti-ghost fields - note that with \be
S_{{SL(2)}} (g, A_z, A_{\bar z}, J^+, \bar J^+) +
(\delta_{\textrm{BRST}} + \bar \delta_{\textrm{BRST}})
\left({k'\over 2\pi} \int_{\Sigma} d^2 z \ {\tilde A}^1_{\bar z} b
+ {\tilde A}^1_{z} \bar b \right),\nonumber \ee one will indeed
have the desired total action, which can be written as
\begin{eqnarray} \label{total desired action for SL(2)} S_{\textrm{WZW}}(g) + {k'\over {2\pi}}\int_{\Sigma} d^2z \ c
\partial_{\bar z} b + \bar c\partial_{z} \bar b - {k'\over
{2\pi}}\int_{\Sigma} d^2z \ \tilde A^1_{\bar z}(J^1_+(z) + M^1_+ -
{\tilde B}) - \tilde A^1_{z}(\bar J^1_+(z) + \bar M^1_+ + {\tilde {\bar B}})\nonumber \\
\hspace{-4.0cm}- {\tilde A}^1_z g {\tilde A}^1_{\bar z} g^{-1} +
{\tilde A}^1_z {\tilde A}^1_{\bar z}. \nonumber \\ \end{eqnarray}
From the equation of motion by varying $\tilde B$, we have the
condition $\tilde A^1_{\bar z} = 0$. From the equation of motion
by varying $\tilde {\bar B}$, we have the condition $\tilde
A^1_{z} = 0$. Together with the equation of motion by varying
${\tilde A}^1_{\bar z}$ and ${\tilde A}^1_{z}$, we have, by
integrating out $\tilde A^1_{\bar z}$ and $\tilde A^1_{z}$ in
(\ref{total desired action for SL(2)}), the relations $J^1_+ +
M^1_+ = {\tilde B}$ and $\bar J^1_+ + \bar M^1_+ = -{\tilde {\bar
B}}$. Thus, the partition function of the $B$-gauged WZW model can
also be expressed as \be Z_{SL(2)} = \int [ g^{-1} dg, db, dc] \
\textrm{exp} \left ( iS_{\textrm{WZW}}(g) + {i k'\over
{2\pi}}\int_{\Sigma} d^2z \ c
\partial_{\bar z} b + \bar c
\partial_{z} \bar b \right), \label{Z_SL(2)} \ee whereby the $\it{holomorphic}$ BRST
variations of the fields that leave the effective action in
$Z_{SL(2)}$ above $\it{invariant}$ are now given by \be
\delta_{\textrm{BRST}} (g) = -ct^+_1 g, \quad
\delta_{\textrm{BRST}} (c) = 0, \quad \delta_{\textrm{BRST}}(b) =
 \left(J^1_+ + M^1_+ \right), \quad \delta_{\textrm{BRST}}(\textrm{others}) =0. \label{BRST variations SL(2)} \ee The holomorphic BRST charge which generates the above
transformations is therefore given by \be Q_{\textrm{BRST}} =
\oint {dz \over {2 \pi i}} \ (J^1_+(z) +
M^1_+)c(z).\label{Q_BRST,WZW SL(2)}\ee

\bigskip\noindent{\it The OPE's of the $B$-Gauged WZW Model on $SL(2)$}

Note that consistent with the presence of the dimension one
operators $J_{\pm}(z)$ and $J_3(z)$ in the holomorphic chiral
algebra of the purely bosonic, $\psi^{\bar j}$-independent sector
of the sigma-model on $SL(2)/B$, which, generate an affine $SL(2)$
OPE algebra, one also has, in the holomorphic BRST-cohomology of
the $B$-gauged WZW model on $SL(2)$, the dimension one currents
$J^1_{\pm}(z)$ and $J^1_{c}(z)$ which generate the following
affine $SL(2)$ OPE algebra:
\begin{eqnarray}
\label{OPE for SL(2) (first) }
{J}^1_c (z) {J}^1_{\pm} (z') & \sim & {{{\pm}{ J}^1_{\pm} (z')} \over z-z'},  \\
{J}^1_c (z) {J}^1_c (z') & \sim &  {{k'/2}\over (z-z')^2}, \\
\label{OPE for SL(2) (last) } J^1_+ (z) J^1_- (z') & \sim &
{2J^1_c\over z-z'} + {k' \over{(z-z')^2}}.
\end{eqnarray}
From standard field-theoretic considerations of the
ghost/anti-ghost kinetic term in the effective action of the
gauged WZW model in (\ref{Z_SL(2)}), one will also have the
following OPE (after  absorbing $k'$ by a trivially re-scaling of
the fields) \be b (z) c(z') \sim {1 \over {z-z'}}. \label{OPEs for
bc in SL(2) WZW} \ee However, the OPE's of the $b(z)$ and $c(z)$
fields with any current in (\ref{OPE for SL(2) (first)
})-(\ref{OPE for SL(2) (last) })are trivial. Finally, one can also
verify the nilpotency of $Q_{\textrm{BRST}}$ by using the OPE's in
(\ref{OPE for SL(2) (first) })-(\ref{OPE for SL(2) (last) }),
(\ref{OPEs for bc in SL(2) WZW}), and its explicit expression in
(\ref{Q_BRST,WZW SL(2)}) - the OPE of the BRST current  $(J^1_+(z)
+ M^1_+)c(z)$ with itself is regular. Moreover,  one can quickly
check using (\ref{OPEs for bc in SL(2) WZW}) that
$Q_{\textrm{BRST}}$ in (\ref{Q_BRST,WZW SL(2)}) will indeed
generate the correct field variations in (\ref{BRST variations
SL(2)}).

\bigskip\noindent{\it The Holomorphic Stress Tensor}

Though we did not make this obvious in our discussion above,
$\tilde B$ must actually vanish - by integrating out $\tilde
A^1_{\bar z}$ in (\ref{total desired action for SL(2)}) and using
the condition $\tilde A^1_{z} =0$, we find that we actually have
the relation $(J^1_+(z) + M^1_+) =0$. This relation (involving the
current associated to the Borel subalgebra $\frak b$ of the group
$B$ that we are modding out by) will lead us directly to the
correct form of the holomorphic stress tensor for the gauged WZW
model without reference to a coset formalism. Let us look at this
more closely.

Since we have an affine $SL(2)$ algebra from the OPE's (\ref{OPE
for SL(2) (first) })-(\ref{OPE for SL(2) (last) }), we can employ
the Sugawara formalism to construct the stress tensor associated
with the $S_{\textrm{WZW}}(g)$ part of the total action, from the
currents in (\ref{OPE for SL(2) (first) })-(\ref{OPE for SL(2)
(last) }). Taking into account the part of the total action
associated to the ghost and anti-ghost fields $c(z)$ and $b (z)$,
the stress tensor should be given by \be
 T_{\textrm{gauged}}(z) =  T_{SL(2)}(z)  +  \partial_z b(z) c(z),
\label{T1}
 \ee
where \be T_{SL(2)} (z) =  {{:d^{ab} J^1_a J^1_b (z):} \over
{(k'+2)}}, \ee and $d^{ab}$ is the inverse of the Cartan-Killing
metric of $\frak {sl}_2$. Note that with respect to
$T_{\textrm{gauged}}(z)$, the currents $J^1_{\pm}(z)$ and $J^1_c
(z)$ have conformal dimension one. This is inconsistent with the
condition $J^1_+(z) = -M^1_+$, as $M^1_+$ is a constant of
conformal dimension zero. This means that one must modify
$T_{\textrm{gauge}} (z)$ so that $J^1_+(z)$ will have conformal
dimension zero. An allowable modification involves adding to
$T_{\textrm{gauge}}(z)$ a term that has conformal dimension two. A
little thought will reveal that the total stress tensor must then
take the form \be T_{\textrm{total}}(z) = T_{SL(2)} (z)  +
\partial_z b(z) c(z) + \partial_zJ^1_c (z). \label{T_{total}
SL(2)} \ee A small computation shows that the BRST current
$(J^1_+(z) + M^1_+)c(z)$ has conformal dimension one under
$T_{\textrm{total}}(z)$, which then means that its
$Q_{\textrm{BRST}}$ charge is a conformal dimension zero scalar as
required. This in turn means that $c(z)$ and $b(z)$ must be of
conformal dimension one and zero respectively, i.e., the field
$b(z)$ and $c(z)$ is a scalar and (holomorphic) one-form on
$\Sigma$. Therefore, one should really rewrite $c(z)$ as  $c_z(z)$
in (\ref{Z_SL(2)}), (\ref{Q_BRST,WZW SL(2)}), (\ref{T1}) and
(\ref{T_{total} SL(2)}). In doing so, we find that these equations
are now fully consistent with regards to conformal dimensions.

Note also that because the BRST current is of conformal dimension
one with respect to the holomorphic stress tensor
$T_{\textrm{total}}(z)$, it must be annihilated by
$Q_{\textrm{BRST}}$; this means that $T_{\textrm{total}}(z)$  is
$Q_{\textrm{BRST}}$-closed. One can also verify that
$T_{\textrm{total}}(z)$ cannot be $Q_{\textrm{BRST}}$-exact, i.e.,
$T_{\textrm{total}}(z)$ lies in the holomorphic BRST-cohomology of
the $B$-gauged WZW model on $SL(2)$. Last but not least, a
soon-to-be relevant point to note is that since quantum
corrections can only annihilate classes in the BRST-cohomology and
not create them, the classical counterpart of the holomorphic
stress tensor $T_{\textrm{total}}(z)$ will be a spin-two field
$T_{\textrm{classical}}(z)$ which lies in the $\it{classical}$,
holomorphic BRST-cohomology (or holomorphic chiral algebra) of the
$B$-gauged WZW model on $SL(2)$, where $T_{\textrm{classical}}(z)$
will generate the classical Virasoro transformations on the
fields.

\bigskip\noindent{\it A Duality of Classical $\cal W$-algebras Underlying a Geometric Langlands Correspondence for $G=SL(2)$}

Note that the observable in the holomorphic BRST-cohomology of the
$B$-gauged WZW model on $SL(2)$ that will correspond to $S(z)$ of
the holomorphic chiral algebra of the purely bosonic, $\psi^{\bar
j}$-independent sector of the twisted sigma-model on $SL(2)/B$,
must have the same spin as $S(z)$. In addition, since $S(z)$
generates a classical Virasoro symmetry on the worldsheet, it will
mean that the corresponding observable ought to be a spin-two
field which exists in the $\it{classical}$, holomorphic
BRST-cohomology of the $B$-gauged WZW model on $SL(2)$, and which
also generates a classical Virasoro symmetry on the worldsheet.

In order to ascertain the classical observable which corresponds
to $S(z)$, first recall that the $\it{quantum}$ definition of
$S(z)$ at $k\neq -2$ is given by $S(z) = (k+2)T(z)$. Notice that
since the stress tensor $T(z)$ also exists in the holomorphic
chiral algebra of the purely bosonic, $\psi^{\bar j}$-independent
sector of the sigma-model on $SL(2)/B$, it will imply that $T(z)$
must correspond to the stress tensor $T_{\textrm{total}}(z)$ of
the $B$-gauged WZW model on $SL(2)$. Thus, at $k\neq -2$, $S(z)$
will correspond to ${\overline T}_{\textrm{total}}(z) =
(k+2)T_{\textrm{total}}(z)$, and at $k=-2$, $S(z)$ will correspond
to the classical counterpart ${\overline
T}_{\textrm{classical}}(z)$ of ${\overline
T}_{\textrm{total}}(z)$. Note that ${\overline
T}_{\textrm{classical}}(z)$ lies in the classical, holomorphic
BRST-cohomology of the $B$-gauged WZW model on $SL(2)$ as required
- at $k=-2$, ${\overline T}_{\textrm{total}}(z)$, which usually
exists as a quantum operator, will act by zero in its OPE's with
any other operator, i.e., it will reduce to its classical
counterpart ${\overline T}_{\textrm{classical}}(z)$ in the
classical, holomorphic BRST-cohomology of the gauged WZW model.
Moreover, since the shift in $2$ in the factor $(k+2)$ is due to a
quantum effect as explained earlier, $S(z)$ will actually
correspond to ${\overline T}_{\textrm{classical}}(z) = -
2T_{\textrm{classical}}(z)$ at $k=-2$. Hence, $S(z)$ will indeed
correspond to a spin-two field ${\overline
T}_{\textrm{classical}}(z)$ in the classical, holomorphic
BRST-cohomology of the $B$-gauged WZW on $SL(2)$ which generates a
classical Virasoro transformation of the fields.

What is the classical algebra generated by the Laurent modes of
${\overline T}_{\textrm{classical}}(z)$? To ascertain this, first
note that from the explicit form of $T_{\textrm{total}}(z)$ in
(\ref{T_{total} SL(2)}), we find that it has to have a central
charge of $c= 13 - {6 / {k'+2}} - 6(k'+2)$. Hence, the Virasoro
modes of $T_{\textrm{total}}(z) = \sum_{n} {\hat L}_n z^{-n-2}$
will obey the following commutator relation \be [{\hat L}_n, {\hat
L}_m ] = (n-m) {\hat L}_{n+m} + {1\over {12}} \left[ 13 - {6 \over
{k'+2}} - 6(k'+2) \right] (n^3 - n) \delta_{n, -m} \label{virasoro
relation for wzw sl(2)} \ee at the quantum level. Therefore, the
commutator relations involving the ${\hat{\overline L}}_n $ modes
of ${\overline T}_{\textrm{total}}(z) = \sum_{n} {\hat {\overline
L}}_n z^{-n-2}$ will be given by \be [{\hat {\overline L}}_n,
{\hat {\overline L}}_m] = (k+2) \left[ (n-m) {\hat{\overline
L}}_{n+m} + (n^3 - n) \delta_{n, -m} \left( {13 (k+2) \over{12}} -
{6 (k+2) \over{12(k'+2)}} - {6 (k+2) (k'+2) \over{12}}\right)
\right]. \label{virasoro relation for wzw sl(2) scaled} \ee At $k
= -2$, ${\overline T}_{\textrm{total}}(z)$ will cease to have a
quantum definition, and it will reduce to its classical
counterpart ${\overline T}_{\textrm{classical}}(z)$. Consequently,
the $k \to -2$ (and $k' \to \infty$) limit of the commutator
relation in (\ref{virasoro relation for wzw sl(2) scaled}), can be
interpreted as its classical limit. Therefore, one can view the
term $(k+2)$ in (\ref{virasoro relation for wzw sl(2) scaled}) as
the parameter $i \hbar$, where $\hbar \to 0$ is equivalent to the
classical limit of the commutator relations. Since in a
quantisation procedure, we go from $\{ {\overline L}_n, {\overline
L}_m \}_{P.B.} \to {1\over {i\hbar}} [{\hat {\overline L}}_n,
{\hat {\overline L}}_m]$, going in reverse would give us the
classical Poisson bracket relation \be \{{{\overline L}}_n,
{{\overline L}}_m \}_{P.B.} =  (n-m) {{\overline L}}_{n+m} - {6
\over 12}(k+2) (k'+2) (n^3 - n) \delta_{n, -m}, \label{virasoro
relation for wzw sl(2) scaled poisson} \ee where ${\overline
T}_{\textrm{classical}}(z) = \sum_{n} {\overline L}_n z^{-n-2}$.
Since the Poisson bracket must be well-defined as $k \to -2$ and
$k' \to \infty$, it will mean that $(k+2) (k'+2)$ must be equal to
a finite constant $l$ in these limits. For different values of
$l$, the Poisson algebra generated by the Laurent modes of
${\overline T}_{\textrm{classical}}(z)$, will be a classical
Virasoro algebra with different central charges. Note at this
point that an equivalence - at the level of the holomorphic chiral
algebra - between the $\psi^{\bar j}$-independent sector of the
twisted sigma-model on $SL(2)/B$ and the $B$-gauged WZW model on
$SL(2)$, will mean that the Laurent modes of ${\overline
T}_{\textrm{classical}}(z)$ and $S(z)$ ought to generate an
isomorphic classical algebra. In other words, the $\overline
L_m$'s ought to generate the same classical Virasoro algebra with
central charge $-6$ that is generated by the Laurent modes $S_m$
of the sigma-model description in (\ref{poisson brackets of S_n}),
i.e., we must have $l=1$, or rather $(k+2) = 1/ (k'+2)$. Thus, the
Poisson algebra generated by the $\overline L_m$'s must be given
by \be \{{{\overline L}}_n, {{\overline L}}_m \}_{P.B.} = (n-m)
{{\overline L}}_{n+m} - {6 \over 12} (n^3 - n) \delta_{n, -m}.
\label{virasoro relation for wzw sl(2) scaled poisson at l=1} \ee
This algebra also coincides with ${\cal W}_{\infty}(\widehat{\frak
{sl}}_2)$, the classical $\cal W$-algebra associated to
$\widehat{\frak {sl}}_2$ at level $k'\to \infty$ obtained via a DS
reduction scheme \cite{book}. Since the $S_m$'s that correspond to
the ${\overline L}_m$'s in (\ref{virasoro relation for wzw sl(2)
scaled poisson at l=1}) span $\frak {z} (\widehat{\frak{sl}}_2)$,
and since for ${\frak g}= {\frak {sl}_2}= {^L\frak g}$, we have
$h^{\vee} = {^Lh}^{\vee} =2$, and $r^{\vee} =1$, where $r^{\vee}$
is the lacing number of $\frak g$, we find that an equivalence -
at the level of the holomorphic chiral algebra - between the
$\psi^{\bar j}$-independent sector of the twisted sigma-model on
$SL(2)/B$ and the $B$-gauged WZW model on $SL(2)$, will imply an
isomorphism of Poisson algebras \be {\frak z} ( \widehat{\frak g})
\cong {\cal W}_{\infty}(^L{\widehat{\frak g}}),\label{isomorphism
of W-algebras SL(2)} \ee and the level relation \be
(k+{h^{\vee}})r^{\vee} = {1\over {(k'+{{^Lh}^{\vee}})}}.
\label{langlands level duality for sl(2)}\ee Note at this point
that the purely bosonic, $\psi^{\bar j}$-independent sector of the
twisted sigma-model on $SL(2)/B$, can be described, via
(\ref{Sbosonic explicit}), by a bosonic string on $SL(2)/B$. On
the other hand, note that since a bosonic string on a group
manifold $G$ can be described as a WZW model on $G$, it will mean
that the $B$-gauged WZW model on $SL(2)$ can be interpreted as a
$B$-gauged bosonic string on $SL(2)$. Thus, we see that an
equivalence, at the level of the holomorphic chiral algebra,
between a bosonic string on $SL(2)/B$ and a $B$-gauged version of
itself on $SL(2)$ - a statement which stems from the ubiquitous
notion that one can always physically interpret a geometrical
symmetry of the target space as a gauge symmetry in the worldsheet
theory - will imply an isomorphism of classical $\cal W$-algebras
and a level relation which underlie a geometric Langlands
correspondence for $G=SL(2)$! Notice that the correspondence
between the $k \to -2$ and $k' \to \infty$ limits (within the
context of the above Poisson algebras) is indeed consistent with
the relation (\ref{langlands level duality for sl(2)}). These
limits define a ``classical'' geometric Langlands correspondence.
A ``quantum'' generalisation of the $SL(2)$ correspondence can be
defined for other values of $k$ and $k'$ that satisfy the relation
(\ref{langlands level duality for sl(2)}), but with the
isomorphism of (\ref{isomorphism of W-algebras SL(2)}) replaced by
an isomorphism of $\it{quantum}$ $\cal W$-algebras (derived from a
DS-reduction scheme) associated to $\widehat{\frak {sl}}_2$ at
levels $k$ and $k'$ respectively \cite{Frenkel}.

\newsubsection{The Twisted Sigma-Model on $SL(3)/B$ and its Classical Holomorphic Chiral Algebra}

Now, let us take $X= SL(3)/B$, where $B$ is the subgroup of upper
triangular matrices of $SL(3)$ with a nilpotent Lie algebra $\frak
b$. Note that $\textrm{dim}_{\mathbb C} X = 3$, and one can cover
$X$ with six open charts $U_w$ where $w=1,2, \dots, 6$, such that
each open chart $U_w$ can be identified with the affine space
$\mathbb {C}^3$. Hence, the sheaf of CDO's in any $U_w$ can be
described by three free $\beta\gamma$ systems with the action \be
I = \sum _{i=1}^3 \ {1\over 2\pi}\int|d^2z| \ \beta_i
\partial_{\bar z} \gamma^i. \label{beta-gamma action for SL(3)/B
on U1} \ee As before, the $\beta_i$'s and $\gamma^i$'s are fields
of dimension $(1,0)$ and $(0,0)$ respectively. They obey the
standard free-field OPE's; there are no singularities in the
operator products $\beta_i(z)\cdot \beta_i(z')$ and
$\gamma^i(z)\cdot\gamma^i(z')$, while \be \beta_i(z)\gamma^j(z')
\sim  -{\delta_i^j\over z-z'}. \ee

Similarly, the sheaf of CDO's in a neighbouring intersecting chart
$U_{w+1}$ is described by three free $\tilde\beta\tilde\gamma$
systems with action \be I= \sum _{i=1}^3\ {1\over 2\pi}\int|d^2z|
\ \tilde \beta_i \partial_{\bar z} \tilde\gamma^i,
\label{beta-gamma action for SL(3)/B on U2} \ee where the $\tilde
\beta_i$ and $\tilde \gamma^i$ fields obey the same OPE's as the
$\beta_i$ and $\gamma^i$ fields. In other words, the non-trivial
OPE's are given by \be \tilde \beta_i(z) \tilde \gamma^j(z')  \sim
-{\delta_i^j\over z-z'}. \ee

In order to describe a globally-defined sheaf of CDO's, one will
need to glue the free conformal field theories with actions
(\ref{beta-gamma action for SL(3)/B on U1}) and (\ref{beta-gamma
action for SL(3)/B on U2}) in the overlap region $U_w \cap
U_{w+1}$ for every $w = 1,2, \dots 6$, where $U_7 = U_1$. To do
so,  one must use the admissible automorphisms of the free
conformal field theories defined in
(\ref{autoCDOgamma})-(\ref{autoCDObeta}) to glue the free-fields
together. In the case of $X = SL(3)/B$, the relation between the
coordinates in $U_{w}$ and $U_{w+1}$ will mean that the $\tilde
\gamma^i$'s in $U_{w+1}$ will be related to the $\gamma^i$'s in
$U_{w}$ via the relation $[\tilde \gamma] = [V_{w+1}]^{-1}
[V_{w}][\gamma]$, where the $3 \times 3$ matrices $[V_{w+1}]$ and
$[V_w]$ are elements of the $S_3$ permutation subgroup of $GL(3)$
matrices associated to the open charts $U_{w+1}$ and $U_w$
respectively, and $[\gamma]$ is a $3 \times1$ column matrix with
the $\gamma^i$'s as entries.  By substituting this relation
between the $\tilde \gamma^i$'s and $\gamma^i$'s in
(\ref{autoCDOgamma})-(\ref{autoCDObeta}), one will have the
admissible automorphisms of the fields, which can then be used to
glue together the local sheaves of CDO's in the overlap region
$U_{w} \cap U_{w+1}$ for every $w=1,2,\dots,6$. These gluing
relations for the free fields can be written as
\begin{eqnarray}
\label{autoCDOgammaSL(3)}
{\tilde \gamma}^i & = & [V^{-1}_{w+1}\cdot V_w] ^i{}_j \ \gamma^j ,\\
\label{autoCDObetaSL(3)} {\tilde \beta}_i  & = &  \beta_k D^k{}_i
+
\partial_z \gamma^j E_{i j},
\end{eqnarray}
where $i,j,k = 1, 2, \dots, 3$. Here, $D$ and $E$ are $3 \times 3$
matrices, whereby $[(D^T)^{-1}]_i{}^k = \partial_i
[V^{-1}_{w+1}\cdot V_w] ^k{}_j \ \gamma^j$ and $[E]_{ij} =
\partial_i B_j$. It can be verified that $\tilde \beta$ and
$\tilde \gamma$ obey the correct OPE's amongst themselves.
Moreover, let $R_w$ represent a transformation of the fields in
going from $U_{w}$ to $U_{w+1}$. One can indeed verify that just
as in the previous case where we considered constructing a sheaf
of CDO's on $SL(2)/B$, there is no anomaly to a global definition
of a sheaf of CDO's on $X =SL(3)/B$ - a careful computation will
reveal that one will get the desired composition maps
$(R_6R_5R_4R_3R_2R_1) \cdot \gamma^j = \gamma^j$ and $
(R_6R_5R_4R_3R_2R_1) \cdot \beta_i = \beta_i$. Again, this is just
a statement that one can always define a sheaf of CDO's on any
flag manifold $SL(N)/B$ \cite{MSV}.

\bigskip\noindent{\it Global Sections of the Sheaf of CDO's on $X=SL(3)/B$}

Since $X=SL(3)/B$ is of complex dimension $3$, the chiral algebra
$\cal A$ will be given by ${\cal A} = \bigoplus_{g_R =0}^{g_R = 3}
H^{g_R}( X, {\widehat {\cal O}^{ch}_X})$ as a vector space. As
before and throughout this paper, it would suffice for our purpose
to concentrate on just the purely bosonic sector of $\cal A$ -
from our $\overline Q_+$-Cech cohomology dictionary, this again
translates to studying only the global sections in $H^0(X,
{\widehat {\cal O}^{ch}_X})$.

According to theorem 5.13 of \cite{MSV}, one can always find
elements in $H^0(M, {\widehat {\cal O}^{ch}_{M}})$ for any flag
manifold $M = SL(N)/B$, that will furnish a module of an affine
$SL(N)$ algebra at the critical level. This means that one can
always find dimension one global sections of  the sheaf ${\widehat
{\cal O}^{ch}_{X}}$ that correspond to $\psi^{\bar i}$-independent
currents $J^a(z)$ for $a =1,2, \dots \textrm{dim}\ {\frak {sl}_3}
=8$, that satisfy the OPE's of an affine $SL(3)$ algebra at the
critical level $k = -3$: \be J_a (z) J_b (z') \sim -{{3 d_{ab}}
\over{(z-z')^2}} + \sum_c f_{ab}{}^c {{J_c(z')}\over {(z-z')}},
\ee where $d_{ab}$ is the Cartan-Killing metric of $\frak
{sl}_3$.\footnote{Note that one can consistently introduce
appropriate fluxes to deform the level away from $-3$ - recall
from our discussion in $\S$A.7 that the $E_{ij}= \partial_i B_j$
term in (\ref{autoCDObetaSL(3)}) is related to the fluxes that
correspond to the moduli of the chiral algebra, and since this
term will determine the level $k$ of the affine $SL(3)$ algebra
via the identification of the global sections $\tilde \beta_i$
with the affine currents valued in the subalgebra of $\frak{sl}_3$
associated to its positive roots, turning on the relevant fluxes
will deform $k$ away from $-3$. Henceforth, whenever we consider
$k\neq -3$, we really mean turning on fluxes in this manner.}
Since these are global sections, it will be true that $\widetilde
J_a(z) = J_a(z)$ on any $U_{w} \cap U_{w+1}$ and $a$. Moreover,
from our $\overline Q_+$-Cech cohomology dictionary, they will be
$\overline Q_+$-closed chiral vertex operators that are
holomorphic in $z$, which means that one can expand them in a
Laurent series that allows an affinisation of the $SL(3)$ algebra
generated by their resulting zero modes. Similar to the $SL(2)/B
\simeq {\mathbb {CP}^1}$ case, the space of these operators obeys
all the physical axioms of a chiral algebra except for
reparameterisation invariance on the $z$-plane or worldsheet
$\Sigma$. We will substantiate this last statement next by showing
that the holomorphic stress tensor fails to exist in the
$\overline Q_+$-cohomology at the quantum level. Again, this
observation will be important in our discussion of a geometric
Langlands correspondence for $G=SL(3)$.

\bigskip\noindent{\it The Segal-Sugawara Tensor and the Classical Holomorphic Chiral Algebra}

Recall that for any affine algebra $\widehat {\frak g}$ at level
$k \neq -h^{\vee}$, where $h^{\vee}$ is the dual Coxeter number of
the Lie algebra $\frak g$, one can construct the corresponding
stress tensor out of the currents of $\widehat {\frak g}$ via a
Segal-Sugawara construction \cite{Ketov}. In the present case of
an affine $SL(3)$ algebra, the stress tensor can be constructed as
\be T(z) = {{: d^{ab}J_aJ_b(z) :} \over {k+3}}, \label{SS def T(z)
for SL(3)} \ee where $d^{ab}$ is the inverse of the Cartan-Killing
metric of $\frak {sl}_3$, and $h^{\vee} = 3$. As required, for
every $k \neq {-3}$, the modes of the Laurent expansion of $T(z)$
will span a Virasoro algebra. In particular, $T(z)$ will generate
holomorphic reparametrisations of the coordinates on the
worldsheet $\Sigma$. Notice that this definition of $T(z)$ in
(\ref{SS def T(z) for SL(3)}) is ill-defined when $k=-3$.
Nevertheless,  one can always associate $T(z)$ with the
Segal-Sugawara operator $S(z)$ that is well-defined at any finite
level, whereby \be S(z) =  (k+3) T(z), \label{S(z) for SL(3)} \ee
and \be S(z) = {: d^{ab} J_a J_b (z):}. \label{s(z) for sl(3)} \ee
From (\ref{S(z) for SL(3)}), we see that $S(z)$ generates, in its
OPE's with other field operators, $(k+3)$ times the
transformations usually generated by the stress tensor $T(z)$.
Therefore, at the level $k= -3$, $S(z)$ generates no
transformations at all - its OPE's with all other field operators
are trivial. This is equivalent to saying that the holomorphic
stress tensor does not exist at the quantum level, since $S(z)$,
which is the only well-defined operator at this level that could
possibly generate the transformation of fields under an arbitrary
holomorphic reparametrisation of the worldsheet coordinates on
$\Sigma$, acts by zero in the OPE's.

Despite the fact that $S(z)$ will cease to exist in the spectrum
of physical operators associated to the twisted sigma-model on
$X=SL(3)/B$ at the quantum level, it will nevertheless exist as a
field in its classical ${\overline Q}_+$-cohomology or holomorphic
chiral algebra. One can convince oneself that this is true as
follows. Firstly, from our ${\overline Q}_+$-Cech cohomology
dictionary, since the $J_a(z)$'s are in $H^0(X, {\widehat{\cal
O}}^{ch}_{X})$, it will mean that they are in the $\overline
Q_+$-cohomology of the sigma-model at the quantum level. Secondly,
since quantum corrections can only annihilate cohomology classes
and not create them, it will mean that the $J_a(z)$'s will be in
the classical $\overline Q_+$-cohomology of the sigma-model, i.e.,
the currents are $\overline Q_+$-closed and are therefore
invariant under the transformations generated by $\overline Q_+$
in the absence of quantum corrections. Hence, one can readily see
that (the classical counterpart of) $S(z)$ in (\ref{s(z) for
sl(3)}) will also be $\overline Q_+$-closed at the classical
level. Lastly, recall from section 2.3 that $[\overline Q _+,
T(z)] = 0$ such that $T(z) \neq \{ {\overline Q}_+, \cdots \}$  in
the absence of quantum corrections to the action of $\overline
Q_+$ in the classical theory. Note also that the integer $3$ in
the factor $(k+3)$ of the expression $S(z)$ in (\ref{S(z) for
SL(3)}), is due to a shift by $h^{\vee}=3$ in the level $k$
because of quantum renormalisation effects \cite{Fuchs}, i.e., the
classical expression of $S(z)$ for a general level $k$ can
actually be written as $S(z) = k T(z)$, and therefore, one will
have $[\overline Q_+, -3 T(z)] = [\overline Q_+, S(z)] = 0$, where
$S(z) \neq \{ {\overline Q}_+, \cdots \}$ in the classical theory.
Therefore, $S(z)$ will be a spin-two field in the classical
holomorphic chiral algebra of the purely bosonic sector of the
twisted sigma-model on $X= SL(3)/B$.  This observation is also
consistent with the fact that $S(z)$ fails to correspond to a
global section of the sheaf ${\cal O}^{ch}_{X}$ of CDO's - note
that in our case, we actually have $S(z) = -3 T(z)$ in the
classical theory, and this will mean that under quantum
corrections to the action of $\overline Q_+$, we will have
$[\overline Q_+, S_{zz}] = -3
\partial_z(R_{i \bar j}
\partial_z \phi^i \psi^{\bar j}) \neq 0$ (since $R_{i\bar j} \neq 0$ for any flag manifold $SL(N)/B$), which corresponds in the Cech
cohomology picture to the expression ${\widetilde{\widehat S}(z)}
- {\widehat S}(z) \neq 0$, i.e., ${\widehat S}(z)$, the Cech
cohomology counterpart to the $S(z)$ operator, will fail to be in
$H^0(X, {\widehat {\cal O}}^{ch}_{X})$. Consequently, one can
always represent $S(z)$ by a classical $c$-number. This point will
again be important when we discuss how one can define Hecke
eigensheaves that will correspond to flat $^LG$-bundles on a
Riemann surface $\Sigma$ in our physical interpretation of the
geometric Langlands correspondence for $G=SL(3)$.

The fact that $S(z)$ acts trivially in any OPE with other field
operators implies that its Laurent modes will commute with the
Laurent modes of any of these other field operators; in
particular, they will commute with the Laurent modes of the
$J_a(z)$ currents - in other words, the Laurent modes of $S(z)$
will span the centre    ${\frak z}(\widehat{\frak{sl}}_3)$ of the
completed universal enveloping algebra of the affine      $SL(3)$
algebra $\widehat{\frak{sl}}_3$ at the critical level $k=-3$
(generated by the Laurent modes of the $J_a(z)$ currents in the
quantum chiral algebra of the twisted sigma-model on $SL(3)/B$).
Notice also that $S(z)$ is $\psi^{\bar j}$-independent and is
therefore purely bosonic in nature. In other words, the local
field $S(z)$ exists only in the $\it{classical}$ holomorphic
chiral algebra of the $\it{purely}$ $\it{bosonic}$ (or $\psi^{\bar
j}$-independent) sector of the twisted sigma-model on $X=
SL(3)/B$.

\bigskip\noindent{\it A Classical Virasoro Algebra}

Note that since $S(z)$ is holomorphic in $z$ and is of conformal
dimension two, one can expand it in terms of a Laurent expansion
as \be S(z) = \sum_{n \in {\mathbb Z}} {\hat S}_n z^{-n-2}.
\label{quantum S(z) for SL(3)} \ee Recall that for the general
case of $k \neq -3$, a quantum definition of $S(z)$ exists, such
that the ${\hat S}_n$ modes of the Laurent expansion can be
related to the $J_{a,n}$ modes of the $\widehat{\frak {sl}}_3$
currents through the quantum commutator relations
\begin{eqnarray}
[{\hat S}_n, J_{a, m}] & = & -(k + 3) m J_{a, n+m},\\
{[{\hat S}_n, {\hat S}_m ]} & = & (k+ 3) \left( (n-m) {\hat
S}_{n+m} + { 8 k \over 12} \  (n^3-n) \ \delta_{n, -m}\right),
\label{S_n for SL(3)}
\end{eqnarray}
where $a = 1, 2, \dots, 8$. If we now let $k = - 3$, we will have
$[{\hat S}_n, J_{a,m}] = [{\hat S}_n, {\hat S}_m] =0$ - the
$S_m$'s thus generate the (classical) centre of the completed
universal enveloping algebra of $\widehat{\frak {sl}}_3$ as
mentioned above.

Since we now understand that $S(z)$ must be a holomorphic
classical field at $k =-3$, let us rewrite the Laurent expansion
of $S(z)$ as \be S(z) = \sum_{n \in {\mathbb Z}} S_n z^{-n-2}
\label{S(z) classical for SL(3)} \ee
 so as to differentiate the classical modes of expansion $S_n$ from their quantum counterpart ${\hat S}_n$ in (\ref{S(z) for SL(3)}). Unlike the ${\hat S}_n$'s which obey the quantum commutator relations in (\ref{S_n for SL(3)}) for an arbitrary level $k \neq -3$, the $S_n$'s, being the modes of a Laurent expansion of a classical field, will instead obey Poisson bracket relations that define a certain classical algebra when $k=-3$.

Based on our arguments thus far, we learn that the quantum version
of $S(z)$ as expressed in (\ref{S(z) for SL(3)}), must reduce to
its classical counterpart as expressed in (\ref{S(z) classical for
SL(3)}), when $k = -3$. In other words, one can see that by taking
$(k+3) \to 0$, we are going to the classical limit. This is
analogous to taking the  ${\hbar} \to 0$ limit in any quantum
mechanical theory whenever one wants to ascertain its classical
counterpart. In fact, by identifying $(k+3)$ with $i \hbar$, and
by noting that one must make the replacement from Possion brackets
to commutators via $\{S_n, S_m \}_{P.B.} \rightarrow {1\over {i
\hbar}} [ {\hat S}_n, {\hat S}_m ]$ in quantising the $S_n$'s into
operators, we can ascertain the classical algebra generated by the
$S_n$'s from (\ref{S_n for SL(3)}) as \be \{S_n, S_m\}_{P.B.} =
(n-m) {S}_{n+m} - {24 \over 12}\ (n^3-n) \ \delta_{n, -m}.
\label{poisson brackets of S_n for SL(3)} \ee Since we have the
classical relation $S(z) \sim T(z)$, it means that we can
interpret the $S_n$ modes as the Virasoro modes of the Laurent
expansion of the classical stress tensor field $T(z)$. In other
words, the $S_n$'s generate a classical Virasoro algebra with
central charge $-24$ as given by (\ref{poisson brackets of S_n for
SL(3)}). This can be denoted mathematically as the Virasoro
Poisson algebra $Sym'(vir_{-24})$.

\bigskip\noindent{\it A Higher-Spin Analog of the Segal-Sugawara Tensor and the Classical Holomorphic Chiral Algebra}

For an affine $SL(N)$ algebra where $N > 2$,  one can generalise
the Sugawara formalism to construct higher-spin analogs of the
holomorphic stress tensor with the currents. These higher-spin
analogs have conformal weights $3, 4, \dots N$. These higher-spin
analogs are called Casimir operators, and were first constructed
in \cite{casimir operators}.

In the context of our affine $SL(3)$ algebra with a module that is
furnished by the global sections of the sheaf of CDO's on
$X=SL(3)/B$, a spin-three analog of the holomorphic stress tensor
will be given by the 3rd-order Casimir operator \cite{review} \be
T^{(3)} (z) = {{:{\tilde d}^{abc} (k) (J_a (J_bJ_c))(z):} \over
{k+3}}, \label{casimir of spin-three for SL(3)} \ee where ${\tilde
d}^{abc}(k)$ is a completely symmetric traceless
$\frak{sl}_3$-invariant tensor of rank 3 that depends on the level
$k$ of the affine $SL(3)$ algebra in question.  ${\tilde
d}^{abc}(k)$ is also well-defined and finite at $k=-3$. The
superscript on $T^{(3)}(z)$ just denotes that it is a spin-three
analog of $T(z)$.

As with $T(z)$ in (\ref{SS def T(z) for SL(3)}), $T^{(3)} (z)$ is
ill-defined when $k = -3$. Nevertheless, one can always make
reference to a higher-spin analog of the Segal-Sugawara tensor
$S^{(3)}(z)$ that is well-defined for any finite value of $k$,
where its relation to $T^{(3)}(z)$ is given by \be S^{(3)}(z) =
(k+3) T^{(3)}(z), \ee and \be S^{(3)}(z) = {:{\tilde d}^{abc} (k)
(J_a (J_bJ_c))(z):}. \label{S^{(3)}(z)} \ee That is, the operator
$S^{(3)}(z)$ generates in its OPE's with all other operators of
the quantum theory,  $(k+3)$ times the field transformations
typically generated by $T^{(3)}(z)$.

Notice however, that at $k=-3$, $S^{(3)}(z)$ acts by zero in its
OPE with any other operator. This is equivalent to saying that
$T^{(3)}(z)$ does not exist as a quantum operator at all, since
the only well-defined operator $S^{(3)}(z)$ which is supposed to
generate the field transformations associated to $T^{(3)}(z)$, act
by zero and thus generate no field transformations at all. From
our $\overline Q_+$-Cech cohomology dictionary, this means that
the $\psi^{\bar i}$-indepedent operator $T^{(3)}(z)$ will fail to
correspond to a dimension three global section of $\widehat{\cal
O}^{ch}_X$. Since we have, at the classical level, the relation
$S^{(3)}(z) = -3 T^{(3)}(z)$, it will mean that $S^{(3)}(z)$ will
also fail to  correspond to a dimension three global section of
$\widehat{\cal O}^{ch}_X$. Thus, $S^{(3)}(z) $ will fail to be an
operator at the quantum level. Is it even a spin-three field in
the classical holomorphic chiral algebra of the twisted
sigma-model on $SL(3)/B$, one might ask. The answer is yes. To see
this, recall that each of the $J_a(z)$'s are separately $\overline
Q_+$-invariant and not $\overline Q_+$-exact at the classical
level. Therefore, the classical counterpart of $S^{(3)}(z)$ in
(\ref{S^{(3)}(z)}) must also be such, which in turn means that it
will be in the classical $\overline Q_+$-cohomology and hence
classical chiral algebra of the twisted sigma-model on $SL(3)/B$.

The fact that $S^{(3)}(z)$ acts trivially in any OPE with other
field operators implies that its Laurent modes will commute with
the Laurent modes of any other operator; in particular, they will
commute with the Laurent modes of the currents $J_a(z)$ for
$a=1,2, \dots, 8$ - in other words, the Laurent modes of
$S^{(3)}(z)$ will span the centre ${\frak
z}(\widehat{\frak{sl}}_3)$ of the completed universal enveloping
algebra of the affine $SL(3)$ algebra $\widehat{\frak{sl}}_3$ at
the critical level $k=-3$ (generated by the Laurent modes of the
$J_a(z)$ currents of the quantum chiral algebra of the twisted
sigma-model on $SL(3)/B$). Last but not least, notice that the
$S^{(3)}(z)$ field is also $\psi^{\bar j}$-independent and is
therefore purely bosonic in nature. In other words, the local
fields $S(z)$  and $S^{(3)}(z)$, whose Laurent modes together
generate ${\frak z}(\widehat{\frak{sl}}_3)$, exist only in the
$\it{classical}$ holomorphic chiral algebra of the $\it{purely}$
$\it{bosonic}$ (or $\psi^{\bar j}$-independent) sector of the
twisted sigma-model on $X=SL(3)/B$.

\bigskip\noindent{\it A Classical ${\cal W}_3$-algebra}

For an affine $SL(3)$ algebra at an arbitrary level $k \neq -3$,
as in the case of $S(z)$ discussed earlier, a quantum definition
of $S^{(3)}(z)$ exists. In fact, consider the following operators
given by ${\overline S^{(3)}}(z)= ( \sqrt{3/200}):d^{abc} (J_a(J_b
J_c))(z):$ and ${\overline S}(z) = (1/4): d^{ab} J_a J_b (z):$,
where $d^{abc}$ is just a rank-three extension of $d^{ab}$. It can
be shown that ${\overline S}^{(3)}(z)$ and ${\overline S}(z)$
together span a $\it{closed}$ Casimir OPE algebra which is
isomorphic to a particular ${\cal W}_3$ OPE algebra \cite{casimir
operators}. This implies that for $k\neq -3$, both ${\overline
S}^{(3)}(z)$ and ${\overline S}(z)$ and therefore $S(z) \sim :
d^{ab} J_a J_b(z):$ and $S^{(3)}(z) \sim :d^{abc} (J_a(J_b
J_c))(z):$, will exist as quantum operators in some cohomology -
the $\overline Q_+$-cohomology in this instance. This will in turn
mean that $S(z) = (k+3) T(z)$ and $S^{(3)}(z) = (k+3) T^{(3)}(z)$
must also span a closed OPE algebra that is equivalent - at the
level of $\overline Q_+$-cohomology - to this Casimir OPE algebra,
when $k \neq -3$. Since we know that for $k \neq -3$, $T(z)$ will
generate a Virasoro subalgebra of a closed ${\cal W}_3$ OPE
algebra with central charge $c = 8k / (k+3)$, it will mean that
$S(z)$ and $S^{(3)}(z)$ will satisfy a rescaled (by a factor of
$(k+3)$) version of a closed ${\cal W}_3$ OPE algebra at $c = 8k
/(k+3)$ for $k\neq -3$. Because $S^{(3)}(z)$ is holomorphic in
$z$, we can Laurent expand it as \be S^{(3)}(z) = \sum_{n \in
{\mathbb Z}} {\hat S}^{(3)}_n z^{-n-3}. \label{S^{(3)}(z) quantum
for SL(3)} \ee At $k \neq -3$, the Laurent modes ${\hat S}^{(3)}_n
$, together with the Laurent modes ${\hat S}_n$ of $S(z)$, will
then obey the following quantum commutator relations \be [{\hat
S}_n, {\hat S}^{(3)}_m] = (k+3) (2n-m){\hat S}^{(3)} _{n+m},
\label{algebra 1} \ee and
\begin{eqnarray}
\label{W3 commutator algebra}
[{\hat S}^{(3)}_m, {\hat S}^{(3)}_n] & = & (k+3) \left [ {8k \over 360} \ m(m^2-1)(m^2-4) \delta_{m,-n} \right]  \nonumber \\
&& \hspace{-1.5cm} + (k+3) \left[ (m-n) \left({1\over 15} (m+n+3) (m+n+2) - {1 \over 6}(m+2)(n+2) \right) {\hat S}_{m+n} \right] \nonumber \\
&&  \hspace{-1.5cm} + (k+3) \left[ {16 \over {62k +66}} (m-n) \left ( \sum_p {\hat S}_{m+n-p} {\hat S}_p  - {3\over 10} (k+3) (m+n+3) (m+n+2) {\hat S}_{m+n} \right) \right]. \nonumber \\
&&
\end{eqnarray}

Now let us consider the case when $k = -3$. From our earlier
explanations about the nature of $S^{(3)}(z)$ and $S(z)$ at
$k=-3$, we find that they will cease to exist as quantum operators
at $k=-3$. Since we understand that $S^{(3)}(z)$, just like
$S(z)$, must be a holomorphic classical field at $k =-3$, we shall
rewrite the Laurent expansion of $S^{(3)}(z)$ as \be S^{(3)}(z) =
\sum_{n \in {\mathbb Z}} S^{(3)}_n z^{-n-3}, \label{S^{(3)}(z)
classical for SL(3)} \ee so as to differentiate the classical
modes of expansion $S^{(3)}_n$ from their quantum counterpart
${\hat S}^{(3)}_n$ in (\ref{S^{(3)}(z) quantum for SL(3)}). Unlike
the ${\hat S}^{(3)}_n$'s which obey the quantum commutator
relations in (\ref{W3 commutator algebra}) for an arbitrary level
$k \neq -3$, the $S^{(3)}_n$'s, being the modes of a Laurent
expansion of a classical field, will instead obey Poisson bracket
relations that define a certain classical algebra when $k=-3$.
Since every ${\hat S}^{(3)}_n$ must reduce to its classical
counterpart ${S}^{(3)}_n$ when $k = -3$,  one can see that by
taking $(k+3) \to 0$, we are actually going to the classical
limit. This is analogous to taking the  ${\hbar} \to 0$ limit in
any quantum mechanical theory whenever one wants to ascertain its
classical counterpart. In fact, by identifying $(k+3)$ with $i
\hbar$, and by noting that one must make the replacement from
Possion brackets to commutators via $\{D_n, D_m \}_{P.B.}
\rightarrow {1\over {i \hbar}} [ {\hat D}_n, {\hat D}_m ]$ in
quantising any classical mode $D_n$ into an operator, we can
ascertain the classical algebra generated by the $S^{(3)}_n$'s and
$S_n$'s from (\ref{algebra 1}) and (\ref{W3 commutator algebra})
as \be \{{S}_n, {S}^{(3)}_m\}_{P.B.} = (2n-m) \ {\hat S}^{(3)}
_{n+m}, \label{S classical algebra} \ee and
\begin{eqnarray}
\label{W3 classical algebra}
\{{S}^{(3)}_m, {S}^{(3)}_n\}_{P.B.} & = &  -{24 \over 360} \ m(m^2-1)(m^2-4) \delta_{m,-n}  -  {4 \over 30} (m-n) \sum_p {S}_{m+n-p}\ {S}_p  \nonumber \\
&& +   (m-n) \left({1\over 15} (m+n+3) (m+n+2) - {1 \over 6}(m+2)(n+2) \right) {S}_{m+n}. \nonumber \\
&&\nonumber \\
&&
\end{eqnarray}
Together with the earlier expression\be \{S_n, S_m\}_{P.B.} =
(n-m) {S}_{n+m} - {24 \over 12}\ (n^3-n) \ \delta_{n, -m},
\label{W3 classical algebra 1} \ee we see that the $S^{(3)}_m$'s
and $S_m$'s generate a classical ${\cal W}_3$-algebra with central
charge $-24$. Note that the algebra is closed amongst the
${S}^{(3)}_n$ and ${{S}_n}$ modes; this is true because both
${S}^{(3)}(z)$ and ${S}(z)$ are in the classical $\overline
Q_+$-cohomology of the sigma-model.\footnote{Note at this point
that if $\cal O$ and $\cal O'$ are non-exact $\overline
Q_+$-closed observables in the (classical) $\overline
Q_+$-cohomology, i.e., ${\{\overline Q_{+}, \cal O\}}={\{\overline
Q_{+}, {\cal O}'\}}= 0$, then $\{\overline Q_{+}, {\cal O}{\cal
O}'\} =0$. Moreover, if $\{\overline Q_{+}, {\cal O}\}=0$, then
${\cal O}\{\overline Q_{+}, W\}= \{\overline Q_{+}, {\cal O}W\}$
for any observable $W$.  These two statements mean that the
cohomology classes of observables that commute with $\overline
Q_{+}$ form a closed and well-defined (classical) algebra.} Thus,
if we denote this classical algebra by ${\cal W}_3(-24)$, we then
have the identification ${\frak z}(\widehat{\frak{sl}}_3) \simeq
{\cal W}_3(-24)$.

\newsubsection{A Gauged WZW Model and the Geometric Langlands Correspondence for $G=SL(3)$}

\bigskip\noindent{\it The $B$-gauged WZW Model on SL(3)}

According to our discussion in $\S$2.2, the classical holomorphic
chiral algebra of the purely bosonic sector of the twisted
sigma-model on $SL(3)/B$ - in which lie the fields $S(z)$ and
$S^{(3)}(z)$ - will be given by the classical, holomorphic
BRST-cohomology of a $B$-gauged WZW model on $SL(3)$ - from which
one ought to find non-trivial classes that are in one-to-one
correspondence with the fields $S(z)$ and $S^{(3)}(z)$
respectively. As such, we shall proceed to specialise the action
$S_{\textrm{B-gauged}} (g, A_z, A_{\bar z}, J^+, \bar J^+)$ of a
(non-dynamically) $B$-gauged WZW model on $\it{any}$ $SL(N)$
defined in (\ref{B-gauged WZW action SL(N)}) of $\S2.2$, to the
case where the target-space is now $SL(3)$.

In the case of $SL(3)$, we have $\textrm{dim} \ {\frak n}_{\pm} =
3$ and $\textrm{dim}\ {\frak c} =2$, so we can write $J(z) =
\sum_{a=1}^{3} J^a_-(z) t^{-}_a + \sum_{a=1}^{2} J^a_c(z) t^{c}_a
+ \sum_{l=1}^{3} J^l_+(z) t^{+}_l$, and $\bar J(\bar z) =
\sum_{a=1}^{3} {\bar J}^a_-(\bar z) t^{-}_a + \sum_{a=1}^{2} \bar
J^a_c(\bar z) t^{c}_a + \sum_{l=1}^{3} \bar J^l_+(\bar z)
t^{+}_l$, where $t^{-}_a \in {\frak n}_-$, $t^{c}_a \in {\frak
c}$, and $t^{+}_a \in {\frak n}_+$. One can also write $M =
\sum_{a=1}^{3} M^a_- t^{-}_a + \sum_{a=1}^{2} M^a_c t^{c}_a  +
\sum_{l=1}^{3} M^l_+ t^{+}_l$, where $M^{a}_{ -; c}$ and $M^l_+$
are arbitrary number constants, and $\bar M = \sum_{a=1}^{3} \bar
M^a_- t^{-}_a + \sum_{a=1}^{2} \bar M^a_c t^{c}_a  +
\sum_{l=1}^{3} \bar M^l_+ t^{+}_l$, where $\bar M^{a}_{ -; c}$ and
$\bar M^l_+$ are arbitrary number constants. In addition, one can
also write $A_{\bar z} = \sum_{l =1}^{3} {\tilde A}_{\bar z}^l
t^+_l$ and $A_{z} = \sum_{l =1}^{3} {\tilde A}_{z}^l t^+_l$. Let
us denote ${J}^+(z) = \sum_{l=1}^{3} {J}^l_+ (z) t^{+}_l$ and
${M}^+ = \sum_{l=1}^{3} {M}^l_+ t^{+}_l$. Let us also denote
${\bar J}^+(\bar z) = \sum_{l=1}^{3} {\bar J}^l_+ (\bar z)
t^{+}_l$ and ${\bar M}^+ = \sum_{l=1}^{3} {\bar M}^l_+ t^{+}_l$.
Hence, one can write the action for the $B$-gauged WZW model on
$SL(3)$ as
\begin{eqnarray} S_{{SL(3)}} (g, A_z, A_{\bar z}, J^+, \bar J^+)&
= & S_{\textrm{WZ}} (g) - {k' \over {2\pi}} \int_{\Sigma} d^2z \
\sum_{l=1}^3 \left[ {\tilde A}^l_{\bar z}( J^l_+(z) + {M}^l_+) -
{\tilde A}^l_{z}( \bar J^l_+(\bar z) + {\bar
M}^l_+) \right]\nonumber \\
&& \hspace{3.5cm} - \textrm {Tr} [ {A}_z g {A}_{\bar z} g^{-1} -
{A}_z {A}_{\bar z}]
 \label{B-gauged WZW action general SL(3)}
\end{eqnarray}

Due to the $B$-gauge invariance of the theory, we must divide the
measure in any path integral computation by the volume of the
$B$-gauge symmetry. That is, the partition function has to take
the form \be Z_{SL(3)} = \int_{\Sigma} { {[g^{-1}dg, d{\tilde
A}^l_{z}, d{\tilde A}^l_{\bar z}]} \over {(\textrm{gauge
volume})}} \ \textrm{exp} \left(i S_{SL(3)}(g, A_z, A_{\bar z},
J^+, {\bar J}^+) \right). \ee One must now fix this gauge
invariance to eliminate the non-unique degrees of freedom. One can
do this by employing the BRST formalism which requires the
introduction of Faddev-Popov ghost fields.

In order to obtain the $\it{holomorphic}$ BRST transformations of
the fields, one simply replaces the position-dependent
infinitesimal gauge parameter $\epsilon^l$ of $h= B=
\textrm{exp}(- \sum^3 _{l=1} \epsilon^l t^+_l)$ in the
corresponding $\textrm{\it left-sector}$ of the gauge
transformations in (\ref{gauge tx}) with the ghost field $c^l$,
which then gives us \be \delta_{\textrm{BRST}}(g) = -c^l t^+_l g,
\quad \delta_{\textrm{BRST}}(\tilde A^l_{\bar z}) = - D_{\bar z}
c^l, \quad \delta_{\textrm{BRST}}(\textrm{others}) =0. \label{BRST
tx SL(3)} \ee The components of the ghost field $c(z) =
\sum_{l=1}^3 c^l (z) t^+_l$ and those of its anti-ghost partner
$b(z) = \sum_{l=1}^3b^l (z) t^+_l$ will transform as \be
\delta_{\textrm{BRST}} (c^l) = - {1\over 2}f_{mk}^l c^mc^k, \quad
\delta_{\textrm{BRST}}(b^l) = {\tilde B}^l, \quad
\delta_{\textrm{BRST}} {(\tilde B^l)} = 0, \ee where the
$f^l_{mk}$'s are the structure constants of the nilpotent
subalgebra ${\frak n}_+$. Also, the ${\tilde B}^l$'s are the
Nakanishi-Lautrup auxiliary fields that are the BRST transforms of
the $b^l$'s. They also serve as a Lagrange multipliers to impose
the gauge-fixing conditions.

In order to obtain the $\it{antiholomorphic}$ BRST transformations
of the fields, one employs the same recipe to the corresponding
$\textrm{\it right-sector}$ of the gauge transformations in
(\ref{gauge tx}) with the infinitesimal position-dependent gauge
parameter now replaced by the ghost field $\bar c^l$, which then
gives us \be \bar \delta_{\textrm{BRST}}(g) = \bar c^l t^+_l g,
\quad \bar \delta_{\textrm{BRST}}({\tilde A}^l_{z}) = - D_{z} \bar
c^l, \quad \bar\delta_{\textrm{BRST}}(\textrm{others})
=0.\label{BRST tx 1 SL(3)}\ee The components of the ghost field
${\bar c}(\bar z) = \sum_{l=1}^3 {\bar c}^l (\bar z) t^+_l$ and
those of its anti-ghost partner $\bar b(\bar z) = \sum_{l=1}^3
{\bar b}^l (\bar z) t^+_l$ will transform as \be
\bar\delta_{\textrm{BRST}} (\bar c^l) = - {1\over 2}f_{mk}^l {\bar
c}^m {\bar c}^k, \quad \bar\delta_{\textrm{BRST}}(\bar b^l) =
{\tilde {\bar B}^l}, \quad \bar\delta_{\textrm{BRST}} {(\tilde
{\bar B}^l)} = 0. \ee In the above, the $\tilde {\bar B}^l$'s are
the Nakanishi-Lautrup auxiliary fields that are the
antiholomorphic BRST transforms of the $\bar b^l$ fields. They
also serve as Lagrange multipliers to impose the gauge-fixing
conditions.

Since the BRST transformations in (\ref{BRST tx SL(3)}) and
(\ref{BRST tx 1 SL(3)})  are just infinitesimal versions of the
gauge transformations in (\ref{gauge tx}), $S_{{SL(3)}} (g, A_z,
A_{\bar z}, J^+, \bar J^+)$ will be invariant under them. An
important point to note at this juncture is that in addition to
$(\delta_{\textrm{BRST}} + \bar\delta_{\textrm{BRST}}) \cdot
(\delta_{\textrm{BRST}} + \bar\delta_{\textrm{BRST}}) = 0$, the
holomorphic and antiholomorphic BRST-variations are also
separately nilpotent, i.e., $\delta^2_{\textrm{BRST}} = 0$ and
$\bar\delta^2_{\textrm{BRST}}=0$. Moreover,
$\delta_{\textrm{BRST}}\cdot \bar\delta_{\textrm{BRST}}= - \bar
\delta_{\textrm{BRST}} \cdot \delta_{\textrm{BRST}}$. This means
that the BRST-cohomology of the $B$-gauged WZW model on $SL(3)$
can be decomposed into $\it{independent}$ holomorphic and
antiholomorphic sectors that are just complex conjugate of each
other, and that it can be computed via a spectral sequence,
whereby the first two complexes will be furnished by its
holomorphic and antiholomorphic BRST-cohomologies respectively.
Since we will only be interested in the holomorphic chiral algebra
of the $B$-gauged WZW model on $SL(3)$ (which as mentioned, is
just identical to its antiholomorphic chiral algebra by a complex
conjugation), we shall henceforth focus on the $\it{holomorphic}$
BRST-cohomology of the $B$-gauged WZW model on $SL(3)$.

By the usual recipe of the BRST formalism, one can fix the gauge
by adding to the BRST-invariant action $S_{{SL(3)}} (g, A_z,
A_{\bar z}, J^+, \bar J^+)$, a BRST-exact term. Since the BRST
transformation by $(\delta_{\textrm{BRST}} + \bar
\delta_{\textrm{BRST}})$  is nilpotent, the new total action will
still be BRST-invariant as required. The choice of the BRST-exact
operator will then define the gauge-fixing conditions. A
consistent choice of the BRST-exact operator that will give us the
requisite action for the ghost and anti-ghost fields is \be
S_{{SL(3)}} (g, A_z, A_{\bar z}, J^+, \bar J^+) +
(\delta_{\textrm{BRST}} + \bar \delta_{\textrm{BRST}})
\left({k'\over 2\pi} \int_{\Sigma} d^2 z \ \sum_{l=1}^3 {\tilde
A}^l_{\bar z} b^l + {\tilde A}^l_{z} {\bar b}^l \right),\nonumber
\ee where one will indeed have the desired total action, which can
be written as \begin{eqnarray} \label{total desired action for
SL(3)} S_{\textrm{WZW}}(g)  -  {k' \over {2\pi}} \int_{\Sigma}
d^2z \ \{ \sum_{l=1}^3 \left[ {\tilde A}^l_{\bar z}( J^l_+(z) +
{M}^l_+ - \tilde B^l) - {\tilde A}^l_{z}( \bar J^l_+(\bar z) +
{\bar M}^l_+ + \tilde {\bar B}^l)
\right]  \nonumber \\
 - \textrm {Tr} [ {A}_z g {A}_{\bar z} g^{-1} - {A}_z {A}_{\bar
z}] \} +  {k'\over {2\pi}}\int_{\Sigma} d^2z \ \sum_{l=1}^3 \left
( c^l
D_{\bar z} b^l + + \bar c^l D_{z} \bar b^l \right). \nonumber \\
\end{eqnarray} From the equations of motion by varying the ${\tilde
B}^l$'s, we have the conditions $\tilde A^l_{\bar z} = 0$ for
$l=1,2,3$. From the equations of motion by varying the $\tilde
{\bar B}^l$'s, we also have the conditions $\tilde A^l_{z} = 0$
for $l=1,2,3$. Thus, the partition function of the $B$-gauged WZW
model can also be expressed as
 \be
Z_{SL(3)} = \int [ g^{-1} dg, db, dc, d\bar b, d\bar c] \
\textrm{exp} \left ( iS_{\textrm{WZW}}(g) + {i k'\over
{2\pi}}\int_{\Sigma} d^2z \ \textrm{Tr} (c \cdot \partial_{\bar z}
b) (z) + \textrm{Tr} (\bar c \cdot \partial_{z} \bar b) (\bar z)
\right), \label{Z_SL(3)} \ee where the $\it{holomorphic}$ BRST
variations of the fields which leave the effective action in
(\ref{Z_SL(3)}) $\it{invariant}$ are now given by \begin{eqnarray}
\delta_{\textrm{BRST}} (g) = -c^mt^+_m g, & \quad
\delta_{\textrm{BRST}} (c^l) = -{1\over 2} f^l_{mk}c^m c^k,
\quad \delta_{\textrm{BRST}}(b^l) =  J^l_+ + M^l_+ - f^l_{mk}b^mc^k,\nonumber \\
\hspace{-2.0cm}\quad \delta_{\textrm{BRST}}(\textrm{others}) = 0.
& \label{BRST variations SL(3)}
 \end{eqnarray}

Though we did not make this obvious in our discussion above, by
integrating out the $\tilde A^l_{\bar z}$'s in (\ref{B-gauged WZW
action general SL(3)}), and using the above conditions $\tilde
A^l_{z} =0$ for $l=1,2,3$, we find that we actually have the
relations $(J^l_+(z) + M^l_+) =0$ for $l=1,2,3$. These relations
(involving the current associated to the Borel subalgebra $\frak
b$ of the group $B$ that we are modding out by) will lead us
directly to the correct form of the holomorphic stress tensor for
the gauged WZW model without reference to a coset formalism. Let
us look at this more closely. Since we have, in the holomorphic
BRST-cohomology of this non-dynamically $B$-gauged WZW model on
$SL(3)$, currents that generate an affine $SL(3)$ OPE algebra,
(consistent with the presence of an affine $SL(3)$ OPE algebra
generated by operators in the holomorphic chiral algebra of the
purely bosonic sector of the sigma-model on $SL(3)/B$), we can
employ the Sugawara formalism to construct the holomorphic stress
tensor associated to the $S_{\textrm{WZW}}(g)$ part of the total
action from the currents, and it can be written as \be T_{SL(3)}
(z) = {{:J(z) \cdot J(z):} \over {(k'+3)}}, \ee where the above
dot product between the currents $J(z)$ is taken with respect to
the Cartan-Killing metric on $\frak{sl}_3$. Note that with respect
to $T_{SL(3)}(z)$, the currents $J^l_{+}(z)$ for $l=1,2,3$ (and in
fact all the other affine $SL(3)$ currents) have conformal
dimension one. This is inconsistent with the condition $J^l_+(z) =
-M^l_+$, as the $M^l_+$'s are constants of conformal dimension
zero. This means that one must modify $T_{SL(3)} (z)$ so that the
$J^l_+(z)$'s will have conformal dimension zero. A physically
consistent modification involves adding to $T_{SL(3)}(z)$ a term
that has conformal dimension two. A little thought will reveal
that the only consistent candidate for the modified stress tensor
of $S_{\textrm{WZW}}(g)$ will be given by \be
T_{\textrm{modified}}(z) = T_{SL(3)} (z)  + \partial_zJ^1_c (z) +
\partial_zJ^2_c (z). \label{T_{total} SL(3)}
 \ee
With respect to $T_{\textrm{modified}}(z)$, the currents $J^1_+
(z)$ and $J^2_+ (z)$ have vanishing conformal dimensions as
required. On the other hand, the current $J^3_+(z)$ will now have
conformal dimension -1. Thus, it must mean that there cannot be
any restriction on the conformal dimension of $J^3_+(z)$, and
therefore, $M^3_+$ must vanish.

In order for the above observations to be consistent with the fact
that the BRST-charge $Q_{\textrm{BRST}}$ generating the variations
$\delta_{\textrm{BRST}}( b^l)$ of (\ref{BRST variations SL(3)})
must be a scalar of conformal dimension zero, we find that $b^l$'s
and hence the $c^l$'s must have the following conformal
dimensions: $(b^1, b^2, b^3)  \leftrightarrow (0,0,-1)$ and $(c^1,
c^2, c^3) \leftrightarrow (1,1,2)$. From the effective action in
(\ref{Z_SL(3)}), we can compute the holomorphic stress tensor of
the left-moving ghost/anti-ghost system. Including this
contribution, we find that the total holomorphic stress tensor can
be written as\begin{eqnarray}
T_{\textrm{total}} (z)   & = & T_{SL(3)} (z)  + \partial_zJ^1_c (z) + \partial_zJ^2_c (z) + \partial_z b^1(z) c^1_z(z) + \partial_z b^2(z) c^2_z(z) \nonumber \\
&& + 2 \partial_z b^{3,z}(z) c_{zz}^3(z) + b^{3,z}(z) \partial_z
c_{zz}^3(z).
\end{eqnarray}

The conserved current associated to the holomorphic
BRST-variations of the fields in (\ref{BRST variations SL(3)}) can
be computed as \be I_{\textrm{BRST}} =  c^1_z(J^1_+(z) + M^1_+) +
c^2_z(J^2_+(z) + M^2_+) + c^3_{zz}J^{3,z}_+(z)  + c^1_z(z)
c^2_z(z) b^{3,z}(z). \ee Apart from a trivial inspection, one can
also verify, from the affine $SL(3)$ OPE algebra and the OPE
algebra between the left-moving ghost/anti-ghost fields, that
$I_{\textrm{BRST}}$ is of conformal dimension one with respect to
$T_{\textrm{total}} (z)$. This means that  as required, one can
define a conformal dimension zero scalar BRST-charge
 \be
Q_{\textrm{BRST}} = \oint {dz \over {2 \pi i}}  \ c^1_z(J^1_+(z) +
M^1_+) + c^2_z(J^2_+(z) + M^2_+) + c^3_{zz}J^{3,z}_+(z)  +
c^1_z(z) c^2_z(z) b^{3,z}(z), \label{Q_BRST,WZW SL(3)} \ee which
generates the correct holomorphic BRST-variation of the fields.
Note that the holomorphic BRST-charge can also be written in its
general form as \be Q_{\textrm{BRST}} = \oint {dz \over {2 \pi i}}
\left( \sum_{l=1}^{\textrm{dim}{\frak n}_+ =3} \ c^l (J^l_+(z) +
M^l_+) -{1\over 2} \sum_{l,m,k=1}^{\textrm{dim}{\frak n}_+ =3}
f^l_{mk}b^mc^lc^k \right). \label{Q_BRST,WZW SL(3)-clean} \ee
Using the free field OPE's that the ghost fields generate, one can
immediately verify that $Q_{\textrm{BRST}}$ as given in
(\ref{Q_BRST,WZW SL(3)-clean}) will indeed generate the field
variations in (\ref{BRST variations SL(3)}).

\bigskip\noindent{\it More about the Holomorphic Stress Tensor and its Higher-Spin Analog}

Since $I_{\textrm{BRST}}$ is of  conformal dimension one with
respect to $T_{\textrm{total}}(z)$, it will mean that
$T_{\textrm{total}}(z)$ will be annihilated by
$Q_{\textrm{BRST}}$. Moreover, one can also check that
$T_{\textrm{total}}(z) \neq \{Q_{\textrm{BRST}}, \cdots\}$. In
other words, $T_{\textrm{total}}(z)$ is a spin-two observable in
the holomorphic BRST-cohomology of the $B$-gauged WZW model on
$SL(3)$. From the explicit expression of $T_{\textrm{total}}(z)$,
we find that its Laurent modes will generate a Virasoro algebra
with central charge $c= 2- 24(k'+2)^2 / (k'+3)$.

Note that one can define a cohomologically equivalent total
holomorphic stress tensor $T_{W_3}(z)$ via \be T_{W_3}(z) =
T_{\textrm{total}} (z)  + \{Q_{\textrm{BRST}}, t(z) \}, \ee
whereby
\begin{eqnarray}
t(z) & = & \gamma_{1,z}(z) \partial_z b^1(z) + \gamma_{2,z} (z) \partial_z b^2(z) + 2\gamma_{3,zz} (z) \partial_z b^{3,z}(z) + \partial_z \gamma_{3,zz}(z) b^{3,z}(z) \nonumber \\
&& - \gamma_{1,z}(z) \partial_z [\gamma_{2,z}(z) b^{3,z}(z)],
\end{eqnarray}
such that \be T_{W_3}(z) = -{1\over 2} \left ( (\partial_z
\varphi_1(z))^2 + (\partial_z\varphi_2(z))^2 + {i 2(k'+2) \over
{\sqrt{2k'+6}}} \ \partial^2_z \varphi_1(z) \right), \ee in which
the $\gamma$'s and $\varphi$'s are just auxillary fields which
satisfy the OPE's of a free $\beta\gamma$ and free scalar system
respectively. It can be shown \cite{Fateev} that $T_{W_3}(z)$,
together with a spin-three field $T^{(3)}_{W_3}(z)$ which is a
higher-spin analog of $T_{W_3}(z)$, will satisfy the free boson
realisation of a closed ${\cal W}_3$ OPE algebra with the same
central charge $c = 2- 24(k'+2)^2 / (k'+3) = 50- 24(k'+3) -  24/
(k'+3)$. This implies that one can always find a spin-three
observable, independent of the $\gamma$'s and $\varphi$'s, and
composed out of the fields in the gauged WZW model only, which is
cohomologically equivalent to $T^{(3)}_{W_3}(z)$, and which is
also non-trivial in the holomorphic BRST-cohomology of the gauged
WZW model.\footnote{Note that within the ${\cal W}_3$ OPE algebra,
we have an OPE of the form $T^{(3)}_{W_3}(z) \cdot
T^{(3)}_{W_3}(z')$, which is equivalent, up to singular terms, to
a sum of the fields $T_{W_3}(z)$ and its partial derivatives only.
These terms in the sum are certainly BRST-closed but non-exact.
Thus, in order to be consistent with the OPE, it implies that
$T^{(3)}_{W_3}(z)$ must also be BRST-closed and non-exact.} Let us
denote this spin-three observable as
$T^{(3)}_{\textrm{total}}(z)$. Note that
$T^{(3)}_{\textrm{total}}(z)$ is just a spin-three analog of
$T_{\textrm{total}}(z)$, and together with
$T_{\textrm{total}}(z)$, it will generate a ${\cal W}_3$ OPE
algebra with central charge $c = 50- 24(k'+3) -  24/ (k'+3)$ for
finite (and therefore non-classical) values of the level $k'$.
This observation will turn out to be consistent with our
discussion in $\S$3.3 where we unravel the role that the
$B$-gauged WZW model on $SL(3)$ plays in a physical realisation of
the DS-reduction scheme of generating a ${\cal
W}_{k'}(\widehat{\frak{sl}}_3)$ OPE algebra.

Last but not least, a soon-to-be relevant point to note is that
since quantum corrections can only annihilate classes in the
BRST-cohomology and not create them, the classical counterparts of
the holomorphic stress tensor $T_{\textrm{total}}(z)$ and its
spin-three analog $T^{(3)}(z)$, will be the spin-two and
spin-three fields $T_{\textrm{classical}}(z)$ and
$T^{(3)}_{\textrm{classical}}(z)$ which lie in the
$\it{classical}$, holomorphic BRST-cohomology (or holomorphic
chiral algebra) of the $B$-gauged WZW model on $SL(3)$, where
$T_{\textrm{classical}}(z)$ will generate a classical Virasoro
transformation on the fields, while
$T^{(3)}_{\textrm{classical}}(z)$ will generate a classical ${\cal
W}_3$ transformation on them.

\bigskip\noindent{\it A Duality of Classical $\cal W$-Algebras Underlying a Geometric Langlands Correspondence for $G=SL(3)$}

Recall that the observables in the holomorphic BRST-cohomology of
the gauged WZW model that should correspond to $S(z)$ and
$S^{(3)}(z)$ of the holomorphic chiral algebra of the purely
bosonic, $\psi^{\bar j}$-independent sector of the twisted
sigma-model on $SL(3)/B$, must have the same spins as $S(z)$ and
$S^{(3)}(z)$. Since $S(z)$ generates a classical Virasoro symmetry
on the worldsheet, it should correspond to a spin-two observable
in the classical, holomorphic BRST-cohomology of the $B$-gauged
WZW model on $SL(3)$ which generates a classical Virasoro symmetry
on the worldsheet. Since $S^{(3)}(z)$ generates a classical ${\cal
W}_3$ symmetry on the worldsheet, it should correspond to a
spin-three observable in the classical, holomorphic
BRST-cohomology of the $B$-gauged WZW model on $SL(3)$ which
generates a classical ${\cal W}_3$ symmetry on the worldsheet.

 In order to ascertain the classical observable which
corresponds to $S(z)$, first recall that the $\it{quantum}$
definition of $S(z)$ at $k\neq -3$ is given by $S(z) = (k+3)T(z)$.
Notice that since the stress tensor $T(z)$ also exists in the
holomorphic chiral algebra of the purely bosonic, $\psi^{\bar
j}$-independent sector of the sigma-model on $SL(3)/B$, it will
imply that $T(z)$ must correspond to the stress tensor
$T_{\textrm{total}}(z)$ of the $B$-gauged WZW model on $SL(3)$.
Thus, at $k\neq -3$, $S(z)$ will correspond to ${\overline
T}_{\textrm{total}}(z) = (k+3)T_{\textrm{total}}(z)$, and at
$k=-3$, $S(z)$ will correspond to the classical counterpart
${\overline T}_{\textrm{classical}}(z)$ of ${\overline
T}_{\textrm{total}}(z)$. Note that ${\overline
T}_{\textrm{classical}}(z)$ lies in the classical, holomorphic
BRST-cohomology of the $B$-gauged WZW model on $SL(3)$ as required
- at $k=-3$, ${\overline T}_{\textrm{total}}(z)$, which usually
exists as a quantum operator, will act by zero in its OPE's with
any other operator, i.e., it will reduce to its classical
counterpart ${\overline T}_{\textrm{classical}}(z)$ in the
classical, holomorphic BRST-cohomology of the gauged WZW model.
Moreover, since the shift in $3$ in the factor $(k+3)$ is due to a
quantum effect as explained earlier, $S(z)$ will actually
correspond to ${\overline T}_{\textrm{classical}}(z) = -
3T_{\textrm{classical}}(z)$ at $k=-3$. Hence, $S(z)$ will indeed
correspond to a spin-two field in the classical, holomorphic
BRST-cohomology of the $B$-gauged WZW on $SL(3)$ which generates a
classical Virasoro transformation of the fields.

In order to ascertain the classical observable which corresponds
to $S^{(3)}(z)$, first recall that the $\it{quantum}$ definition
of $S^{(3)}(z)$ at $k\neq -3$ is given by $S^{(3)}(z) =
(k+3)T^{(3)}(z)$. Notice that since $T^{(3)}(z)$ also exists as a
spin-three analog of $T(z)$ in the holomorphic chiral algebra of
the purely bosonic, $\psi^{\bar j}$-independent sector of the
sigma-model on $SL(3)/B$, it will imply that $T^{(3)}(z)$ must
correspond to the spin-three operator $T_{\textrm{total}}(z)$ of
the $B$-gauged WZW model on $SL(3)$. Thus, at $k\neq -3$,
$S^{(3)}(z)$ will correspond to ${\overline
T}^{(3)}_{\textrm{total}}(z) = (k+3)T^{(3)}_{\textrm{total}}(z)$,
and at $k=-3$, $S(z)$ will correspond to the classical counterpart
${\overline T}^{(3)}_{\textrm{classical}}(z)$ of ${\overline
T}^{(3)}_{\textrm{total}}(z)$. Note that ${\overline
T}^{(3)}_{\textrm{classical}}(z)$ lies in the classical,
holomorphic BRST-cohomology of the $B$-gauged WZW model on $SL(3)$
as required - at $k=-3$, ${\overline
T}^{(3)}_{\textrm{total}}(z)$, which usually exists as a quantum
operator, will act by zero in its OPE's with any other operator,
i.e., it will reduce to its classical counterpart ${\overline
T}^{(3)}_{\textrm{classical}}(z)$ in the classical, holomorphic
BRST-cohomology of the gauged WZW model. Moreover, since the shift
in $3$ in the factor $(k+3)$ is due to a quantum effect as
explained earlier, $S^{(3)}(z)$ will actually correspond to
${\overline T}^{(3)}_{\textrm{classical}}(z) = -
3T^{(3)}_{\textrm{classical}}(z)$ at $k=-3$. Hence, $S^{(3)}(z)$
will indeed correspond to a spin-three field in the classical,
holomorphic BRST-cohomology of the $B$-gauged WZW on $SL(3)$ which
generates a classical ${\cal W}_3$ transformation of the fields.

What is the classical algebra generated by the Laurent modes of
${\overline T}_{\textrm{classical}}(z)$? To ascertain this, first
recall that the Laurent modes of $T_{\textrm{total}}(z)$ generate
a Virasoro algebra of central charge $c = 50- 24(k'+3) - 24/
(k'+3)$. Hence, the Virasoro modes of $T_{\textrm{total}}(z) =
\sum_{n} {\hat L}_n z^{-n-2}$ will obey the following quantum
commutator relation \be [{\hat L}_n, {\hat L}_m ] = (n-m) {\hat
L}_{n+m} + {1\over {12}} \left[ 50- 24(k'+3) -  24/ (k'+3) \right]
(n^3 - n) \delta_{n, -m}. \label{virasoro relation for wzw sl(3)}
\ee Therefore, the commutator relations involving the
${\hat{\overline L}}_n $ modes of ${\overline
T}_{\textrm{total}}(z) = \sum_{n} {\hat {\overline L}}_n z^{-n-2}$
will be given by
 \be [{\hat {\overline
L}}_n, {\hat {\overline L}}_m] = (k+3) \left[ (n-m)
{\hat{\overline L}}_{n+m} + (n^3 - n) \delta_{n, -m} \left(
{50\over 12} (k+3) -{{24 (k+3) (k'+3)}\over {12}} - {{2 (k+3)}
\over {(k'+3)}}  \right) \right].
 \label{virasoro relation for wzw
sl(3) scaled}
 \ee
At $k = -3$, ${\overline T}_{\textrm{total}}(z)$ will cease to
have a quantum definition, and it will reduce to its classical
counterpart ${\overline T}_{\textrm{classical}}(z)$. Hence, the $k
\to -3$ (and $k' \to \infty$) limit of the commutator relation in
(\ref{virasoro relation for wzw sl(3) scaled}), can be interpreted
as its classical limit. Therefore, one can view the term $(k+3)$
in (\ref{virasoro relation for wzw sl(3) scaled}) as the parameter
$i \hbar$, where $\hbar \to 0$ is equivalent to the classical
limit of the commutator relations. Since in a quantisation
procedure, we go from $\{ {\overline L}_n, {\overline L}_m
\}_{P.B.} \to {1\over {i\hbar}} [{\hat {\overline L}}_n, {\hat
{\overline L}}_m]$, going in reverse would give us the classical
Poisson bracket relation \be \{{{\overline L}}_n, {{\overline
L}}_m \}_{P.B.} =  (n-m) {{\overline L}}_{n+m} - {24 \over
12}(k+3) (k'+3) (n^3 - n) \delta_{n, -m}, \label{virasoro relation
for wzw sl(3) scaled poisson} \ee where ${\overline
T}_{\textrm{classical}}(z) = \sum_{n} {\overline L}_n z^{-n-2}$.
Since the Poisson bracket must be well-defined as $k \to -3$ and
$k' \to \infty$, it will mean that $(k+3) (k'+3)$ must be equal to
a finite constant $q$ in these limits. For different values of
$q$, the Poisson algebra generated by the Laurent modes of
${\overline T}_{\textrm{classical}}(z)$, will be a classical
Virasoro algebra with different central charge. Note at this point
that an equivalence - at the level of the holomorphic chiral
algebra - between the $\psi^{\bar j}$-independent sector of the
twisted sigma-model on $SL(3)/B$ and the $B$-gauged WZW model on
$SL(3)$, will mean that the Laurent modes of ${\overline
T}_{\textrm{classical}}(z)$ and $S(z)$ ought to generate an
isomorphic classical algebra. This in turn means that the
$\overline L_m$'s must generate the same classical Virasoro
algebra $Sym'(vir_{-24})$ with central charge $-24$ that is
generated by the Laurent modes $S_m$ of the purely bosonic sector
of the sigma-model on $SL(3)/B$. Thus, we must have $q=1$, or
rather $(k+3) = 1/ (k'+3)$.

What is the classical algebra generated by the Laurent modes of
${\overline T}^{(3)}_{\textrm{classical}}(z)$? To ascertain this,
first recall that it was argued that the Laurent modes of
$T^{(3)}_{\textrm{total}}(z)$ and $T_{\textrm{total}}(z)$ will
together generate a quantum ${\cal W}_3$-algebra of central charge
$c = 50- 24(k'+3) -  24/ (k'+3)$. Hence, the Laurent modes ${\hat
L}^{(3)}_n$ of $T^{(3)}_{\textrm{total}}(z) = \sum_{n} {\hat
L}^{(3)}_n z^{-n-3}$ and the ${\hat L}_n$'s  will obey the
following quantum commutator relations
 \be
[{\hat L}_n, {\hat L}^{(3)}_m] =  (2n-m){\hat L}^{(3)} _{n+m},
\label{LL} \ee and
\begin{eqnarray}
\label{W3 commutator algebra for WZW side}
[{\hat L}^{(3)}_m, {\hat L}^{(3)}_n] & = &   {{50- 24(k'+3) -  24 / (k'+3)} \over 360 } \ \ m(m^2-1)(m^2-4) \delta_{m,-n}   \nonumber \\
&& \hspace{-0.5cm} +  (m-n) \left({1\over 15} (m+n+3) (m+n+2) - {1 \over 6}(m+2)(n+2) \right) {\hat {\overline L}}_{m+n}  \nonumber \\
&& \hspace{-0.5cm}  +  \left({16 \over {22 + 5 [50-24(k'+3) -  24 / (k'+3)]}}\right) (m-n) \left ( \sum_p {\hat {\overline L}}_{m+n-p} {\hat {\overline L}}_p \right)  \nonumber \\
&& \hspace{-1.0cm}  -\left({16 \over {22 + 5 [50-24(k'+3) -  24 / (k'+3)]}}\right) (m-n) \left({3\over 10} (m+n+3) (m+n+2) {\hat {\overline L}}_{m+n} \right).  \nonumber \\
 \end{eqnarray}
Therefore, from (\ref{LL}) and (\ref{W3 commutator algebra for WZW
side}), the commutator relations involving the ${\hat{\overline
L}}_n $ modes of ${\overline T}_{\textrm{total}}(z) = \sum_{n}
{\hat {\overline L}}_n z^{-n-2}$ and the ${\hat{\overline
L}}^{(3)}_n $ modes of ${\overline T}^{(3)}_{\textrm{total}}(z) =
\sum_{n} {\hat {\overline L}}^{(3)}_n z^{-n-2}$, will be given by
\be [{\hat {\overline L}}_n, {\hat{\overline L}}^{(3)}_m] =
(k+3)(2n-m){\hat {\overline L}}^{(3)} _{n+m}, \label{LL rescaled}
\ee and
\begin{eqnarray}
\label{W3 commutator algebra for WZW side rescaled}
[{\hat {\overline L}}^{(3)}_m, {\hat {\overline L}}^{(3)}_n] & = & \nonumber \\
&& \nonumber \\
&&\hspace{-2.0cm} (k+3) \left[ {{50(k+3) - 24(k+3)(k'+3) -  {24(k+3) \over (k'+3)}} \over 360 } \ \ m(m^2-1)(m^2-4) \delta_{m,-n} \right]  \nonumber \\
&& \hspace{-2.0cm} + (k+3)\left[ (m-n) \left({1\over 15} (m+n+3) (m+n+2) - {1 \over 6}(m+2)(n+2) \right) {\hat {\overline L}}_{m+n} \right] \nonumber \\
&& \hspace{-2.0cm}  + (k+3) \left[  \left({16 \over {22(k+3) + 5 [50 (k+3) -24 (k+3) (k'+3) -  {24(k+3) \over (k'+3)}]}}\right) (m-n) \left ( \sum_p {\hat {\overline L}}_{m+n-p} {\hat {\overline L}}_p \right) \right]  \nonumber \\
&& \hspace{-2.0cm}  - (k+3)\left[ \left({16 \over {22 + 5 [50-24(k'+3) -  24 / (k'+3)]}}\right) (m-n) \left({3\over 10} (m+n+3) (m+n+2) {\hat {\overline L}}_{m+n} \right) \right]. \nonumber \\
&& \nonumber \\
\end{eqnarray}
At $k = -3$, ${\overline T}^{(3)}_{\textrm{total}}(z)$ will cease
to have a quantum definition, and it will reduce to its classical
counterpart ${\overline T}^{(3)}_{\textrm{classical}}(z)$. Hence,
the $k \to -3$ (and $k' \to \infty$) limit of the commutator
relations in (\ref{LL rescaled}) and (\ref{W3 commutator algebra
for WZW side rescaled}), can be interpreted as their classical
limits. Therefore, one can view the term $(k+3)$ in (\ref{LL
rescaled}) and  (\ref{W3 commutator algebra for WZW side
rescaled}) as the parameter $i \hbar$, where $\hbar \to 0$ is
equivalent to the classical limit of the commutator relations.
Since in a quantisation procedure, we go from $\{ {\overline D}_n,
{\overline D}_m \}_{P.B.} \to {1\over {i\hbar}} [{\hat {\overline
D}}_n, {\hat {\overline D}}_m]$ for any classical observable
${{\overline D}}_n$, going in reverse would give us the classical
Poisson bracket relations \be \{{{\overline L}}_n, {{\overline
L}}^{(3)}_m\}_{P.B.} = (2n-m){{\overline L}}^{(3)} _{n+m},
\label{LL rescaled classical} \ee and
\begin{eqnarray}
\label{W3 commutator algebra for WZW side rescaled classical}
\{{{\overline L}}^{(3)}_m, {{\overline L}}^{(3)}_n\}_{P.B.} & = & \left[ {{ - 24(k+3)(k'+3) } \over 360 } \ \ m(m^2-1)(m^2-4) \delta_{m,-n} \right]  \nonumber \\
&& \hspace{-0.0cm} + \left[ (m-n) \left({1\over 15} (m+n+3) (m+n+2) - {1 \over 6}(m+2)(n+2) \right) {{\overline L}}_{m+n} \right] \nonumber \\
&& \hspace{-0.0cm}  + \left[  \left({-4 \over {  30 (k+3) (k'+3)}}\right) (m-n) \left ( \sum_p {{\overline L}}_{m+n-p} {{\overline L}}_p \right) \right],  \nonumber \\
&&
\end{eqnarray}
where ${\overline T}^{(3)}_{\textrm{classical}}(z) = \sum_{n}
{\overline L}^{(3)}_n z^{-n-3}.$ Since the Poisson bracket in
(\ref{W3 commutator algebra for WZW side rescaled classical}) must
be well-defined as $k \to -3$ and $k' \to \infty$, it will mean
that $(k+3)(k'+3)$ must be equal to a finite constant $q$ in these
limits. For different values of $q$, the total Poisson algebra
generated by the Laurent modes of ${\overline
T}_{\textrm{classical}}(z)$ and ${\overline
T}^{(3)}_{\textrm{classical}}(z)$ in (\ref{virasoro relation for
wzw sl(3) scaled poisson}), (\ref{LL rescaled classical}) and
(\ref{W3 commutator algebra for WZW side rescaled classical}),
will be a classical ${\cal W}_3$-algebra with different central
charges. Note at this point that an equivalence - at the level of
the holomorphic chiral algebra - between the $\psi^{\bar
j}$-independent sector of the twisted sigma-model on $SL(3)/B$ and
the $B$-gauged WZW model on $SL(3)$, will mean that the Laurent
modes of $({\overline T}_{\textrm{classical}}(z), {\overline
T}^{(3)}_{\textrm{classical}}(z))$ and $(S(z), S^{(3)}(z))$ ought
to generate an isomorphic classical algebra. This means that we
must have $q=1$, or rather $(k+3) = 1/ (k'+3)$, so that the total
Poisson algebra will be given by \be \{{{\overline L}}_n,
{{\overline L}}_m \}_{P.B.} = (n-m) {{\overline L}}_{n+m} - {24
\over 12}(n^3 - n) \delta_{n, -m}, \label{virasoro relation for
wzw sl(3) scaled poisson at q=1} \ee \be \{{{\overline L}}_n,
{{\overline L}}^{(3)}_m\}_{P.B.} = (2n-m){{\overline L}}^{(3)}
_{n+m}, \label{LL rescaled classical at q=1} \ee and
\begin{eqnarray}
\label{W3 commutator algebra for WZW side rescaled classical at
q=1}
\{{{\overline L}}^{(3)}_m, {{\overline L}}^{(3)}_n\}_{P.B.} & = &  -{24 \over 360 } \ \ m(m^2-1)(m^2-4) \delta_{m,-n}   \nonumber \\
&& \hspace{-0.0cm} +  (m-n) \left({1\over 15} (m+n+3) (m+n+2) - {1 \over 6}(m+2)(n+2) \right) {{\overline L}}_{m+n}  \nonumber \\
&& \hspace{-0.0cm}    -{4 \over 30 } (m-n) \left ( \sum_p {{\overline L}}_{m+n-p} {{\overline L}}_p \right),  \nonumber \\
&&
\end{eqnarray}
the classical ${\cal W}_3$-algebra with central charge $-24$
generated by the Laurent modes $S_m$ and $S^{(3)}_m$ in in
(\ref{W3 classical algebra 1}), (\ref{S classical algebra}) and
(\ref{W3 classical algebra}). Note also that this algebra
coincides with ${\cal W}_{\infty}(\widehat{\frak {sl}}_3)$, the
classical $\cal W$-algebra associated to $\widehat{\frak {sl}}_3$
at level $k' \to \infty$ obtained via a DS reduction scheme
\cite{book}. Since the $S_m$'s and $S^{(3)}_m$'s which correspond
respectively to the ${{\overline L}}_n$'s and ${{\overline
L}}^{(3)}_m$'s span $\frak {z} (\widehat{\frak{sl}}_3)$, and since
for ${\frak g}= {\frak {sl}_3} = {^L\frak g}$, we have $h^{\vee} =
{^Lh}^{\vee} =3$, and $r^{\vee} =1$, where $r^{\vee}$ is the
lacing number of $\frak g$,  we see that an equivalence - at the
level of the holomorphic chiral algebra - between the $\psi^{\bar
j}$-independent sector of the twisted sigma-model on $SL(3)/B$ and
the $B$-gauged WZW model on $SL(3)$, will imply an isomorphism of
Poisson algebras \be {\frak z} ( \widehat{\frak g}) \cong {\cal
W}_{\infty}(^L{\widehat{\frak g}}), \label{isomorphism of
W-algebras SL(3)} \ee and the level relation \be
(k+{h^{\vee}})r^{\vee} = {1\over {(k'+{{^Lh}^{\vee}})}}.
\label{langlands level duality for sl(3)} \ee  Note at this point
that the purely bosonic, $\psi^{\bar j}$-independent sector of the
twisted sigma-model on $SL(3)/B$, can be described, via
(\ref{Sbosonic explicit}), by a bosonic string on $SL(3)/B$. On
the other hand, note that since a bosonic string on a group
manifold $G$ can be described as a WZW model on $G$, it will mean
that the $B$-gauged WZW model on $SL(3)$ can be interpreted as a
$B$-gauged bosonic string on $SL(3)$. Thus, we see that an
equivalence, at the level of the holomorphic chiral algebra,
between a bosonic string on $SL(3)/B$ and a $B$-gauged version of
itself on $SL(3)$ - a statement which stems from the ubiquitous
notion that one can always physically interpret a geometrical
symmetry of the target space as a gauge symmetry in the worldsheet
theory - will imply an isomorphism of classical $\cal W$-algebras
and a level relation which underlie a geometric Langlands
correspondence for $G=SL(3)$! Notice also that the correspondence
between the $k \to -3$ and $k' \to \infty$ limits (within the
context of the above Poisson algebras) is indeed consistent with
the relation (\ref{langlands level duality for sl(3)}). These
limits define a ``classical'' geometric Langlands correspondence.
A ``quantum'' generalisation of the correspondence for $SL(3)$ can
be defined for other values of $k$ and $k'$ that satisfy the
relation (\ref{langlands level duality for sl(3)}), but with the
isomorphism of (\ref{isomorphism of W-algebras SL(3)}) replaced by
an isomorphism of $\it{quantum}$ $\cal W$-algebras (derived from a
DS-reduction scheme) associated to $\widehat{\frak {sl}}_3$ at
levels $k$ and $k'$ respectively \cite{Frenkel}.

\newsection{An Equivalence of Classical Holomorphic Chiral Algebras and the Geometric Langlands Correspondence for $G= SL(N)$}

We shall now proceed to show that our observations in $\S$2 for
$G=SL(2)$ and $SL(3)$ can be extended to any $G=SL(N)$.

To this end, we shall first discuss the twisted sigma-model on
$SL(N)/B$, and elaborate on the higher-order Casimir invariant
operators which generalise the Segal-Sugawara tensor $S(z)$ to
higher-spin analogs of itself that we will denote as $S^{(3)}(z),
S^{(4)}(z), \dots, S^{(N)}(z)$. We shall show that $S(z),
S^{(3)}(z), S^{(4)}(z), \dots, S^{(N)}(z)$ have Laurent modes
which generate the centre $\frak z(\widehat{\frak {sl}}_N)$ of the
completed universal enveloping algebra of $\widehat{\frak {sl}}_N$
at the critical level $k=-h^{\vee}$, where a module of
$\widehat{\frak {sl}}_N$ at $k=-h^{\vee}$ is always furnished by
the global sections of the sheaf of CDO's on $SL(N)/B$
corresponding to local operators in the quantum holomorphic chiral
algebra of the twisted sigma-model on $SL(N)/B$. We shall also
show that $S(z), S^{(3)}(z), S^{(4)}(z), \dots, S^{(N)}(z)$ exist
only in the classical holomorphic chiral algebra of the purely
bosonic sector of the twisted sigma-model on $SL(N)/B$, and that
moreover, their Laurent modes which span $\frak z(\widehat{\frak
{sl}}_N)$ - an ingredient which furnishes the left-hand side of
the $\cal W$-algebra duality - will generate a classical ${\cal
W}_N$-algebra.

Next, we shall show that the holomorphic BRST-cohomology of the
$B$-gauged WZW model on $SL(N)$ at level $k'$ introduced in
$\S$2.2 actually furnishes a physical realisation of the algebraic
DS reduction scheme that defines the set of fields with spins
$2,3, \dots N$ whose Laurent modes generate ${\cal
W}_{k'}(\widehat{\frak {sl}}_N)$, the ${\cal W}$-algebra
associated to some affine Lie algebra $\widehat{\frak {sl}}_N$ at
level $k'$. Consequently, the classical, holomorphic
BRST-cohomology of the $B$-gauged  WZW model on $SL(N)$ will
reproduce the holomorphic classical fields with spins $2,3,\dots,
N$ that have Laurent modes which generate the Poisson algebra
${\cal W}_{\infty}(\widehat{\frak {sl}}_N)$, the classical $\cal
W$-algebra from the DS-reduction scheme in the limit of $k' \to
\infty$ - an ingredient which furnishes the right-hand side of the
$\cal W$-algebra duality.

Finally, we shall show that that an equivalence, at the level of
the holomorphic chiral algebra, between a bosonic string on
$SL(N)/B$ and a $B$-gauged version of itself on $SL(N)$ - a
statement which stems from the ubiquitous notion that one can
always physically interpret a geometrical symmetry of the target
space as a gauge symmetry in the worldsheet theory - will imply an
isomorphism of classical $\cal W$-algebras and a level relation
which underlie a geometric Langlands correspondence for $G=SL(N)$.

\newsubsection{The Twisted Sigma-Model on $SL(N)/B$ and its Classical Holomorphic Chiral Algebra}

Now, let us take $X= SL(N)/B$, where $B$ is the subgroup of upper
triangular matrices of $SL(3)$ with a nilpotent Lie algebra $\frak
b$. Note that one can cover $X$ with $N!$ open charts $U_w$ where
$w=1,2, \dots, N!$, such that each open chart $U_w$ can be
identified with the affine space $\mathbb {C}^{N(N-1)/2}$. Hence,
the sheaf of CDO's in any $U_w$ can be described by $N(N-1)/2$
free $\beta\gamma$ systems with the action \be I = \sum
_{i=1}^{N(N-1)/2} \ {1\over 2\pi}\int|d^2z| \ \beta_i
\partial_{\bar z} \gamma^i. \label{beta-gamma action for SL(N)/B
on U1} \ee As before, the $\beta_i$'s and $\gamma^i$'s are fields
of dimension $(1,0)$ and $(0,0)$ respectively. They obey the
standard free-field OPE's; there are no singularities in the
operator products $\beta_i(z)\cdot \beta_i(z')$ and
$\gamma^i(z)\cdot\gamma^i(z')$, while \be \beta_i(z)\gamma^j(z')
\sim  -{\delta_i^j\over z-z'}. \ee

Similarly, the sheaf of CDO's in a neighbouring intersecting chart
$U_{w+1}$ is described by $N(N-1)/2$ free
$\tilde\beta\tilde\gamma$ systems with action \be I= \sum
_{i=1}^{N(N-1)/2}\ {1\over 2\pi}\int|d^2z| \ \tilde \beta_i
\partial_{\bar z} \tilde\gamma^i, \label{beta-gamma action for
SL(N)/B on U2} \ee where the $\tilde \beta_i$ and $\tilde
\gamma^i$ fields obey the same OPE's as the $\beta_i$ and
$\gamma^i$ fields. In other words, the non-trivial OPE's are given
by \be \tilde \beta_i(z) \tilde \gamma^j(z')  \sim
-{\delta_i^j\over z-z'}. \ee

In order to describe a globally-defined sheaf of CDO's, one will
need to glue the free conformal field theories with actions
(\ref{beta-gamma action for SL(N)/B on U1}) and (\ref{beta-gamma
action for SL(N)/B on U2}) in the overlap region $U_w \cap
U_{w+1}$ for every $w = 1,2, \dots N!$, where $U_{1+ N!} = U_1$.
To do so, one must use the admissible automorphisms of the free
conformal field theories defined in
(\ref{autoCDOgamma})-(\ref{autoCDObeta}) to glue the free-fields
together. In the case of $X = SL(N)/B$, the relation between the
coordinates in $U_{w}$ and $U_{w+1}$ will mean that the $\tilde
\gamma^i$'s in $U_{w+1}$ will be related to the $\gamma^i$'s in
$U_{w}$ via the relation $[\tilde \gamma] = [V_{w+1}]^{-1}
[V_{w}][\gamma]$, where the $[N(N-1)/ 2]\times [N(N-1)/ 2]$
matrices $[V_{w+1}]$ and $[V_w]$ are realisations of the $S_N$
permutation subgroup of $GL(N)$ associated to the open charts
$U_{w+1}$ and $U_w$ respectively, and $[\gamma]$ is a $[N(N-1)/ 2]
\times1$ column matrix with the $\gamma^i$'s as entries.  By
substituting this relation between the $\tilde \gamma^i$'s and
$\gamma^i$'s in (\ref{autoCDOgamma})-(\ref{autoCDObeta}), one will
have the admissible automorphisms of the fields, which can then be
used to glue together the local sheaves of CDO's in the overlap
region $U_{w} \cap U_{w+1}$ for every $w=1,2,\dots,N!$. These
gluing relations for the free fields can be written as
\begin{eqnarray}
\label{autoCDOgammaSL(N)}
{\tilde \gamma}^i & = & [V^{-1}_{w+1}\cdot V_w] ^i{}_j \ \gamma^j ,\\
\label{autoCDObetaSL(N)} {\tilde \beta}_i  & = &  \beta_k D^k{}_i
+
\partial_z \gamma^j E_{i j},
\end{eqnarray}
where $i,j,k = 1, 2, \dots, N(N-1)/ 2$. Here, $D$ and $E$ are
$[N(N-1)/ 2] \times [N(N-1)/ 2]$ matrices, whereby
$[(D^T)^{-1}]_i{}^k = \partial_i [V^{-1}_{w+1}\cdot V_w] ^k{}_j \
\gamma^j$ and $[E]_{ij} = \partial_i B_j$. It can be verified that
$\tilde \beta$ and $\tilde \gamma$ obey the correct OPE's amongst
themselves. Moreover, let $R_w$ represent a transformation of the
fields in going from $U_{w}$ to $U_{w+1}$. One can indeed verify
that  there is no anomaly to a global definition of a sheaf of
CDO's on $X =SL(N)/B$  - a careful computation will reveal that
one will get the desired composition maps $(R_{N!} \dots
R_4R_3R_2R_1) \cdot \gamma^j = \gamma^j$ and $ (R_{N!}\dots
R_4R_3R_2R_1) \cdot \beta_i = \beta_i$. Again, this is just a
statement that one can always define a sheaf of CDO's on any flag
manifold $SL(N)/B$ \cite{MSV}.

\bigskip\noindent{\it Global Sections of the Sheaf of CDO's on $X=SL(N)/B$}

\vspace{0.2cm}

Since $X=SL(N)/B$ is of complex dimension $N(N-1)/2$, the chiral
algebra $\cal A$ will be given by ${\cal A} = \bigoplus_{g_R
=0}^{g_R = {N(N-1)/2} } H^{g_R}( X, {\widehat {\cal O}^{ch}_X})$
as a vector space. As before, it would suffice for our purpose to
concentrate on just the purely bosonic sector of $\cal A$ - from
our $\overline Q_+$-Cech cohomology dictionary, this again
translates to studying only the global sections in $H^0(X,
{\widehat {\cal O}^{ch}_X})$.

According to theorem 5.13 of \cite{MSV}, one can always find
elements in $H^0(M, {\widehat {\cal O}^{ch}_{M}})$ for any flag
manifold $M = SL(N)/B$, that will furnish a module of an affine
$SL(N)$ algebra at the critical level. This means that one can
always find dimension one global sections of  the sheaf ${\widehat
{\cal O}^{ch}_{X}}$ that correspond to $\psi^{\bar i}$-independent
currents $J^a(z)$ for $a =1,2, \dots \textrm{dim}\ {\frak
{sl}_N}$, that satisfy the OPE's of an affine $SL(N)$ algebra at
the critical level $k = -h^{\vee}$: \be
 J_a (z) J_b (z') \sim -{{h^{\vee} d_{ab}}\over{(z-z')^2}} + \sum_c f_{ab}{}^c {{J_c(z')}\over {(z-z')}},
\ee where  $h^{\vee}$ is the dual Coxeter number of the Lie
algebra $\frak {sl}_N$, and $d_{ab}$ is its Cartan-Killing
metric.\footnote{Note that one can consistently introduce
appropriate fluxes to deform the level away from $-h^{\vee}$ -
recall from our discussion in $\S$A.7 that the $E_{ij}= \partial_i
B_j$ term in (\ref{autoCDObetaSL(N)}) is related to the fluxes
that correspond to the moduli of the chiral algebra, and since
this term will determine the level $k$ of the affine $SL(N)$
algebra via the identification of the global sections $\tilde
\beta_i$ with the affine currents valued in the subalgebra of
$\frak{sl}_N$ associated to its positive roots, turning on the
relevant fluxes will deform $k$ away from $-h^{\vee}$. Henceforth,
whenever we consider $k\neq -h^{\vee}$, we really mean turning on
fluxes in this manner. This point of departure will be important
in a forthcoming paper where we aim to investigate the physical
interpretation of a``quantum'' geometric Langlands correspondence
in the context of CDO's.} Since these are global sections, it will
be true that $\widetilde J_a(z) = J_a(z)$ on any $U_{w} \cap
U_{w+1}$ for all $a$. Moreover, from our $\overline Q_+$-Cech
cohomology dictionary, they will be $\overline Q_+$-closed chiral
vertex operators that are holomorphic in $z$, which means that one
can expand them in a Laurent series that allows an affinisation of
the $SL(N)$ algebra generated by their resulting zero modes. The
space of these operators obeys all the physical axioms of a chiral
algebra except for reparameterisation invariance on the $z$-plane
or worldsheet $\Sigma$. We will substantiate this last statement
next by showing that the holomorphic stress tensor fails to exist
in the $\overline Q_+$-cohomology at the quantum level. Again,
this observation will be important in our discussion of a
geometric Langlands correspondence for $G=SL(N)$.

\bigskip\noindent{\it The Segal-Sugawara Tensor and the Classical Holomorphic Chiral Algebra}

Recall that for any affine algebra $\widehat {\frak {sl}}_N$ at
level $k \neq -h^{\vee}$, one can construct the corresponding
stress tensor out of the currents of $\widehat {\frak {sl}}_N$ via
a Segal-Sugawara construction \cite{Ketov}: \be T(z) = {{:
d^{ab}J_aJ_b(z) :} \over {k+h^{\vee}}}, \label{SS def T(z) for
SL(N)}. \ee As required, for every $k \neq {-h^{\vee}}$, the modes
of the Laurent expansion of $T(z)$ will span a Virasoro algebra.
In particular, $T(z)$ will generate holomorphic reparametrisations
of the coordinates on the worldsheet $\Sigma$. Notice that this
definition of $T(z)$ in (\ref{SS def T(z) for SL(N)}) is
ill-defined when $k=-h^{\vee}$. Nevertheless, one can always
associate $T(z)$ with the Segal-Sugawara operator $S(z)$ that is
well-defined at any finite level, whereby \be S(z) =  (k+h^{\vee})
T(z), \label{S(z) for SL(N)} \ee and \be S(z) = {: d^{ab} J_a J_b
(z):}. \label{s(z) for sl(N)} \ee From (\ref{S(z) for SL(N)}), we
see that $S(z)$ generates, in its OPE's with other field
operators, $(k+h^{\vee})$ times the transformations usually
generated by the stress tensor $T(z)$. Therefore, at the level $k=
-h^{\vee}$, $S(z)$ generates no transformations at all - its OPE's
with all other field operators are trivial. This is equivalent to
saying that the holomorphic stress tensor does not exist at the
quantum level, since $S(z)$, which is the only well-defined
operator at this level that could possibly generate the
transformation of fields under an arbitrary holomorphic
reparametrisation of the worldsheet coordinates on $\Sigma$, acts
by zero in the OPE's.

Despite the fact that $S(z)$ will cease to exist in the spectrum
of physical operators associated to the twisted sigma-model on
$X=SL(N)/B$ at the quantum level, it will nevertheless exist as a
field in its classical ${\overline Q}_+$-cohomology or holomorphic
chiral algebra. One can convince oneself that this is true as
follows. Firstly, from our ${\overline Q}_+$-Cech cohomology
dictionary, since the $J_a(z)$'s are in $H^0(X, {\widehat{\cal
O}}^{ch}_{X})$, it will mean that they are in the $\overline
Q_+$-cohomology of the sigma-model at the quantum level. Secondly,
since quantum corrections can only annihilate cohomology classes
and not create them, it will mean that the $J_a(z)$'s will be in
the classical $\overline Q_+$-cohomology of the sigma-model, i.e.,
the currents are $\overline Q_+$-closed and are therefore
invariant under the transformations generated by $\overline Q_+$
in the absence of quantum corrections. Hence, one can readily see
that $S(z)$ in (\ref{s(z) for sl(N)}) will also be $\overline
Q_+$-closed at the classical level. Lastly, recall from section
2.3 that $[\overline Q _+, T(z)] = 0$ such that $T(z) \neq \{
{\overline Q}_+, \cdots \}$  in the absence of quantum corrections
to the action of $\overline Q_+$ in the classical theory. Note
also that the integer $h^{\vee}$ in the factor $(k+h^{\vee})$ of
the expression $S(z)$ in (\ref{S(z) for SL(N)}), is due to a shift
by $h^{\vee}$ in the level $k$ because of quantum renormalisation
effects \cite{Fuchs}, i.e., the classical expression of $S(z)$ for
a general level $k$ can actually be written as $S(z) = k T(z)$,
and therefore, one will have $[\overline Q_+, -h^{\vee} T(z)] =
[\overline Q_+, S(z)] = 0$, where $S(z) \neq \{ {\overline Q}_+,
\cdots \}$ in the classical theory. Therefore, $S(z)$ will be a
spin-two field in the classical holomorphic chiral algebra of the
purely bosonic sector of the twisted sigma-model on $X= SL(N)/B$.
This observation is also consistent with the fact that $S(z)$
fails to correspond to a global section of the sheaf ${\cal
O}^{ch}_{X}$ of CDO's - note that in our case, we actually have
$S(z) = -h^{\vee} T(z)$ in the classical theory, and this will
mean that under quantum corrections to the action of $\overline
Q_+$, we will have $[\overline Q_+, S_{zz}] = -h^{\vee}
\partial_z(R_{i \bar j}
\partial_z \phi^i \psi^{\bar j}) \neq 0$ (since $R_{i\bar j} \neq 0$ for any flag manifold $SL(N)/B$), which corresponds in the Cech
cohomology picture to the expression ${\widetilde{\widehat S}(z)}
- {\widehat S}(z) \neq 0$, i.e., ${\widehat S}(z)$, the Cech
cohomology counterpart to the $S(z)$ operator, will fail to be in
$H^0(X, {\widehat {\cal O}}^{ch}_{X})$. Consequently, one can
always represent $S(z)$ by a classical $c$-number. This point will
be important when we discuss how one can define Hecke eigensheaves
that will correspond to flat $^LG$-bundles on a Riemann surface
$\Sigma$ in our physical interpretation of the geometric Langlands
correspondence for $G=SL(N)$.

The fact that $S(z)$ acts trivially in any OPE with other field
operators implies that its Laurent modes will commute with the
Laurent modes of any of these other field operators; in
particular, they will commute with the Laurent modes of the
$J_a(z)$ currents - in other words, the Laurent modes of $S(z)$
will span the centre    ${\frak z}(\widehat{\frak{sl}}_N)$ of the
completed universal enveloping algebra of the affine      $SL(N)$
algebra $\widehat{\frak{sl}}_N$ at the critical level
$k=-h^{\vee}$ (generated by the Laurent modes of the $J_a(z)$
currents in the quantum chiral algebra of the twisted sigma-model
on $SL(N)/B$). Notice also that $S(z)$ is $\psi^{\bar
j}$-independent and is therefore purely bosonic in nature. In
other words, the local field $S(z)$ exists only in the
$\it{classical}$ holomorphic chiral algebra of the $\it{purely}$
$\it{bosonic}$ (or $\psi^{\bar j}$-independent) sector of the
twisted sigma-model on $X= SL(N)/B$.

\bigskip\noindent{\it A Classical Virasoro Algebra}

Note that since $S(z)$ is holomorphic in $z$ and is of conformal
dimension two, one can expand it in terms of a Laurent expansion
as \be S(z) = \sum_{n \in {\mathbb Z}} {\hat S}_n z^{-n-2}.
\label{quantum S(z) for SL(N)} \ee Recall that for the general
case of $k \neq -h^{\vee}$, a quantum definition of $S(z)$ exists,
such that the ${\hat S}_n$ modes of the Laurent expansion can be
related to the $J_{a,n}$ modes of the $\widehat{\frak {sl}}_N$
currents through the quantum commutator relations
\begin{eqnarray}
[{\hat S}_n, J_{a, m}] & = & -(k + h^{\vee}) m J_{a, n+m},\\
{[{\hat S}_n, {\hat S}_m ]} & = & (k+ h^{\vee}) \left( (n-m) {\hat
S}_{n+m} + { k \ \textrm{dim}\ {\frak{sl}_N}  \over 12} \  (n^3-n)
\ \delta_{n, -m}\right), \label{S_n for SL(N)}
\end{eqnarray}
where $a = 1, 2, \dots, \textrm{dim}\ {\frak{sl}_N}$. If we now
let $k = - h^{\vee}$, we will have $[{\hat S}_n, J_{a,m}] = [{\hat
S}_n, {\hat S}_m] =0$ - the $S_m$'s thus generate the (classical)
centre of the completed universal enveloping algebra of
$\widehat{\frak {sl}}_N$ as mentioned above.

Since we now understand that $S(z)$ must be a holomorphic
classical field at $k =-h^{\vee}$, let us rewrite the Laurent
expansion of $S(z)$ as \be S(z) = \sum_{n \in {\mathbb Z}} S_n
z^{-n-2} \label{S(z) classical for SL(N)} \ee
 so as to differentiate the classical modes of expansion $S_n$ from their quantum counterpart ${\hat S}_n$ in (\ref{S(z) for SL(N)}). Unlike the ${\hat S}_n$'s which obey the quantum commutator relations in (\ref{S_n for SL(N)}) for an arbitrary level $k \neq -h^{\vee}$, the $S_n$'s, being the modes of a Laurent expansion of a classical field, will instead obey Poisson bracket relations that define a certain classical algebra when $k=-h^{\vee}$.

Based on our arguments thus far, we learn that the quantum version
of $S(z)$ as expressed in (\ref{S(z) for SL(N)}), must reduce to
its classical counterpart as expressed in (\ref{S(z) classical for
SL(N)}), when $k = -h^{\vee}$. In other words, one can see that by
taking $(k+h^{\vee}) \to 0$, we are going to the classical limit.
This is analogous to taking the  ${\hbar} \to 0$ limit in any
quantum mechanical theory whenever one wants to ascertain its
classical counterpart. In fact, by identifying $(k+h^{\vee})$ with
$i \hbar$, and by noting that one must make the replacement from
Possion brackets to commutators via $\{S_n, S_m \}_{P.B.}
\rightarrow {1\over {i \hbar}} [ {\hat S}_n, {\hat S}_m ]$ in
quantising the $S_n$'s into operators, we can ascertain the
classical algebra generated by the $S_n$'s from (\ref{S_n for
SL(N)}) as \be \{S_n, S_m\}_{P.B.} = (n-m) {S}_{n+m} - {h^{\vee}
(\textrm{dim}{\frak{sl}_N}) \over 12}\ (n^3-n) \ \delta_{n, -m}.
\label{poisson brackets of S_n for SL(N)} \ee Since we have the
classical relation $S(z) \sim T(z)$, it means that we can
interpret the $S_n$ modes as the Virasoro modes of the Laurent
expansion of the classical stress tensor field $T(z)$. In other
words, the $S_n$'s generate a classical Virasoro algebra with
central charge $- h^{\vee} (\textrm{dim} {\frak{sl}_N}) $ as given
by (\ref{poisson brackets of S_n for SL(N)}). This is can be
denoted mathematically as the Virasoro Poisson algebra
$Sym'(vir_{- h^{\vee}\cdot \textrm{dim}{\frak{sl}_N} })$.

\bigskip\noindent{\it Higher-Spin Analogs of the Segal-Sugawara Tensor and the Classical Holomorphic Chiral Algebra}

For an affine $SL(N)$ algebra where $N > 2$,  one can generalise
the Sugawara formalism to construct higher-spin analogs of the
holomorphic stress tensor with the currents.  These higher-spin
analogs are called Casimir operators, and were first constructed
in \cite{casimir operators}.

In the context of an affine $SL(N)$ algebra with a module that is
furnished by the global sections of the sheaf of CDO's on
$X=SL(N)/B$, a spin-$s_i$ analog of the holomorphic stress tensor
will be given by the ${s_i}^{th}$-order Casimir operator
\cite{review}
 \be
T^{(s_i)} (z) = {{:{\tilde d}^{a_1 a_2 a_3 \dots a_{s_i}} (k)
(J_{a_1} J_{a_2}\dots J_{a_{s_i}})(z):} \over {k+h^{\vee}}},
\label{casimir of spin-three for SL(N)} \ee where ${\tilde
d}^{a_1a_2 a_3 \dots a_{s_i}}(k)$ is a completely symmetric
traceless $\frak{sl}_N$-invariant tensor of rank $s_i$ that
depends on the level $k$ of the affine $SL(N)$ algebra. It is also
well-defined and finite at $k=-h^{\vee}$. The superscript on
$T^{(s_i)}(z)$ just denotes that it is a spin-$s_i$ analog of
$T(z)$. Note that $i = 1,2, \dots, \textrm{rank}(\frak{sl}_N)$,
and  the spins $s_i$ can take the values $1+ e_i$, where $e_i = 1,
2, \dots, N-1$. Thus, one can have $\textrm{rank}(\frak{sl}_N)$ of
these Casimir operators, and the spin-2 Casimir operator is just
the holomorphic stress tensor from the usual Sugawara
construction.

As with $T(z)$ in (\ref{SS def T(z) for SL(N)}), $T^{(s_i)} (z)$
is ill-defined when $k = -h^{\vee}$. Nevertheless, one can always
make reference to a spin-$s_i$ analog of the Segal-Sugawara tensor
$S^{(s_i)}(z)$ that is well-defined for any finite value of $k$,
where its relation to $T^{(s_i)}(z)$ is given by \be S^{(s_i)}(z)
= (k+h^{\vee}) T^{(s_i)}(z), \ee and \be S^{(s_i)}(z) = :{\tilde
d}^{a_1 a_2 a_3 \dots a_{s_i}}(k) (J_{a_1} J_{a_2}\dots
J_{a_{s_i}})  (z):. \label{S^{(s_i)}(z)} \ee That is, the operator
$S^{(s_i)}(z)$ generates in its OPE's with all other operators of
the quantum theory,  $(k+h^{\vee})$ times the field
transformations typically generated by $T^{(s_i)}(z)$.

Notice  however, that at $k=-h^{\vee}$, $S^{(s_i)}(z)$ acts by
zero in its OPE with any other operator. This is equivalent to
saying that $T^{(s_i)}(z)$ does not exist as a quantum operator at
all, since the only well-defined operator $S^{(s_i)}(z)$ which is
supposed to generate the field transformations associated to
$T^{(s_i)}(z)$, act by zero and thus generate no field
transformations at all. From our $\overline Q_+$-Cech cohomology
dictionary, this means that the $\psi^{\bar i}$-independent
operator $T^{(s_i)}(z)$ will fail to correspond to a dimension
$s_i$ global section of $\widehat{\cal O}^{ch}_X$. Since we have,
at the classical level, the relation $S^{(s_i)}(z) = -h^{\vee}
T^{(s_i)}(z)$, it will mean that $S^{(s_i)}(z)$ will also fail to
correspond to a dimension $s_i$ global section of $\widehat{\cal
O}^{ch}_X$. Thus, $S^{(s_i)}(z) $ will fail to be an operator at
the quantum level. Is it even a spin-$s_i$ field in the classical
holomorphic chiral algebra of the twisted sigma-model on
$SL(N)/B$, one might ask. The answer is yes. To see this, recall
that each of the $J_{a_k}(z)$'s are separately $\overline
Q_+$-invariant and not $\overline Q_+$-exact at the classical
level. Therefore, the classical counterpart of $S^{(s_i)}(z)$ in
(\ref{S^{(s_i)}(z)}) must also be such, which in turn means that
it will be in the classical $\overline Q_+$-cohomology and hence
classical chiral algebra of the twisted sigma-model on $SL(N)/B$.

The fact that the $S^{(s_i)}(z)$'s act trivially in any OPE with
other field operators implies that their Laurent modes will
commute with the Laurent modes of any other operator; in
particular, they will commute with the Laurent modes of the
currents $J_a(z)$ for $a=1,2, \dots, \textrm{dim} {\frak{sl}_N}$ -
in other words, the Laurent modes of all $\textrm{rank}(
{\frak{sl}_N})$ of the $S^{(s_i)}(z)$ fields will span fully the
centre ${\frak z}(\widehat{\frak{sl}}_N)$ of the completed
universal enveloping algebra of $\widehat{\frak{sl}}_N$ at the
critical level $k=-h^{\vee}$ (which is in turn generated by the
Laurent modes of the $J_a(z)$ currents of the quantum chiral
algebra of the twisted sigma-model on $SL(N)/B$). Last but not
least, notice that the $S^{(s_i)}(z)$ fields are also $\psi^{\bar
j}$-independent and are therefore purely bosonic in nature. In
other words, the local fields $S^{(s_i)}(z)$, for $i=1,2,\dots
\textrm{rank}(\frak{sl}_N)$, whose Laurent modes will together
generate ${\frak z}(\widehat{\frak{sl}}_N)$, exist only in the
$\it{classical}$ holomorphic chiral algebra of the $\it{purely}$
$\it{bosonic}$ (or $\psi^{\bar j}$-independent) sector of the
twisted sigma-model on $X=SL(N)/B$.


\bigskip\noindent{\it A Classical ${\cal W}_N$-algebra}

For an affine $SL(N)$ algebra at an arbitrary level $k \neq
-h^{\vee}$, as in the case of $S(z)$ discussed earlier, a quantum
definition of $S^{(s_i)}(z)$ exists. In fact, consider the
following operators given by ${\overline S^{(s_i)}}(z)=
\eta^{(s_i)}(N):d^{a_1a_2 \dots a_{s_i}} (J_{a_1}J_{a_2} \dots
J_{a_{s_i}})(z):$ for $i=1, 2, \dots, \textrm{rank}(\frak
{sl}_N)$, where $\eta^{(s_i)}(N) $ is just a normalisation that
depends on $N$, and $d^{a_1a_2 \dots a_{s_i}}$ is just a
rank-$s_i$ extension of $d^{ab}$. It can be shown that the
${\overline S}^{(s_i)}(z)$'s generate a $\it{closed}$ Casimir OPE
algebra which is isomorphic to a particular ${\cal W}_N$ OPE
algebra \cite{casimir operators}. This implies that for $k\neq
-h^{\vee}$, every ${\overline S}^{(s_i)}(z)$ and therefore every
$S^{(s_i)}(z) \sim :d^{a_1a_2 \dots a_{s_i}} (J_{a_1}J_{a_2 }
\dots J_{a_{s_i}})(z):$, will exist as a quantum operator in some
cohomology - the $\overline Q_+$-cohomology in this instance. This
will in turn mean that the set of $S^{(s_i)}(z) = (k+h^{\vee})
T^{(s_i)}(z)$ operators must also span a closed OPE algebra that
is equivalent - at the level of $\overline Q_+$-cohomology - to
this Casimir OPE algebra, when $k \neq -h^{\vee}$. Since we know
that for $k \neq -h^{\vee}$, $T^{(2)}(z) =T(z)$ will generate a
Virasoro subalgebra of a closed ${\cal W}_N$ OPE algebra with
central charge $c = k \ \textrm{dim}(\frak{sl}_N)  /
(k+h^{\vee})$, it will imply that the  $S^{(s_i)}(z)$'s will
satisfy a rescaled (by a factor of $(k+h^{\vee})$) version of a
closed ${\cal W}_N$ OPE algebra at $c = k \
\textrm{dim}(\frak{sl}_N)  / (k+h^{\vee})$ for $k\neq -h^{\vee}$.
Because each $S^{(s_i)}(z)$ is holomorphic in $z$, we can Laurent
expand it as \be S^{(s_i)}(z) = \sum_{n \in {\mathbb Z}} {\hat
S}^{(s_i)}_n z^{-n-{s_i}}. \label{S^{(s_i)}(z) quantum for SL(N)}
\ee At $k \neq -h^{\vee}$, since $S^{(2)}(z) = S(z) $, the Laurent
modes ${\hat S}^{(2)}_n$ of $S^{(2)}(z)$, will then generate the
Virasoro algebra with the following quantum commutator relations
given in (\ref{S_n for SL(N)}): \be {[{\hat S}^{(2)}_n, {\hat
S}^{(2)}_m ]}  =  (k+ h^{\vee}) \left( (n-m) {\hat S}^{(2)}_{n+m}
+ { k \ \textrm{dim}\ {\frak{sl}_N}  \over 12} \  (n^3-n) \
\delta_{n, -m}\right). \label{S^{(2)}(z) quantum for SL(N)} \ee
Likewise, the Laurent modes of the other spin-$s_i$ operators
$S^{(s_i)}(z)$ will generate (up to a factor of $(k+h^{\vee})$
like in (\ref{S^{(2)}(z) quantum for SL(N)})) the quantum
commutator relations of a ${\cal W}_N$ algebra. Since we shall not
need to refer explicitly to these relations in our following
discussion, we shall omit them for brevity, as they tend to get
very elaborate for $N \geq 4$.

Now let us consider the case when $k = -h^{\vee}$. From our
earlier explanations about the nature of the $S^{(s_i)}(z)$
operators, we find that they will cease to exist as quantum
operators at $k=-h^{\vee}$. Since we understand that the
$S^{(s_i)}(z)$'s must be  holomorphic classical fields at $k
=-h^{\vee}$, we shall rewrite the Laurent expansion of
$S^{(s_i)}(z)$ as \be S^{(s_i)}(z) = \sum_{n \in {\mathbb Z}}
S^{(s_i)}_n z^{-n-{s_i}}, \label{S^{(s_i)}(z) classical for SL(N)}
\ee so as to differentiate the classical modes of expansion
$S^{(s_i)}_n$ from their quantum counterpart ${\hat S}^{(s_i)}_n$
in (\ref{S^{(s_i)}(z) quantum for SL(N)}). Unlike the ${\hat
S}^{(s_i)}_n$'s which obey the quantum commutator relations of a
${\cal W}_N$-algebra for an arbitrary level $k \neq -h^{\vee}$,
the $S^{(s_i)}_n$'s, being the modes of a Laurent expansion of a
classical field, will instead obey Poisson bracket relations that
define a certain classical algebra at $k=-h^{\vee}$. Since every
${\hat S}^{(s_i)}_n$ must reduce to its classical counterpart
${S}^{(s_i)}_n$ at $k = -h^{\vee}$,  one can see that by taking
$(k+h^{\vee}) \to 0$, we are actually going to the classical
limit. This is analogous to taking the  ${\hbar} \to 0$ limit in
any quantum mechanical theory whenever one wants to ascertain its
classical counterpart. In fact, by identifying $(k+h^{\vee})$ with
$i \hbar$, and by noting that one must make the replacement from
Possion brackets to commutators via $\{E^{(s_i)}_n, E^{(s_j)}_m
\}_{P.B.} \rightarrow {1\over {i \hbar}} [ {\hat E}^{(s_i)}_n,
{\hat E}^{(s_j)}_m ]$ in quantising any classical mode
$E^{(s_i)}_n$ into an operator  ${\hat E}^{(s_i)}_n$, we can
ascertain the classical algebra generated by the $S^{(s_i)}_n$'s
from the ${\cal W}_N$-algebra commutator relations that they
satisfy. Since the $S^{(s_i)}(z)$ fields all lie in the classical
$\overline Q_+$-cohomology of the twisted sigma-model on
$SL(N)/B$, it will mean that their Laurent modes $S^{(s_i)}_n$
will generate a closed, classical algebra as well.\footnote{Note
at this point that if $\cal O$ and $\cal O'$ are non-exact,
$Q$-closed observables in some $Q$-cohomology, i.e., ${\{Q, \cal
O\}}={\{Q, {\cal O}'\}}= 0$, then $\{Q, {\cal O}{\cal O}'\} =0$.
Moreover, if $\{Q, {\cal O}\}=0$, then ${\cal O}\{Q, W\}= \{Q,
{\cal O}W\}$ for any observable $W$.  These two statements mean
that the cohomology classes of observables that commute with $Q$
form a closed and well-defined algebra.} In fact, they will
generate a closed classical ${\cal W}_N$-algebra. In order to
ascertain the central charge of this classical ${\cal
W}_N$-algebra, it suffices to determine the central charge of its
classical Virasoro subalgebra generated by the $S^{(2)}_m$'s. From
(\ref{S^{(2)}(z) quantum for SL(N)}), we find  that as $k \to
-h^{\vee}$, the $S^{(2)}_m$'s satisfy \be \{{S}^{(2)}_n,
{S}^{(2)}_m \}_{P.B.}  =   (n-m) {S}^{(2)}_{n+m} - { h^{\vee} \
\textrm{dim}\ {\frak{sl}_N}  \over 12} \  (n^3-n) \ \delta_{n,
-m}, \label{S^{(2)}(z) classical for SL(N)} \ee the classical
Virasoro algebra with central charge $c = - h^{\vee}
\textrm{dim}(\frak {sl}_N)$. Hence, the $S^{(s_i)}_n$'s will
generate a classical ${\cal W}_N$-algebra with central charge $c =
- h^{\vee} \textrm{dim}(\frak {sl}_N)$. Indeed for the case of
$\frak g= \frak{sl}_2$ as analysed in $\S$2.2, the modes
$S^{(2)}_m$ was shown to generate a classical ${\cal W}_2$-algebra
with central charge $c= - h^{\vee} \textrm{dim}(\frak {sl}_2)
=-6$, where $h^{\vee} =2$ and $ \textrm{dim}(\frak {sl}_2) =3$.
Likwise for the case of $\frak g= \frak{sl}_3$ as analysed in
$\S$2.3, the modes $S^{(2)}_m$ and $S^{(3)}_m$ was shown to
generate a classical ${\cal W}_3$-algebra with central charge $c=
- h^{\vee} \textrm{dim}(\frak {sl}_2) =-24$, where $h^{\vee} = 3$
and $ \textrm{dim}(\frak {sl}_3) =8$.

Last but not least, recall that the Laurent modes of the
$S^{(s_i)}(z)$ fields for $i=1, 2, \dots \newline
\textrm{rank}(\frak{sl}_N)$, will together generate
$\frak{z}(\widehat{\frak{sl}}_N)$, the centre of the completed
universal enveloping algebra of the affine $SL(N)$ algebra
$\widehat{\frak{sl}}_N$ at the critical level $k = -h^{\vee}$. If
we denote the classical ${\cal W}_N$-algebra with central charge
$c= - h^{\vee} \textrm{dim}(\frak {sl}_N)$ as ${\cal W}_N(-
h^{\vee} \textrm{dim}(\frak {sl}_N))$, we will have an
identification of Poisson algebras
$\frak{z}(\widehat{\frak{sl}}_N) \simeq {\cal W}_N(- h^{\vee}
\textrm{dim}(\frak {sl}_N))$.

\newsubsection{$\cal W$-Algebras from an Algebraic Drinfeld-Sokolov Reduction Scheme}

We shall now review a purely algebraic approach to generating
${\cal W}_{k'}(\widehat{\frak{sl}}_N)$, the $\cal W$-algebra
associated to the affine $SL(N)$ algebra $\widehat{\frak{sl}}_N$
at level $k'$. This approach is known as the quantum
Drinfeld-Sokolov (DS) reduction scheme \cite{Frenkel,
DS-reduction}.

In general, the quantum DS-reduction scheme can be summarised as
the following steps. Firstly, one starts with a triple
$(\widehat{\frak g}, \widehat{\frak g}', \chi)$, where
$\widehat{\frak g}'$ is an affine subalgebra of $\widehat{\frak
g}$ at level $k'$, and $\chi$ is a 1-dimensional representation of
$\widehat{\frak g}'$. Next, one imposes the first class
constraints $ g \sim \chi(g)$ , $\forall g \in {\widehat{\frak
g}'}$, via a BRST procedure. The cohomology of the BRST operator
$Q$ on the set of normal-ordered expressions in currents, ghosts
and their derivatives, is what is called the Hecke algebra
$H^i_Q(\widehat{\frak g}, \widehat{\frak g}', \chi)$ of the triple
$(\widehat{\frak g}, \widehat{\frak g}', \chi)$. For generic
values of $k'$, the Hecke algebra vanishes for $i \neq 0$, and the
existing zeroth cohomology $H^0_Q (\widehat{\frak g},
\widehat{\frak g}', \chi)$, is just spanned by a set of local
operators associated to the triple $(\widehat{\frak g},
\widehat{\frak g}', \chi)$, whose Laurent modes generate a closed
$\cal W$-algebra. We shall denote the $\cal W$-algebra associated
with this set of operators  as ${\cal W}_{DS}[\widehat{\frak g},
\widehat{\frak g}', \chi]$. Note that ${\cal
W}_{DS}[\widehat{\frak {sl}}_N, \widehat{\frak {sl}}_N', \chi]$ is
just ${\cal W}_{k'}(\widehat{\frak {sl}}_N)$, the $\cal W$-algebra
associated to $\widehat{\frak {sl}}_N$ at level $k'$ whose quantum
and classical limits we encountered in $\S$2 for $N=2,3$. Let us
be more explicit about how one can go about defining ${\cal
W}_{DS}[\widehat{\frak g}, \widehat {\frak g}', \chi]$ and
therefore ${\cal W}_{k'}(\widehat{\frak {sl}}_N)$,  now that we
have sketched the general idea behind the DS-reduction scheme.

In order for ${\cal W}_{DS} [\widehat{\frak g}, \widehat{\frak
g}', \chi]$ to be a ${\cal W}$-algebra, one has to suitably choose
the triple $(\widehat{\frak g}, \widehat{\frak g}',\chi)$. A
suitable triple can be obtained by considering a principal
$\frak{sl}_2$ embedding in $\frak g$. Let us now describe this
embedding. Suppose we have an $\frak {sl}_2$ subalgebra $\{t_3,
t_+, t_-\}$ of $\frak g$. The adjoint representation of $\frak g$
decomposes into $\frak{sl}_2$ representations of spin $j_k$ , $k =
1,\dots, s$, for example. Then, one may write the $\widehat {\frak
g}$ current $J(z) = \sum^{\textrm{dim}{\frak g}}_a J^a (z) t_a$ as
\be J(z) = \sum^s_{k=1} \sum^{j_k}_{m= -j_k}  J^{k,m} (z) t_{k,m}
\ee where $t_{k,m}$ corresponds to the generator of spin $j_k$ and
isospin $m$ under the $\frak{sl}_2$ subalgebra. In particular, we
have the correspondences $t_{1,1} = t_+$, $t_{1,0} = t_3$ , and
$t_{1,-1} = t_-$. The $\frak {sl}_2$ subalgebra ${t_3, t_+, t_-}$
can be characterized by a ``dual Weyl vector'' $\rho^{\vee}$,
i.e., for $\alpha \in \vartriangle_+$, where $\vartriangle_+$ is
the set of positive roots of $\frak g$, we have $(\rho^{\vee},
\alpha) =1$ if and only if $\alpha$ is a simple root of $\frak g$.
The $\frak{sl}_2$ root $\hat \alpha$ is given by $\hat \alpha =
\rho/(\rho, \rho)$, and $t_3 = \rho \cdot {\frak c}$, where $\frak
c$ is the Cartan sublagebra of $\frak g$.

Take $\widehat {\frak g}'$ to be the affine Lie subalgebra
$\widehat {\frak n}_+$ generated by all $J^{k,m}(z), m > 0$.
Denoting the currents corresponding to positive roots $\alpha$ by
$J^{\alpha}(z)$, and choosing $t_{1,1} = \sum_i e^{\alpha_i}$, one
can then impose the condition (which realises the required
first-class constraint $g \sim \chi(g)$) \be
\chi_{DS}(J^{\alpha}(z)) = 1 \ (\textrm{for simple roots} \
\alpha_i, ), \quad \chi(J^{\alpha}(z)) =0 \  (\textrm{otherwise}).
\label{constraint} \ee

Next, we introduce pairs of ghost fields $(b^{\alpha}(z),
c_{\alpha}(z))$, one for every positive root $\alpha \in
\vartriangle_+$. By definition, they obey the OPE $b_{\alpha}(z)
c_{\beta} (z') \sim \delta_{\alpha \beta} / (z-z')$, where the
$\alpha, \beta$ (and $\gamma$) indices run over the basis of
$\frak n_+$. The BRST operator that is consistent with
(\ref{constraint}) will then be given by $Q = Q_0 + Q_1$, where
\be Q_0 = \oint {dz \over {2\pi i}} \ \left (J^{\alpha}(z)
c_{\alpha}(z) - {1\over 2}f^{\alpha \beta}_{\gamma} (b^{\gamma}
c_{\alpha}c_{\beta})(z) \right) \label{Q_0} \ee is the standard
differential associated to $\widehat{\frak n}_+$, $f^{\alpha
\beta}_{\gamma}$ are the structure constants of ${\frak n}_+$, and
\be Q_1 =   -\oint {dz \over {2\pi i}}\ \chi_{DS}(J^{\alpha}(z))
c_{\alpha}(z). \label{Q_1} \ee They satisfy \be Q^2 = Q^2_0 =
Q^2_1 = \{Q_0, Q_1\} = 0. \label{Q relations} \ee The resulting
$Q$-cohomology is just the Hecke algebra $H^0_Q (\widehat{\frak
g}, \widehat{\frak g}', \chi)$, which is spanned by a set of local
operators whose Laurent modes generate ${\cal
W}_{DS}[\widehat{\frak g}, \widehat{\frak g}', \chi] = {\cal
W}_{k'}(\widehat{\frak g})$. Note that (\ref{Q relations}) implies
that one can compute the Hecke algebra via a spectral sequence of
a double complex with differentials being $Q_0$ and $Q_1$
accordingly - this strategy has indeed been employed in \cite{de
Boer} to compute explicitly the generators of the ${\cal W}_2 =
{\cal W}_{k'}(\widehat{\frak {sl}}_2)$ and ${\cal W}_3 = {\cal
W}_{k'}(\widehat{\frak {sl}}_3)$ OPE algebras with central charges
$c =13-6(k'+2) - 6/(k'+2)$ and $c =50 - 24(k'+3) - 24/(k'+3)$
respectively. We shall have more to say about these $\cal
W$-algebras shortly.

The variation of the various fields under the action of $Q$ can
also be computed using the OPE's of the affine algebra
$\widehat{\frak g}$, the OPE's of the ghost fields, and the
explicit forms of $Q_0$ and $Q_1$ in (\ref{Q_0}) and (\ref{Q_1})
above, and they are given by
 \begin{eqnarray}
 \label{field variations for DS-1 }
\delta c_{\alpha}(z) & = & -\frac{1}{2}f^{\beta \gamma}_{\alpha}(
c_{\beta}c_{\gamma}) (z),  \\
\label{field variations for DS-2 } \delta b^{\alpha}(z) & = &
J^{\alpha} (z) - \chi_{DS} (J^{\alpha}(z)) - f^{\alpha
\beta}_{\gamma} (b^{\gamma} c_{\beta})(z).
\end{eqnarray}

Note also that ${\cal W}_{DS}[\widehat{\frak g}, \widehat{\frak
g}', \chi]$ and thus ${\cal W}_{k'}(\widehat{\frak g})$, will at
least contain the Virasoro algebra. The explicit form of the
stress tensor whose Laurent modes will generate the Virasoro
algebra is (after omitting the normal-ordering symbol)\be
T_{DS}(z) = {{d_{ab}J^a(z)J^b(z)} \over {(k'+h^{\vee})}} +
\sum^{\textrm{dim}\frak c}_{c=1} \partial_z J^c(z) + \sum_{\alpha
\in \vartriangle_+} ((\rho^{\vee}, \alpha) -1)b^{\alpha}\partial_z
c_{\alpha}(z) + (\rho^{\vee}, \alpha)(\partial_z b^{\alpha}
c_{\alpha})(z), \label{stress tensor of DS-reduction}\ee where the
$J^c(z)$'s are just the affine currents that are valued in the
Cartan subalgebra $\frak c$ of the Lie algebra $\frak g$. Note
that with respect to $T_{DS}(z)$, the conformal dimensions of the
pair $(b^{\alpha}(z), c_{\alpha}(z))$ will be given by
$(1-(\rho^{\vee}, \alpha), (\rho^{\vee}, \alpha))$. The central
charge of this Virasoro subalgebra and therefore that of ${\cal
W}_{k'}(\widehat{\frak g})$, will be given by \be c =
{{k'\textrm{dim} {\frak g}}\over {(k'+ h^{\vee})}} -
12k'|\rho^{\vee}|^2 - 2 \sum_{\alpha \in {\vartriangle}_+} \left(
6(\rho^{\vee}, \alpha)^2 - 6(\rho^{\vee}, \alpha)
+1\right).\label{central charge of DS-reduction}\ee

\newsubsection{The $B$-Gauged WZW Model on $SL(N)$ and the Algebraic Drinfeld-Sokolov Reduction Scheme}

\smallskip\noindent{\it The ${\cal W}_{k'}({\widehat
{\frak {sl}}_2})$ Algebra from the DS-Reduction Scheme}

\smallskip

 Note that from (\ref{central charge of DS-reduction}),
for $\frak g = {\frak{sl}_2}$, where $\textrm{dim}({\frak{sl}_2})
=3$, $h^{\vee}=2$, $\rho^{\vee} = \rho$, $|\rho^{\vee}|^2 = 1/2$,
and $(\rho^{\vee}, \alpha)=1$, we find that the central charge of
the resulting algebra ${\cal W}_{k'}(\widehat {\frak{sl}}_2)$
generated by the Laurent modes of the local operators that span
the $Q$-cohomology, will be given by $c = 3k' / (k'+2) - 6k' -2 =
13-6(k'+2) - 6/(k'+2)$. This is exactly the $\cal W$-algebra that
the Laurent modes of the $T_{\textrm{total}}(z)$ operator which
span the holomorphic BRST-cohomology of the $B$-gauged WZW model
on $SL(2)$, generate.

In addition, for $\frak g = \frak{sl}_2$, we have from
(\ref{stress tensor of DS-reduction}), the stress tensor \be
T_{DS}(z) = T_{SL(2)} +
\partial_z J^1(z) + (\partial_z b^1)(c_1)(z), \ee where the conformal
dimensions of $(b^1, c_1)$ are $(0,1)$ respectively. Thus, we see
that $T_{DS}(z)$ is exactly $T_{\textrm{total}}(z)$ of the
$B$-gauged WZW model on $SL(2)$, and that moreover, the ghost
fields $(b^1, c_1)$ of the DS-reduction scheme have the same
conformal dimensions as the $(b^1, c^1_z)$ ghost fields of the WZW
model.

The field variations (\ref{field variations for DS-1
})-(\ref{field variations for DS-2 }) can in this case be written,
(after noting that $\frak n_+ \in \frak{sl}_2$ is abelian and
hence has vanishing structure constants), as
\begin{eqnarray}
 \label{field variations for DS for SL(2)-1}
\delta c_1(z) & = & 0,  \\
\label{field variations for DS for SL(2)-2} \delta b^1(z) & = &
J^1(z) - \chi_{DS} (J^1(z)).
\end{eqnarray}
These variations coincide exactly with the holomorphic
BRST-variations in (\ref{BRST variations SL(2)}) of the $B$-gauged
WZW model on $SL(2)$ after one makes an identification between the
$\it{arbitrary}$ constant $M^1_+$ and $-\chi_{DS}(J^1(z)) = -1$.
Moreover, the BRST-charge $Q = Q_0 +Q_1$ which generates the
variations in (\ref{field variations for DS for
SL(2)-1})-(\ref{field variations for DS for SL(2)-2}) will be
given by \be Q= \oint {dz \over {2 \pi i}}\left(J^1 (z) -
\chi_{DS}(J^1(z))\right)c_1(z). \ee Notice that $Q$ also coincides
with $Q_{\textrm{BRST}}$ of (\ref{Q_BRST,WZW SL(2)}) - the
holomorphic BRST-charge of the $B$-gauged WZW model on $SL(2)$.

In summary, we find that the holomorphic BRST-cohomology of the
$B$-gauged WZW model on $SL(2)$, furnishes a physical realisation
of the purely algebraic DS-reduction scheme of generating the
Hecke algebra associated to ${\cal W}_{k'}(\widehat {\frak
{sl}}_2)$. The classical limit of ${\cal W}_{k'}(\widehat {\frak
{sl}}_2)$ - given by ${\cal W}_{\infty}(\widehat {\frak {sl}}_2)$
- is indeed the classical $\cal W$-algebra generated by the
Laurent modes of the field ${\overline T}_{\textrm{classical}}(z)$
in the classical, holomorphic BRST-cohomology of the $B$-gauged
WZW model on $SL(2)$.

\bigskip\noindent{\it The ${\cal W}_{k'}({\widehat
{\frak {sl}}_3})$ Algebra from the DS-Reduction Scheme}

\smallskip

Likewise, note that from (\ref{central charge of DS-reduction}),
for $\frak g = {\frak{sl}_3}$, where $\textrm{dim}({\frak{sl}_2})
=8$, $h^{\vee}=3$, $\rho^{\vee} = \rho$, $|\rho^{\vee}|^2 = 2$,
$(\rho^{\vee}, \alpha_1)=1$, $(\rho^{\vee}, \alpha_2)=1$, and
$(\rho^{\vee}, \alpha_3)=2$, we find that the central charge of
the resulting algebra ${\cal W}_{k'}(\widehat {\frak{sl}}_3)$
generated by the Laurent modes of local operators in the
$Q$-cohomology, will be given by $c = 8k'/(k'+3) -24k' -30 = 50 -
24(k'+3) - 24/(k'+3)$. This is exactly the $\cal W$-algebra that
the Laurent modes of the $T_{\textrm{total}}(z)$ and
$T^{(3)}_{\textrm{total}}(z)$ operators in the holomorphic
BRST-cohomology of the $B$-gauged WZW model on $SL(3)$, generate.

In addition, for $\frak g = \frak{sl}_3$, we have from
(\ref{stress tensor of DS-reduction}), the stress tensor \be
T_{DS}(z) = T_{SL(3)} + \partial_z J^1(z) + \partial_z J^2(z) +
(\partial_z b^1)(c_1)(z) + (\partial_z b^2)(c_2)(z) + 2(\partial_z
b^3)(c_3)(z) + (b^3)(\partial_z c_3)(z), \ee where the conformal
dimensions of $(b^1, c_1)$, $(b^2,c_2)$ and $(b^3,c_3)$ are
$(0,1)$, $(0,1)$ and $(-1,2)$, respectively. Thus, we see that
$T_{DS}(z)$ is exactly $T_{\textrm{total}}(z)$ of the $B$-gauged
WZW model on $SL(3)$, and that moreover, the ghost fields $(b^1,
c_1)$, $(b^2,c_2)$ and $(b^3,c_3)$ of the DS-reduction scheme have
the same conformal dimensions as the $(b^1, c^1_z)$, $(b^2,c^2_z)$
and $(b^{3,z},c^3_{zz})$ ghost fields of the WZW model.

The field variations (\ref{field variations for DS-1
})-(\ref{field variations for DS-2 }) in this case can be written
as
\begin{eqnarray}
 \label{field variations for DS for SL(3)-1}
\delta c_{\alpha}(z) & = & -\frac{1}{2}f^{\beta \gamma}_{\alpha}(
c_{\beta}c_{\gamma}) (z),  \\
\label{field variations for DS for SL(3)-2} \delta b^{\alpha}(z)
&= & J^{\alpha} (z) - \chi_{DS} (J^{\alpha}(z)) - f^{\alpha
\beta}_{\gamma} (b^{\gamma} c_{\beta})(z),
\end{eqnarray}
where $\alpha, \beta, \gamma = 1,2,3$. Notice that these
variations coincide exactly with the holomorphic BRST-variations
in (\ref{BRST variations SL(3)}) of the $B$-gauged WZW model on
$SL(3)$ after one makes an identification between the
$\it{arbitrary}$ constants $M^{\alpha}_+$ and
$-\chi_{DS}(J^{\alpha}(z))$. Moreover, the BRST-charge $Q = Q_0
+Q_1$ which generates the variations in (\ref{field variations for
DS for SL(3)-1})-(\ref{field variations for DS for SL(3)-2}) will
be given by \be Q = \oint {dz \over {2\pi i}} \ \sum_{\alpha,
\beta, \gamma = 1}^3 \left( (J^{\alpha}(z) -
\chi_{DS}(J^{\alpha}(z))) c_{\alpha}(z) - {1\over 2}f^{\alpha
\beta}_{\gamma} (b^{\gamma} c_{\alpha}c_{\beta})(z) \right).\ee
Notice that $Q$ also coincides with $Q_{\textrm{BRST}}$ of
(\ref{Q_BRST,WZW SL(3)-clean}) - the holomorphic BRST-charge of
the $B$-gauged WZW model on $SL(3)$.

In summary, we find that the holomorphic BRST-cohomology of the
$B$-gauged WZW model on $SL(3)$, furnishes a physical realisation
of the purely algebraic DS-reduction scheme of generating the
Hecke algebra associated to ${\cal W}_{k'}(\widehat {\frak
{sl}}_3)$. The classical limit of ${\cal W}_{k'}(\widehat {\frak
{sl}}_3)$ - given by ${\cal W}_{\infty}(\widehat {\frak {sl}}_3)$
- is indeed the classical $\cal W$-algebra generated by the
Laurent modes of the fields ${\overline
T}_{\textrm{classical}}(z)$ and ${\overline
T}^{(3)}_{\textrm{classical}}(z)$ in the classical, holomorphic
BRST-cohomology of the $B$-gauged WZW model on $SL(3)$.

\bigskip\noindent{\it The $B$-Gauged WZW Model on $SL(N)$ and the ${\cal W}_{k'}(\widehat{\frak{sl}}_N)$ Algebra}

As one might have already guessed, the above observations about
the physical realisation of the algebraic DS-reduction scheme via
the holomorphic BRST-cohomology of a $B$-gauged WZW model on
$SL(N)$, is actually valid for all $N$, not just $N=2,3$. Let us
substantiate this statement now with a discussion of the
BRST-quantisation of the $B$-gauged WZW model on $SL(N)$, which,
for the cases of $SL(2)$ and $SL(3)$, we have already described
explicitly in $\S$2.

Recall from (\ref{B-gauged WZW action SL(N)}) in $\S$2.2 that the
action of the $B$-gauged WZW model on $SL(N)$ takes the form
\begin{eqnarray}
S_{\textrm{B-gauged}} (g, A_{z}, A_{\bar z}, J^+, {\bar J}^+)& = &
S_{\textrm{WZ}} (g) - {k' \over {2\pi}} \int_{\Sigma} d^2z \
\textrm{Tr} [ A_{\bar z}( J^+(z) + {M}^+)- A_{z} ( {\bar J}^+(\bar
z) + {\bar M}^+)\nonumber \\ && \hspace{4.0cm} - A_z g A_{\bar z}
g^{-1} + A_z A_{\bar z}]. \label{B-gauged WZW action SL(N)section
4}\end{eqnarray} As explained in $\S$2.2, with respect to the
Cartan decomposition $\frak {sl}_N = {\frak n}_- \oplus \frak c
\oplus {\frak n}_+$, one can write $J(z) =
\sum_{a=1}^{\textrm{dim}{{\frak n}_-}} J^a_- (z) t^{-}_a +
\sum_{a=1}^{\textrm{dim}{{\frak c}}} J^a_c(z) t^{c}_a +
\sum_{a=1}^{\textrm{dim}{{\frak n}_+}} J^a_+(z) t^{+}_a$, ${\bar
J}(\bar z) = \sum_{a=1}^{\textrm{dim}{{\frak n}_-}} {\bar J}^a_-
(\bar z) t^{-}_a + \sum_{a=1}^{\textrm{dim}{{\frak c}}} {\bar
J}^a_c(\bar z) t^{c}_a + \sum_{a=1}^{\textrm{dim}{{\frak n}_+}}
{\bar J}^a_+(\bar z) t^{+}_a$, $A_{z}= \sum_{a
=1}^{\textrm{dim}{\frak n}_+} {\tilde A}_{z}^a t^+_a$ and $A_{\bar
z}= \sum_{a =1}^{\textrm{dim}{\frak n}_+}  {\tilde A}_{\bar z}^a
t^+_a$, where $t^{-}_a \in {\frak n}_-$, $t^{c}_a \in {\frak c}$,
and $t^{+}_a \in {\frak n}_+$. One can also write $M =
\sum_{a=1}^{\textrm{dim}{{\frak n}_-}} M^a_- t^{-}_a +
\sum_{a=1}^{\textrm{dim}{{\frak c}}} M^a_c t^{c}_a +
\sum_{a=1}^{\textrm{dim}{{\frak n}_+}} M^a_+ t^{+}_a$, where
$M^a_{\pm ; c}$ are arbitrary number constants, and one can also
write $\bar M = \sum_{a=1}^{\textrm{dim}{{\frak n}_-}} \bar M^a_-
t^{-}_a + \sum_{a=1}^{\textrm{dim}{{\frak c}}} \bar M^a_c t^{c}_a
+ \sum_{a=1}^{\textrm{dim}{{\frak n}_+}} \bar M^a_+ t^{+}_a$,
where $\bar M^a_{\pm ; c}$ are arbitrary number constants. Then,
one can write (\ref{B-gauged WZW action SL(N)section 4}) as
\begin{eqnarray} S_{{SL(N)}} (g, A_z, A_{\bar z}, J^+, \bar J^+)&
= & S_{\textrm{WZ}} (g) - {k' \over {2\pi}} \int_{\Sigma} d^2z \
\sum_{l=1}^{\textrm{dim}{\frak n}_+} \left[ {\tilde A}^l_{\bar z}(
J^l_+(z) + {M}^l_+) - {\tilde A}^l_{z}( \bar J^l_+(\bar z) + {\bar
M}^l_+) \right]\nonumber \\
&& \hspace{3.5cm} - \textrm {Tr} [ {A}_z g {A}_{\bar z} g^{-1} -
{A}_z {A}_{\bar z}] \label{B-gauged WZW action SL(N)section 4
effective}
\end{eqnarray}

Due to the $B$-gauge invariance of the theory, we must divide the
measure in any path integral computation by the volume of the
$B$-gauge symmetry. That is, the partition function has to take
the form \be Z_{SL(N)} = \int_{\Sigma} { {[g^{-1}dg, d{\tilde
A}^l_{z}, d{\tilde A}^l_{\bar z}]} \over {(\textrm{gauge
volume})}} \ \textrm{exp} \left(i S_{SL(N)}(g, A_z, A_{\bar z},
J^+, \bar J^+) \right). \ee One must now fix this gauge invariance
to eliminate the non-unique degrees of freedom. One can do this by
employing the BRST formalism which requires the introduction of
Faddev-Popov ghost fields. In order to obtain the
$\it{holomorphic}$ BRST transformations of the fields, one simply
replaces the infinitesimal position-dependent parameters
${\epsilon}^l$ of $h= B = \textrm{exp}( -
\sum_{l=1}^{\textrm{dim}{\frak n}_+} \epsilon^l t^+_m)$ in the
corresponding $\textrm{\it left-sector}$ of the gauge
transformations in (\ref{gauge tx}) with the ghost fields $c^l$,
which then gives us \be \delta_{\textrm{BRST}}(g) = -c^l t^+_l g,
\quad \delta_{\textrm{BRST}}(\tilde A^l_{\bar z}) = - D_{\bar z}
c^l, \quad \delta_{\textrm{BRST}}(\textrm{others}) =0. \label{BRST
tx SL(N)} \ee The components of the ghost field $c(z) =
\sum_{l=1}^{\textrm{dim}{\frak n}_+} c^l (z) t^+_l$ and those of
its anti-ghost partner $b(z) = \sum_{l=1}^{\textrm{dim}{\frak
n}_+} b^l (z) t^+_l$ will transform as \be \delta_{\textrm{BRST}}
(c^l) = - {1\over 2}f_{mk}^l c^mc^k, \quad
\delta_{\textrm{BRST}}(b^l) = {\tilde B}^l, \quad
\delta_{\textrm{BRST}} {(\tilde B^l)} = 0, \ee where the
$f^l_{mk}$'s are the structure constants of the nilpotent
subalgebra ${\frak n}_+$. Also, the ${\tilde B}^l$'s are the
Nakanishi-Lautrup auxiliary fields that are the BRST transforms of
the $b^l$'s. They also serve as Lagrange multipliers to impose the
gauge-fixing conditions.

In order to obtain the $\it{antiholomorphic}$ BRST transformations
of the fields, one employs the same recipe to the corresponding
$\textrm{\it right-sector}$ of the gauge transformations in
(\ref{gauge tx}) with the infinitesimal position-dependent gauge
parameter now replaced by the ghost field $\bar c^l$, which then
gives us \be \bar \delta_{\textrm{BRST}}(g) = \bar c^l t^+_l g,
\quad \bar \delta_{\textrm{BRST}}({\tilde A}^l_{z}) = - D_{z} \bar
c^l, \quad \bar\delta_{\textrm{BRST}}(\textrm{others})
=0.\label{BRST tx 1 SL(N)}\ee The components of the ghost field
${\bar c}(\bar z) = \sum_{l=1}^{\textrm{dim}{\frak n}_+} {\bar
c}^l (\bar z) t^+_l$ and those of its anti-ghost partner $\bar
b(\bar z) = \sum_{l=1}^{\textrm{dim}{\frak n}_+} {\bar b}^l (\bar
z) t^+_l$ will transform as \be \bar\delta_{\textrm{BRST}} (\bar
c^l) = - {1\over 2}f_{mk}^l {\bar c}^m {\bar c}^k, \quad
\bar\delta_{\textrm{BRST}}(\bar b^l) = {\tilde {\bar B}^l}, \quad
\bar\delta_{\textrm{BRST}} {(\tilde {\bar B}^l)} = 0. \ee In the
above, the $\tilde {\bar B}^l$'s are the Nakanishi-Lautrup
auxiliary fields that are the antiholomorphic BRST transforms of
the $\bar b^l$ fields. They also serve as Lagrange multipliers to
impose the gauge-fixing conditions.

Since the BRST transformations in (\ref{BRST tx SL(N)}) and
(\ref{BRST tx 1 SL(N)}) are just infinitesimal versions of the
gauge transformations in (\ref{gauge tx}), $S_{{SL(N)}} (g, A_z,
A_{\bar z}, J^+, \bar J^+)$ will be invariant under them. As in
the $SL(2)$ and $SL(3)$ cases, in addition to
$(\delta_{\textrm{BRST}} + \bar\delta_{\textrm{BRST}}) \cdot
(\delta_{\textrm{BRST}} + \bar\delta_{\textrm{BRST}}) = 0$, the
holomorphic and antiholomorphic BRST-variations are also
separately nilpotent, i.e., $\delta^2_{\textrm{BRST}} = 0$ and
$\bar\delta^2_{\textrm{BRST}}=0$. Moreover,
$\delta_{\textrm{BRST}}\cdot \bar\delta_{\textrm{BRST}}= - \bar
\delta_{\textrm{BRST}} \cdot \delta_{\textrm{BRST}}$. This means
that the BRST-cohomology of the $B$-gauged WZW model on $SL(N)$
can be decomposed into $\it{independent}$ holomorphic and
antiholomorphic sectors that are just complex conjugate of each
other, and that it can be computed via a spectral sequence,
whereby the first two complexes will be furnished by its
holomorphic and antiholomorphic BRST-cohomologies respectively.
Since we will only be interested in the holomorphic chiral algebra
of the $B$-gauged WZW model on $SL(N)$ (which as mentioned, is
just identical to its antiholomorphic chiral algebra by a complex
conjugation), we shall henceforth focus on the $\it{holomorphic}$
BRST-cohomology of the $B$-gauged WZW model on $SL(N)$.

By the usual recipe of the BRST formalism, one can fix the gauge
by adding to the BRST-invariant action $S_{{SL(N)}} (g, A_z,
A_{\bar z}, J^+, \bar J^+)$, a BRST-exact term. Since the BRST
transformation by $(\delta_{\textrm{BRST}} + \bar
\delta_{\textrm{BRST}})$  is nilpotent, the new total action will
still be BRST-invariant as required. The choice of the BRST-exact
operator will then define the gauge-fixing conditions. A
consistent choice of the BRST-exact operator that will give us the
requisite action for the ghost and anti-ghost fields is \be
S_{{SL(N)}} (g, A_z, A_{\bar z}, J^+, \bar J^+) +
(\delta_{\textrm{BRST}} + \bar \delta_{\textrm{BRST}})
\left({k'\over 2\pi} \int_{\Sigma} d^2 z \
\sum_{l=1}^{\textrm{dim}{\frak n}_+} {\tilde A}^l_{\bar z} b^l +
{\tilde A}^l_{z} {\bar b}^l \right),\nonumber \ee where one will
indeed have the desired total action, which can be written as
\begin{eqnarray} \label{total desired action for SL(N)}
S_{\textrm{WZW}}(g)  -  {k' \over {2\pi}} \int_{\Sigma} d^2z \ \{
\sum_{l=1}^{\textrm{dim}{\frak n}_+} \left[ {\tilde A}^l_{\bar z}(
J^l_+(z) + {M}^l_+ - \tilde B^l) - {\tilde A}^l_{z}( \bar
J^l_+(\bar z) + {\bar M}^l_+ + \tilde {\bar B}^l)
\right]  \nonumber \\
 - \textrm {Tr} [ {A}_z g {A}_{\bar z} g^{-1} - {A}_z {A}_{\bar
z}] \} +  {k'\over {2\pi}}\int_{\Sigma} d^2z \
\sum_{l=1}^{\textrm{dim}{\frak n}_+} \left ( c^l
D_{\bar z} b^l + + \bar c^l D_{z} \bar b^l \right). \nonumber \\
\end{eqnarray} From the equations of motion by varying the ${\tilde
B}^l$'s, we have the conditions $\tilde A^l_{\bar z} = 0$ for
$l=1,\dots, {\textrm{dim}{\frak n}_+}$. From the equations of
motion by varying the $\tilde {\bar B}^l$'s, we also have the
conditions $\tilde A^l_{z} = 0$ for $l=1,\dots,{\textrm{dim}{\frak
n}_+}$. Thus, the partition function of the $B$-gauged WZW model
can also be expressed as
 \be
Z_{SL(N)} = \int [ g^{-1} dg, db, dc, d\bar b, d\bar c] \
\textrm{exp} \left ( iS_{\textrm{WZW}}(g) + {i k'\over
{2\pi}}\int_{\Sigma} d^2z \ \textrm{Tr} (c \cdot \partial_{\bar z}
b) (z) + \textrm{Tr} (\bar c \cdot \partial_{z} \bar b) (\bar z)
\right), \label{Z_SL(N)} \ee where the $\it{holomorphic}$ BRST
variations of the fields which leave the effective action in
(\ref{Z_SL(N)}) $\it{invariant}$ are now given by \begin{eqnarray}
\delta_{\textrm{BRST}} (g) = -c^mt^+_m g, & \quad
\delta_{\textrm{BRST}} (c^l) = -{1\over 2} f^l_{mk}c^m c^k,
\quad \delta_{\textrm{BRST}}(b^l) =  J^l_+ + M^l_+ - f^l_{mk}b^mc^k,\nonumber \\
\hspace{-2.0cm}\quad \delta_{\textrm{BRST}}(\textrm{others}) = 0.
& \label{BRST variations SL(N)}
 \end{eqnarray}

The holomorphic BRST-charge generating the field variations in
(\ref{BRST variations SL(N)}) will be given by \be
Q_{\textrm{BRST}} = \oint {dz \over {2 \pi i}} \left(
\sum_{l=1}^{\textrm{dim}{\frak n}_+} \ c^l (z) (J^l_+(z) + M^l_+)
-{1\over 2} \sum_{l=1}^{\textrm{dim}{\frak n}_+} f^l_{mk}b^mc^lc^k
(z) \right). \label{Q_BRST,WZW SL(N)} \ee The free-field action of
the left-moving ghost fields in (\ref{Z_SL(N)}) implies that we
have the usual OPE's of ($\textrm{dim}{\frak n}_+$) free $bc$
systems. From these free $bc$ OPE's, one can verify that
$Q_{\textrm{BRST}}$ in (\ref{Q_BRST,WZW SL(N)}) will indeed
generate the field variations in (\ref{BRST variations SL(N)}).

Again, though we did not make this obvious in our discussion
above, by integrating out the $\tilde A^l_{\bar z}$'s in
(\ref{B-gauged WZW action SL(N)section 4 effective}), and using
the above conditions $\tilde A^l_{z} =0$ for $l=1,\dots,
\textrm{dim}{\frak n}_+$, we find that we actually have the
relations $(J^l_+(z) + M^l_+) =0$ for $l=1,\dots,
\textrm{dim}{\frak n}_+$. These relations (involving the current
associated to the Borel subalgebra $\frak b$ of the group $B$ that
we are modding out by) will lead us directly to the correct form
of the holomorphic stress tensor for the gauged WZW model without
reference to a coset formalism.

Note that as in the $SL(2)$ and $SL(3)$ cases of $\S$2, physically
consistent with the holomorphic chiral algebra of the purely
bosonic sector of the twisted sigma-model on $SL(N)/B$, there are
currents $J^a(z)$ in the holomorphic BRST-cohomology of the
non-dynamically $B$-gauged WZW model on $SL(N)$, where $a=1,2,
\dots, \textrm{dim}{\frak{sl}}_N$, that generate an affine $SL(N)$
OPE algebra at level $k'$. As such, one can construct a
holomorphic stress tensor using the Sugawara formalism as \be
T_{SL(N)}(z) = {: d_{ab}(J^a J^b) (z):\over {k'+ h^{\vee}}}. \ee
However, as shown above, one will have the conditions $J^l_+ =
-M^l_+$ for $l=1,2,\dots, \textrm{dim}{\frak n}_+$. In order that
the conformal dimensions of the $J^l_+$'s be compatible with these
conditions, one must define a modified holomorphic stress tensor:
\be T_{\textrm{modified}}(z) = T_{SL(N)}(z) + \vec{l} \cdot
\partial\vec{J_c}(z),\ee where $\vec{J_c}(z)$ is an
$(N-1)$-dimensional vector with components being the $J^l_c$
currents associated to the Cartan subalgebra $\frak c$, and
$\vec{l}$ is a sum of simple, positive roots of $\frak{sl}_N$. In
order for the above conditions on the $J^l_+$'s be compatible with
the fact that $Q_{\textrm{BRST}}$ generating the holomorphic
variations $\delta_{\textrm{BRST}}(b^l)$ of (\ref{BRST variations
SL(N)}) must be a scalar of dimension zero, the
$(\textrm{dim}{\frak n}_+)$-set of left-moving ghost systems
$(b^l, c^l)$ must have conformal dimensions $(1-h, h)$ for
$l=1,2,\dots, \textrm{dim}{\frak n}_+$, where $h$ is the conformal
dimension of the corresponding $J^l_+$ current  under
$T_{\textrm{modified}}(z)$. With all these points in mind, and by
including the holomorphic stress tensor contribution from the
action of the left-moving ghost fields, we can write the total
holomorphic stress tensor of the $B$-gauged WZW model on $SL(N)$
as \be T_{\textrm{total}}(z)  = {: d_{ab}(J^a J^b) (z):\over {k'+
h^{\vee}}} + \sum_{a=1}^{\textrm{dim}{\frak c}}
\partial_z J^a_c (z) +  \sum_{l
\in \vartriangle_+} ((\rho^{\vee}, l) -1)b^{l}\partial_z c^{l}(z)
+ (\rho^{\vee}, l)(\partial_z b^{l} c^{l})(z),\ee where
$\vartriangle_+$ is the set of positive roots of $\frak{sl}_N$,
and $\rho^{\vee}$ is the ``dual Weyl vector'' of $\frak{sl}_N$.
Notice that $T_{\textrm{total}}(z)$ is just $T_{DS}(z)$ in
(\ref{stress tensor of DS-reduction}) for $\widehat{\frak g} =
\widehat{\frak{sl}}_N$. Moreover, the central charge of
$T_{\textrm{total}}(z)$ will be given by \be c = N^4 -1 - N(N^2
-1)\left ({1\over {k'+N}} + k' + N\right) - (N^4 - 2N^3 +N), \ee
which can be rewritten as \be c = {{k'\textrm{dim} ({\frak
{sl}_N})}\over {(k'+ h^{\vee})}} - 12k'|\rho^{\vee}|^2 - 2 \sum_{l
\in {\vartriangle}_+} \left( 6(\rho^{\vee}, l)^2 - 6(\rho^{\vee},
l) +1\right),\ee since $h^{\vee} = N$ for $\frak{sl}_N$, and
$(N^2-1) = \textrm{dim}(\frak {sl}_N)$. This coincides with the
central charge of $T_{DS}(z)$ in (\ref{central charge of
DS-reduction}) for $\frak g = \frak{sl}_N$.

Note also that for any $J^l_+$ with $h\neq 0$, the corresponding
$M^l_+$ constant must be set to zero for consistency. This means
that we can identify $M^l_+$ with $-\chi_{DS}(J^l_+(z))$. Hence,
we can write the holomorphic BRST-variations in (\ref{BRST
variations SL(N)}) as \be \delta_{\textrm{BRST}} (g) = -c^mt^+_m
g,  \quad \delta_{\textrm{BRST}} (c^l) = -{1\over 2} f^l_{mk}c^m
c^k, \quad \delta_{\textrm{BRST}}(b^l) =  J^l_+ -\chi_{DS}(J^I_+)
- f^l_{mk}b^mc^k,\ee which just coincides with the BRST-variations
of the DS-reduction scheme  in (\ref{field variations for DS-1
})-(\ref{field variations for DS-2 }) for $\frak g = \frak{sl}_N$.
Last but not least, the holomorphic BRST-charge which generates
these field transformations can also be written as \be
Q_{\textrm{BRST}} = \oint {dz \over {2 \pi i}} \left(
\sum_{l=1}^{\textrm{dim}{\frak n}_+} \ c^l (z) \left(J^l_+(z) -
\chi_{DS}(J^l_+(z))\right) -{1\over 2}
\sum_{l=1}^{\textrm{dim}{\frak n}_+} f^l_{mk}b^mc^lc^k (z)
\right). \ee This just coincides with the sum of $Q_0$ and $Q_1$
in (\ref{Q_0}) and (\ref{Q_1}), i.e., it coincides with $Q$ of the
DS-reduction scheme for $\frak g = \frak{sl}_N$.

In summary, we find that the holomorphic BRST-cohomology of the
$B$-gauged WZW model on $SL(N)$, will indeed furnish a physical
realisation of the purely algebraic DS-reduction scheme of
generating the Hecke algebra that is spanned by local operators
whose Laurent modes generate a ${\cal W}_{k'}(\widehat {\frak
{sl}}_N)$ algebra with central charge (\ref{central charge of
DS-reduction}). Consequently, the classical limit of ${\cal
W}_{k'}(\widehat {\frak {sl}}_N)$, i.e., ${\cal
W}_{\infty}(\widehat {\frak {sl}}_N)$, will be given by the
Poisson $\cal W$-algebra generated by the Laurent modes of the
classical counterparts of the local operators which lie in the
classical, holomorphic  BRST-cohomology of the $B$-gauged WZW
model on $SL(N)$. We shall discuss this set of classical fields
next, and their role in an isomorphism of classical $\cal
W$-algebras and a level relation that underlie a geometric
Langlands correspondence for any $G=SL(N)$.

\newsubsection{An Equivalence of Classical Holomorphic Chiral Algebras and a Geometric Langlands Correspondence for Any $SL(N)$}

Via a straightforward extension of our arguments in $\S$2 on the
$SL(2)$ and $SL(3)$ cases to all $SL(N)$,  we find that the
equivalence - at the level of the holomorphic chiral algebra -
between the purely bosonic sector of the twisted sigma-model on
$SL(N)/B$ and the $B$-gauged WZW model on $SL(N)$, will imply an
isomorphism between the closed classical algebra generated by the
Laurent modes of the $S^{(s_i)}(z)$ fields in the classical,
holomorphic chiral algebra of the $\psi^{\bar j}$-independent
sector of the twisted sigma-model on $SL(N)/B$, and the closed
classical algebra generated by the Laurent modes of the
$\it{corresponding}$ ${\overline
T}^{(s_i)}_{\textrm{classical}}(z) = -h^{\vee}\cdot
{T}^{(s_i)}_{\textrm{classical}}(z)$ fields in the classical,
holomorphic BRST-cohomology of the $B$-gauged WZW model on
$SL(N)$. Here, the ${\overline T}^{(s_i)}_{\textrm{classical}}(z)
= \sum_n {\overline L}^{(s_i)}_n z^{-n-s_i}$ fields are just the
classical counterparts of the ${\overline
T}^{(s_i)}_{\textrm{total}}(z) = (k+h^{\vee})\
{T}^{(s_i)}_{\textrm{total}}(z)$ operators that exist in the
quantum, holomorphic BRST-cohomology of the gauged WZW model at
$k\neq-h^{\vee}$, whereby the Laurent modes of the
${T}^{(s_i)}_{\textrm{total}}(z)$ operators will generate the
${\cal W}_{k'}(\widehat{\frak {sl}}_N)$-algebra discussed above.

 Recall from our earlier discussion that
the Laurent modes of the $S^{s_i}(z)$ fields will generate a
classical ${\cal W}_N$-algebra that contains a Virasoro subalgebra
given by \be \{{S}^{(2)}_n, {S}^{(2)}_m \}_{P.B.}  = (n-m)
{S}^{(2)}_{n+m} - { h^{\vee} \ \textrm{dim}\ {\frak{sl}_N} \over
12} \  (n^3-n) \ \delta_{n, -m}.\label{classical algebra 1}\ee
This classical Virasoro subalgebra has central charge $c = -
h^{\vee} \textrm{dim}(\frak {sl}_N)$. Hence, the Laurent modes
$S^{(s_i)}_n$ of the $S^{(s_i)}(z)$ fields will generate a
classical ${\cal W}_N$-algebra with central charge $c = - h^{\vee}
\textrm{dim}(\frak {sl}_N)$, which, we had denoted earlier as
${\cal W}_N(-h^{\vee}\textrm{dim}(\frak {sl}_N))$.

On the other hand, the Laurent modes of the ${\overline
T}^{(s_i)}_{\textrm{classical}}(z)$ fields will generate a
classical ${\cal W}_{\infty}(\widehat{\frak{sl}}_N)$-algebra,
which, consistent with the equivalence of the holomorphic chiral
algebras of the $\psi^{\bar j}$-independent sector of the twisted
sigma-model on $SL(N)/B$ and the $B$-gauged WZW model on $SL(N)$,
is also a classical ${\cal W}_N$-algebra. Likewise, the central
charge of this classical ${\cal W}_{\infty}$-algebra will be given
by the central charge of its classical Virasoro subalgebra. Its
classical Virasoro subalgebra is given by \be \{{{\overline
L}}^{(2)}_n, {{\overline L}}^{(2)}_m \}_{P.B.} = (n-m) {{\overline
L}}^{(2)}_{n+m} + {{c(k,k')_{k\to -h^{\vee}, k'\to \infty}} \over
12} \ (n^3 - n) \delta_{n, -m}, \label{classical algebra 2} \ee
where \be c(k,k') = (k+h^{\vee})\left(N^4 -1 - N(N^2 -1)\left
({1\over {(k'+h^{\vee})}} + (k' + h^{\vee})\right) - (N^4 - 2N^3
+N)\right),\ee and ${\overline L}^{(2)}_n = -h^{\vee}\cdot
L^{(2)}_n$ corresponds to $S^{(2)}_n$, while $L^{(2)}_n$ is a
Virasoro mode of the classical counterpart
$T^{(2)}_{\textrm{classical}}(z)$ of
$T^{(2)}_{\textrm{total}}(z)$. Therefore, the central charge of
the ${\cal W}_{\infty}(\widehat{\frak{sl}}_N)$-algebra generated
by the Laurent modes of the ${\overline
T}^{(s_i)}_{\textrm{classical}}(z)$ fields will be given by
$c(k,k')$ where $k\to -h^{\vee}$ and $k'\to \infty$.

An isomorphism between the classical $\cal W$-algebras generated
by the $S^{(s_i)}_n$'s and the ${\overline L}^{(s_i)}_n$'s
necessarily implies an isomorphism between the classical
subalgebras generated by the $S^{(2)}_n$'s and ${\overline
L}^{(2)}_n$'s in (\ref{classical algebra 1}) and (\ref{classical
algebra 2}) respectively. This in turn means that we must have the
relation \be c(k,k')_{k\to -h^{\vee}, k'\to \infty} = -{{h^{\vee}
\textrm{dim}\ \frak{sl}_N}}.\ee In the examples studied in $\S2$
where $N=2,3$, we saw that the above relation would hold if and
only $(k+h^{\vee})(k'+h^{\vee}) =1$. One can quickly verify that
this would also be the case for $\it{any}$ $N$: notice that the
surviving term in $c(k,k')_{k\to -h^{\vee}, k'\to \infty}$ is just
$-N (N^2 -1)(k+h^{\vee})(k'+h^{\vee})$, and since $N = h^{\vee}$
and $(N^2-1) = \textrm{dim}({\frak {sl}}_N)$, we will have
$c(k,k')_{k\to -h^{\vee}, k'\to \infty} = -h^{\vee}
\textrm{dim}({\frak {sl}}_N)$ $\textrm{\it if and only if}$
$(k+h^{\vee})(k'+h^{\vee})=1$, whence the classical ${\cal
W}_N(-h^{\vee}\textrm{dim}(\frak {sl}_N))$-algebra will be
isomorphic to the ${\cal
W}_{\infty}(\widehat{\frak{sl}}_N)$-algebra with central charge
$c(k,k')_{k\to -h^{\vee}, k'\to \infty} = -h^{\vee} \textrm{dim}
(\frak{sl}_N)$. Since ${\cal W}_N(-h^{\vee} \textrm{dim}
(\frak{sl}_N)) \simeq \frak z(\widehat{\frak{sl}}_N)$, and since
for ${\frak g} = \frak{sl}_N = {^L\frak g}$, $h^{\vee} =
{^Lh^{\vee}}$, and $r^{\vee}=1$, we thus see that an equivalence -
at the level of the holomorphic chiral algebra - between the
$\psi^{\bar j}$-independent sector of the twisted sigma-model on
$SL(N)/B$ and the $B$-gauged WZW model on $SL(N)$, would imply an
isomorphism of Poisson algebras \be \frak z(\widehat{\frak g})
\simeq {\cal W}_{\infty}(^L{\widehat{\frak g}}),\label{isomorphism
of W-algebras SL(N)}\ee and the level relation \be
(k+h^{\vee})r^{\vee} = {1 \over {(k' +
{^Lh^{\vee}})}}.\label{langlands level duality for sl(N)}\ee Note
at this point that the purely bosonic, $\psi^{\bar j}$-independent
sector of the twisted sigma-model on $SL(N)/B$, can be described,
via (\ref{Sbosonic explicit}), by a bosonic string on $SL(N)/B$.
On the other hand, note that since a bosonic string on a group
manifold $G$ can be described as a WZW model on $G$, it will mean
that the $B$-gauged WZW model on $SL(N)$ can be interpreted as a
$B$-gauged bosonic string on $SL(N)$. Thus, we see that an
equivalence, at the level of the holomorphic chiral algebra,
between a bosonic string on $SL(N)/B$ and a $B$-gauged version of
itself on $SL(N)$ - a statement which stems from the ubiquitous
notion that one can always physically interpret a geometrical
symmetry of the target space as a gauge symmetry in the worldsheet
theory - will imply an isomorphism of classical $\cal W$-algebras
and a level relation that underlie a geometric Langlands
correspondence for $G=SL(N)$! Note that the correspondence between
the $k \to -h^{\vee}$ and $k' \to \infty$ limits (within the
context of the above Poisson algebras) is indeed consistent with
the relation (\ref{langlands level duality for sl(N)}). These
limits define a ``classical'' geometric Langlands correspondence.
A ``quantum'' generalisation of the $SL(N)$ correspondence can be
defined for other values of $k$ and $k'$ that satisfy the relation
(\ref{langlands level duality for sl(N)}), but with the
isomorphism of (\ref{isomorphism of W-algebras SL(N)}) replaced by
an isomorphism of $\it{quantum}$ $\cal W$-algebras (derived from a
DS-reduction scheme) associated to $\widehat{\frak {sl}}_N$ at
levels $k$ and $k'$ respectively \cite{Frenkel}.

\newsection{About the Hecke Eigensheaves and Hecke Operators}

We shall now  demonstrate, via the isomorphism of classical $\cal
W$-algebras discussed in $\S$3, how one can derive a
correspondence between flat holomorphic $^LG$-bundles on the
worldsheet $\Sigma$ and Hecke eigensheaves on the moduli space
$\textrm{Bun}_G$ of holomorphic $G$-bundles on $\Sigma$, where
$G=SL(N)$. In the process, we will be able to physically interpret
the Hecke eigensheaves and Hecke operators of the geometric
Langlands program for $G=SL(N)$, in terms of the correlation
functions of purely bosonic local operators in the holomorphic
chiral algebra of the twisted $(0,2)$ sigma-model on the complex
flag manifold $SL(N)/B$.

\newsubsection{Hecke Eigensheaves on $\textrm{Bun}_{SL(N)}$ and Flat $^LSL(N)$-Bundles on $\Sigma$}

\bigskip\noindent{\it Local Primary Field Operators}

As we will explain shortly, the correlation functions of local
primary field operators can be associated to the sought-after
Hecke eigensheaves. As such, let us begin by describing these
operators in the twisted $(0,2)$ sigma-model on a complex flag
manifold $X=SL(N)/B$. By definition, the holomorphic primary field
operators $\Phi^{\lambda}_s(z)$ of any theory with an affine
$SL(N)$ OPE algebra obey \cite{CFT textbook Fransceco et al.} \be
J^a(z)\Phi^{\lambda}_r (z') \sim - \sum_{s}
{{(t^a_{\lambda})_{rs}\
\Phi^{\lambda}_s}(z')\over{z-z'}},\label{primary field OPE's}\ee
where $t^a_{\lambda}$ is a matrix in the $\lambda$ representation
of ${\frak{sl}}_N$, $r,s = 1, \dots, \textrm{dim}|\lambda|$, and
$a=1,\dots, \textrm{dim}(\frak{sl}_N)$.

Since the $\Phi^{\lambda}_s (z)$'s obey OPE relations with the
quantum operators $J^a(z)$, it will mean that they, like the
$J^a(z)$'s, must exist as $\it{quantum}$ bosonic operators of the
sigma-model on $X$. And moreover, since (\ref{primary field
OPE's}) and the affine $SL(N)$ OPE algebra at the critical level
generated by the $J^a(z)$'s in the $\overline Q_+$-cohomology of
the quantum sigma-model together form a closed OPE algebra, it
will mean that the $\Phi^{\lambda}_s (z)$'s are also local
operators in the $\overline Q_+$-cohomology of the sigma-model on
$X$ at the quantum level. From our $\overline Q_+$-Cech cohomology
dictionary, this means that the $\Phi^{\lambda}_s (z)$'s will
correspond to classes in $H^0(X, {\cal O}^{ch}_X)$, i.e., the
global sections of the sheaf ${\cal O}^{ch}_X$ of  CDO's on $X$.
Note that this observation is also consistent with (\ref{primary
field OPE's}) - one can generate other global sections of the
sheaf ${\cal O}^{ch}_X$ from the OPE's of existing global
sections. (Recall that we did this to generate the $J_3(z)$
current from the OPE of the $J_-(z)$ and $J_+(z)$ currents of the
affine $SL(2)$ OPE algebra  when we studied the sigma-model on
$SL(2)/B$ in $\S$2).

The fact that these operators can be described by global sections
of the sheaf of CDO's on $X$ means that they reside within the
purely bosonic sector of the holomorphic chiral algebra of the
underlying sigma-model on $X$. As we shall see, this observation
will serve as a platform for a physical interpretation of the
Hecke eigensheaves.

\bigskip\noindent{\it Space of Coinvariants}

Associated to the correlation functions of the above-described
local primary field operators, is the concept of a space of
coinvariants, which, in its interpretation as a sheaf over the
moduli space of holomorphic $G$-bundles on $\Sigma$ that we will
clarify below, is directly related to the Hecke eigensheaves that
we are looking for. Hence, let us now turn our attention to
describing this space of coinvariants.

Notice that if the twisted sigma-model were to be conformal, i.e.,
$\{\overline Q_+, T_{zz}\} = 0$ even after quantum corrections, we
would have a CFT operator-state isomorphism, such that any primary
field operator $\Phi^{\lambda}_s (z)$ would correspond to a state
$|\Phi^{\lambda}_s \rangle$ in the highest-weight representation
of $\widehat{\frak g} = \widehat{\frak {sl}}_N$. However, since
the twisted sigma-model on a complex flag manifold $SL(N)/B$ lacks
a holomorphic stress tensor and is thus non-conformal, a
$\Phi^{\lambda}_s (z)$ operator will not have a one-to-one
correspondence with a state $|\Phi^{\lambda}_s \rangle$. Rather,
the states just furnish a module of the chiral algebra spanned by
the operators themselves.

Nevertheless, in the axiomatic CFT framework of a theory with an
affine algebra $\widehat {\frak g}$, the operator-state
isomorphism is an axiom that is defined at the outset, and
therefore, any primary field operator will be axiomatically
associated to a state in the highest-weight representation of
$\widehat{\frak g}$. Bearing this in mind, now consider a general
correlation function of $n$ primary field operators such as
$\left< \Phi^{\lambda_1}_s (z_1) \dots \Phi^{\lambda_n}_s
(z_n)\right>$. Note that it can be viewed, in the axiomatic CFT
sense, as a map from a tensor product of $n$ highest-weight
representations of $\widehat{\frak {sl}}_N$ to a complex number.
Next, consider a variation of the correlation function under a
$\it{global}$ $SL(N)$-transformation, i.e., $\delta_{\omega}\left<
\Phi^{\lambda_1}_s (z_1) \dots \Phi^{\lambda_n}_s (z_n)\right> =
\oint_{C} dz \sum_a \omega^a \left< J^a(z) \Phi^{\lambda_1}_s
(z_1) \dots \Phi^{\lambda_n}_s (z_n)\right>$, where $\omega^a$ is
a $\it{position}$-$\it{independent}$ scalar transformation
parameter, and where $C$ is a contour that encircles all the
points $z_1, \dots, z_n$ on $\Sigma$. Since all the $J^a(z)$'s are
dimension-one conserved currents in the $\overline Q_+$-cohomology
of the twisted sigma-model on $SL(N)/B$, they will generate a
symmetry of the theory. In other words, we will have
$\delta_{\omega}\left< \Phi^{\lambda_1}_s (z_1) \dots
\Phi^{\lambda_n}_s (z_n)\right> =0$, which is simply a statement
of the global $SL(N)$-invariance of any theory with an affine
$SL(N)$ algebra. This last statement, together with the one
preceding it, means that a general correlation function of $n$
primary field operators $\left< \Phi^{\lambda_1}_s (z_1) \dots
\Phi^{\lambda_n}_s (z_n)\right>$ will define a ``conformal block''
in the axiomatic CFT sense \cite{Frenkel}. Proceeding from this
mathematical definition of a ``conformal block'', the collection
of operators $\Phi^{\lambda_1}_s (z_1) \dots \Phi^{\lambda_n}_s
(z_n)$ will define a vector $\Phi$ in the dual space of
coinvariants $H_{\frak{sl}_N}(\Phi^{\lambda_1}_s (z_1) \dots
\Phi^{\lambda_n}_s (z_n))$, whereby the ``conformal block'' or
correlation function $\left< \Phi^{\lambda_1}_s (z_1) \dots
\Phi^{\lambda_n}_s (z_n)\right>$ can be computed as the square
$|\Phi|^2$ of length of $\Phi$ with respect to a hermitian inner
product on $H_{\frak{sl}_N} (\Phi^{\lambda_1}_s (z_1) \dots
\Phi^{\lambda_n}_s (z_n))$ \cite{Frenkel}. All correlation
functions of primary field operators can be computed once this
inner product is determined.

\bigskip\noindent{\it Sheaf of Coinvariants on
$\textrm{Bun}_{SL(N)}$}

As mentioned above, what will be directly related to the Hecke
eigensheaves is the sheaf of coinvariants on the moduli space
$\textrm{Bun}_G$ of holomorphic $G$-bundles on the worldsheet
$\Sigma$. Let us now describe how this sheaf of coinvariants
arises. However, before we proceed, let us first explain how
holomorphic $G$-bundles on $\Sigma$ can be consistently defined in
the presence of an affine $G$-algebra in the sigma-model on
$X=SL(N)/B$, where $G = SL(N)$ in our case.

Recall that for the sigma-model on $X = SL(N)/B$, we have the OPE
\be J_a (z) J_b (w) \sim -{{N d_{ab}} \over{(z-w)^2}} + \sum_c
f_{ab}{}^c {{J_c(w)}\over {(z-w)}}, \ee where $d_{ab}$ is the
Cartan-Killing metric of $\frak {sl}_N$. Note also that since the
above dimension-one current operators are holomorphic in $\Sigma$,
they can be expanded in a Laurent expansion around the point $w$
on $\Sigma$ as \be J_a(z) = \sum_n{ {J^n_{a}(w)}
{(z-w)^{-n-1}}}.\ee Consequently, from the above OPE, we will have
the commuator relation \be [J^n_a (w), J^m_b (w)] = \sum_c
f_{ab}{}^c J^{n+m}_c (w) - (N d_{ab})\  n \ \delta_{n+m, 0},\ee
such that the Lie algebra $\frak g = \frak{sl}_N$ generated by the
zero-modes of the currents will be given by \be [J^0_a (w), J^0_b
(w)] = \sum_c f_{ab}{}^c J^{0}_c (w).\ee One can then exponentiate
the above generators that span $\frak{sl}_N$ to define an element
of $G = SL(N)$, and since these generators depend on the point $w$
in $\Sigma$, it will mean that one can, via this exponential map,
consistently define a non-trivial principal $G$-bundle on
$\Sigma$. Moreover, this bundle will be holomorphic as the
underlying generators only vary holomorphically in $w$ on the
worldsheet $\Sigma$.

Let us label the above-described holomorphic $SL(N)$-bundle on
$\Sigma$ as $\cal P$.  Then, the space $H_{\frak{sl}_N}
(\Phi^{\lambda_1}_s (z_1) \dots \Phi^{\lambda_n}_s (z_n))$ of
coinvariants will vary non-trivially under infinitesimal
deformations of $\cal P$. As such, one can define a sheaf on
coinvariants over the space of all holomorphic $SL(N)$-bundles on
$\Sigma$, i.e., $\textrm{Bun}_{SL(N)}$. Let us justify this
statement next.

Firstly, note that with our description of $\cal P$ via the affine
$SL(N)$-algebra of the sigma-model on $X$, there is a mathematical
theorem \cite{book} which states that $\textrm{Bun}_{SL(N)}$ is
locally uniformized by the affine $SL(N)$-algebra. What this means
is that the tangent space $T_{\cal P} \textrm{Bun}_{SL(N)}$ to the
point in $\textrm{Bun}_{SL(N)}$ which corresponds to an
$SL(N)$-bundle on $\Sigma$ labelled by $\cal P$, will be
isomorphic to the space $H^1(\Sigma, \textrm{End} {\cal P})$
\cite{book}. Moreover, deformations of $\cal P$, which correspond
to displacements from this point in $\textrm{Bun}_{SL(N)}$, are
generated by an element $\eta(z) = J^a \eta_a(z)$ of the loop
algebra of $\frak{sl}_N$, where $\eta_a(z)$ is a
$\it{position}$-$\it{dependent}$ scalar deformation parameter (see
$\S$17.1 of \cite {book} and $\S$7.3 of \cite{Frenkel}). With this
in mind, let us again consider the $n$-point correlation function
$\left< \Phi^{\lambda_1}_s (z_1) \dots \Phi^{\lambda_n}_s (z_n)
\right>$. By inserting $\eta(z)$ into this correlation function,
and computing the contour integral around the points $z_1, \dots,
z_n$, we have $\delta_{\eta} \left < \Phi^{\lambda_1}_s (z_1)
\dots \Phi^{\lambda_n}_s (z_n)) \right> = \left< \oint_C dz \
\sum_a \eta_a(z) J^a (z) \Phi^{\lambda_1}_s (z_1) \dots
\Phi^{\lambda_n}_s (z_n) \right>$, where $C$ is a contour which
encircles the points $z_1, \dots, z_n$ on $\Sigma$, and
$\delta_{\eta} \left < \Phi^{\lambda_1}_s (z_1) \dots
\Phi^{\lambda_n}_s (z_n) \right>$ will be the variation of $\left<
\Phi^{\lambda_1}_s (z_1) \dots \Phi^{\lambda_n}_s (z_n) \right>$
under an infinitesimal deformation of $\cal P$ generated by
$\eta(z)$ (see eqn. (7.9) of \cite{Frenkel} and also
\cite{Eguchi}). Note that this variation does not vanish, since
$\eta_a(z)$, unlike $\omega$ earlier, is a position-dependent
parameter of a $\it{local}$ $SL(N)$-transformation. Therefore, as
explained above, since the correlation function $\left<
\Phi^{\lambda_1}_s (z_1) \dots \Phi^{\lambda_n}_s (z_n) \right>$
is associated to $\Phi$ in the dual space of coinvariants
$H_{\frak{sl}_N} (\Phi^{\lambda_1}_s (z_1) \dots
\Phi^{\lambda_n}_s (z_n))$, one can see that $\Phi$ must vary in
$H_{\frak{sl}_N} (\Phi^{\lambda_1}_s (z_1) \dots
\Phi^{\lambda_n}_s (z_n))$ as one moves infinitesimally along a
path in $\textrm{Bun}_{SL(N)}$. Since $\Phi$ is just a vector in
some basis of $H_{\frak{sl}_N} (\Phi^{\lambda_1}_s (z_1) \dots
\Phi^{\lambda_n}_s (z_n))$, one could instead interpret $\Phi$ to
be fixed, while $H_{\frak{sl}_N} (\Phi^{\lambda_1}_s (z_1) \dots
\Phi^{\lambda_n}_s (z_n))$ varies as one moves infinitesimally
along a path in $\textrm{Bun}_{SL(N)}$, as $\cal P$ is subjected
to infinitesimal deformations. Consequently, we have an
interpretation of a $\textrm {\it sheaf of coinvariants}$ on
$\textrm{Bun}_{SL(N)}$, where the fibre of this sheaf over each
point in $\textrm{Bun}_{SL(N)}$ is just the space $H_{\frak{sl}_N}
(\Phi^{\lambda_1}_s (z_1) \dots \Phi^{\lambda_n}_s (z_n))$ of
coinvariants corresponding to a particular bundle $\cal P$ that
one can consistently define over $\Sigma$ using the affine
$SL(N)$-algebra of the sigma-model on $X=SL(N)/B$. Note howeover,
that since we are dealing with an affine $SL(N)$ algebra at the
critical level $k = -h^{\vee}$, the dimension of the space of
coinvariants will be different over different points in
$\textrm{Bun}_{SL(N)}$. In other words, the sheaf of coinvariants
on $\textrm{Bun}_{SL(N)}$ does not have a structure of a vector
bundle, since the fibre space of a vector bundle must have a fixed
dimension over different points on the base. Put abstractly, this
is because $\widehat{\frak {sl}}_N$-modules at the critical level
may only be exponentiated to a subgroup of the Kac-Moody group
$\widehat {SL(N)}$. Nevertheless, the sheaf of coinvariants is a
twisted $\cal D$-module on $\textrm{Bun}_{SL(N)}$ \cite{Frenkel}.

From the above discussion, one can also make the following
physical observation. Notice that the variation $\delta_{\eta}
\left < \Phi^{\lambda_1}_s (z_1) \dots \Phi^{\lambda_n}_s (z_n))
\right> = \left< \oint_C dz \ \sum_a \eta_a(z) J^a (z)
\Phi^{\lambda_1}_s (z_1) \dots \Phi^{\lambda_n}_s (z_n) \right>$
in the correlation function as one moves along
$\textrm{Bun}_{SL(N)}$, can be interpreted, at the lowest order in
sigma-model perturation theory, as a variation in the correlation
function due to a $\it{marginal}$ deformation of the sigma-model
action by the term $\oint dz \ \eta(z)$. Since a deformation of
the action by the dimensionless term $\oint dz \ \eta(z)$ is
tantamount to a displacement in the moduli space of the
sigma-model itself, it will mean that $\delta_{\eta} \left <
\Phi^{\lambda_1}_s (z_1) \dots \Phi^{\lambda_n}_s (z_n)) \right>$
is also the change in the correlation function as one varies the
moduli of the sigma-model. This implies that
$\textrm{Bun}_{SL(N)}$ will at least span a subspace of the entire
moduli space of the sigma-model on $X = SL(N)/B$. This should come
as no surprise since $\cal P$ is actually associated to the affine
$SL(N)$-algebra of the sigma-model on $X = SL(N)/B$ as explained
above, and moreover, the affine $SL(N)$- algebra does depend on
the moduli of the sigma-model as mentioned in $\S$2 and $\S$3.

 Last but not least, note that the sheaf of coinvariants can also be obtained purely mathematically \cite{Frenkel} via a localisation functor $\Delta$,
  which maps the set $V_{\chi}$ --- consisting of all polynomials
$F({\cal J}(z))$ (which exist in the chiral algebra of the twisted
sigma-model on $SL(N)/B$) that are defined over the field of
complex numbers and the $c$-number operators $S^{(s_i)}(z)$, and
that are of arbitrary positive degree in the quantum operator
${\cal J}(z) = {1 \over {(-n_1 -1)! \dots (-n_m
-1)!}}:\partial_z^{-n_1 -1}J^{a_1}(z) \dots \partial_z^{-n_m
-1}J^{a_m}(z):$ --- to the corresponding sheaf $\Delta(V_{\chi})$
of coinvaraints on $\textrm{Bun}_{SL(N)}$, where $\chi$ denotes a
parameterisation of $V_{\chi}$ that depends on the choice of the
set of $S^{(s_i)}(z)$ fields for $i=1,\dots, \textrm{rank}(\frak
{sl}_N)$. In other words, the sheaf of coinvariants will be
parameterised by $\chi$.\footnote{Note that in order to be
consistent with the notation used in the mathematical literature,
we have chosen to use the symbol $\chi$ to label the
parameterisation of $V_{\chi}$. Hopefully, $\chi$ that appears
here and henceforth will not be confused with the one-dimensional
representation $\chi$ of $\widehat{\frak g}'$ in $\S$3.} This
observation is pivotal in the mathematical description of the
correspondence between Hecke eignesheaves on
$\textrm{Bun}_{SL(N)}$ and flat holomorphic $^LSL(N)$-bundles on
$\Sigma$ via the algebraic CFT approach to the geometric Langlands
program \cite{Frenkel}. As we will explain below, this
parameterisation of the sheaf of coinvariants on
$\textrm{Bun}_{SL(N)}$ by the set of $S^{s_i}(z)$ fields can be
shown to arise physically in the sigma-model as well.

\bigskip\noindent{\it A $\frak{z}(\widehat{\frak {sl}}_N)$-Dependent Realisation of the
Affine $SL(N)$ Algebra at the Critical Level}

Before one can understand how, within the context of the
sigma-model on $X = SL(N)/B$,  the sheaf of coinvariants can be
parameterised by a choice of the set of $S^{s_i}(z)$ fields for $i
= 1,\dots, \textrm{rank}(\frak{sl}_N)$, it will be necessary for
us to understand how one can achieve a $\frak{z}(\widehat{\frak
{sl}}_N)$-dependent realisation of the affine $SL(N)$ OPE algebra
at $k=-h^{\vee}$ spanned by the set of $J^a(z)$ currents that
correspond to classes in $H^0(X,{\cal O}^{ch}_X)$, where
$X=SL(N)/B$.

To this end, let us start with the case of the affine $SL(2)$ OPE
algebra at level $k=-2$, spanned by the currents $\{J_+, J_-,
J_3\}$ in the holomorphic chiral algebra of the twisted
sigma-model on $X= SL(2)/B$, that correspond to classes in
$H^0(X,{\cal O}^{ch}_X)$. Recall that the set $\{J_+, J_-, J_3\}$
can be expressed in terms of the fields of the free $\beta\gamma$
system associated to the sheaf ${\cal O}^{ch}_X$ of CDO's on
$X=SL(2)/B$ in (\ref{J_+}), (\ref{J_-}) and (\ref{J_3})
respectively. As explained, these are classes in $H^0(X,{\cal
O}^{ch}_X)$, i.e., if the set $\{J_+, J_-, J_3\}$ were to be
defined on the North pole of $X= SL(2)/B \simeq \mathbb{P}^1$,
while the set $\{{\widetilde J}_+, {\widetilde J}_-, {\widetilde
J}_3\}$ were to defined their corresponding counterparts on the
South pole of $X= SL(2)/B \cong \mathbb{P}^1$, one will have
$\widetilde J_+ - J_+ = 0$, $\widetilde J_- - J_- = 0$ and
$\widetilde J_3 - J_3 = 0$.

Let us now modify the expressions of $\{J_+, J_-, J_3\}$ as
follows: \begin{eqnarray} \label{J'_+} J'_+ (z) & = & - :\gamma^2
(z) \beta (z): + 2\partial_z \gamma (z) + {1 \over 2}
\gamma(z) c(z), \\
\label{J'_-}
J'_-(z) & = & \beta(z),\\
\label{J'_3} J'_3(z) & = & - :\gamma(z) \beta (z): + {1\over 2}
c(z),
\end{eqnarray}
where $c(z)$ is a classical $c$-number field that is holomorphic
in $z$ and of conformal dimension one, i.e., it has a Laurent
expansion given by $c(z)= \sum_{n \in \mathbb Z} c_n z^{-n-1}$.
Since $c(z)$ is a classical field, it will not participate as an
interacting quantum field in any of the OPE's amongst the quantum
operators $\{J'_+, J'_-, J'_3\}$. Rather, it will just act as a
simple multiplication on the $\gamma(z)$ and $\beta(z)$ fields, or
functions thereof. Moreover, this means that $c(z)$, like $S(z)$,
must also be trivial in the $\overline Q_+$-cohomology of the
twisted sigma-model on $SL(2)/B$ at the quantum level, i.e., it
can be expressed as a $\overline Q_+$-exact term $\{\overline Q_+,
\dots \}$ in the $\it{quantum}$ theory. Now, recall that we had
the (non quantum-corrected) geometrical gluing relation $\gamma =
1/ {\widetilde \gamma}$, where $\gamma$ and $\widetilde \gamma$
are defined on the North and South poles of $X = SL(2)/B \simeq
\mathbb P^1$ respectively. This expression means that $\gamma$
defines a global section of the sheaf $\widehat{\cal O}^{ch}_X$.
From our $\overline Q_+$-Cech cohomology dictionary, this will
mean that $\gamma(z)$ must correspond to an operator in the
twisted sigma-model on $X$ that is annihilated by the quantum
action of $\overline Q_+$. This, together with the fact that
$c(z)$ can be expressed as $\{\overline Q_+, \dots\}$, will mean
that the term ${1 \over 2} \gamma(z) c(z)$ in $J'_+ (z)$ of
(\ref{J'_+}) above, can be written as a $\overline Q_+$-exact term
$\{\overline Q_+, \dots \}$. Likewise, the term ${1\over 2} c(z)$
in $J'_3 (z)$ of (\ref{J'_3}) can also be written as a $\overline
Q_+$-exact term $\{\overline Q_+, \dots \}$. Consequently, since
${\overline Q}^2_+ =0$ even at the quantum level, $\{J'_+, J'_-,
J'_3\}$ continues to be a set of quantum operators that are
$\overline Q_+$-closed and non-$\overline Q_+$-exact, which
therefore correspond to classes in $H^0(X, {\widehat{\cal
O}^{ch}_X})$. Since the OPE's of $\overline Q_+$-exact terms such
as ${1 \over 2} \gamma(z) c(z)$ and ${1\over 2} c(z)$ with the
other $\overline Q_+$-closed terms $\{-\gamma^2 \beta + 2
\partial_z \gamma , \beta, -\gamma\beta\}$ that correspond respectively to the
set of original operators $\{J_+, J_-, J_3\}$ must again result in
$\overline Q_+$-exact terms that are trivial in $\overline
Q_+$-cohomology, they can be discarded in the OPE's involving the
set of operators $\{J'_+, J'_-, J'_3\}$, i.e., despite being
expressed differently from the set of original operators $\{J_+,
J_-, J_3\}$, the set of operators $\{J'_+, J'_-, J'_3\}$ will
persist to generate an affine $SL(2)$ OPE algebra at the critical
level $k=-2$. In other words, via the set of modified operators
$\{J'_+, J'_-, J'_3\}$ and their corresponding Laurent modes, we
have a different realisation of the affine $SL(2)$ algebra at the
critical level $k=-2$.

Obviously, from (\ref{J'_+})-(\ref{J'_3}), we see that the above
realisation depends on the choice of $c(z)$. What determines
$c(z)$ then? To answer this, let us first recall that the
Segal-Sugawara tensor $S'(z)$ in the context of the modified
operators $\{J'_+, J'_-, J'_3\}$, can be expressed as $S'(z) =
:(J'_+ J'_- + J^{'2}_3) (z):$ in the quantum theory. However,
recall also that the original Segal-Sugawara tensor given by $S(z)
= :(J_+ J_- + J^{2}_3) (z):$ acts by zero in the quantum theory.
This means that the non-vanishing contributions to $S'(z)$ come
only from terms that involve the $c(z)$ field. A simple
computation will show that $S'(z) = {1\over 4} c(z)^2 - {1\over 2}
\partial_z c(z)$. As required, $S'(z)$ is a classical
holomorphic field of dimension-two. Clearly, a unique choice of
$S'(z)$ will determine a unique $c(z)$. In summary, we can
generate different realisations of the affine $SL(2)$ OPE algebra
at level $k=-h^{\vee}$ via the set of operators $\{J'_+, J'_-,
J'_3\}$, that are parameterised by the choice of the corresponding
Segal-Sugawara tensor $S'(z)$ in the classical holomorphic chiral
algebra of the purely bosonic sector of the twisted $(0,2)$
sigma-model on $X=SL(2)/B$. Since the Laurent modes of $S'(z)$
span the centre $\frak z(\widehat{\frak {sl}}_2)$ of the completed
universal enveloping algebra of $\widehat{\frak {sl}}_2$ at the
critical level $k=-2$, we effectively have a $\frak
z(\widehat{\frak {sl}}_2)$-dependent realisation of the affine
$SL(2)$ (OPE) algebra at the critical level.

The above arguments can actually be extended to $\it{any}$
$SL(N)$, not just $SL(2)$, i.e., for a twisted sigma-model on
$X=SL(N)/B$, one can always find different realisations of an
affine $SL(N)$ OPE algebra at the critical level $k=-h^{\vee}$
that are spanned by the local operators in the holomorphic chiral
algebra of the sigma-model which correspond to classes in $H^0(X,
{\widehat{\cal O}^{ch}_X})$, that are $\frak z(\widehat{\frak
{sl}}_N)$-dependent. We shall now import some important results in
\cite{frenkel lectures wakimoto} to demonstrate this. Firstly,
consider the set of local operators composed out of the $N(N-1)/2$
(i.e. $\textrm{dim}_{\mathbb C}X$)  free $\beta_i(z)$ and
$\gamma^i(z)$ fields of the $N(N-1)/2$ linear $\beta\gamma$
systems associated to the sheaf of CDO's on $X$:
\begin{eqnarray}
J^i_-(z) &= & \beta^{\alpha_i}(z) + \sum_{\varphi \in \Delta_+}
: P^i_\varphi(\gamma^\alpha(z)) \beta^\varphi(z) :, \\
J^k_c(z) &= & - \sum_{\varphi \in \Delta_+} \varphi(h^k) :
\gamma^\varphi(z)
\beta^\varphi(z) :, \\
J^i_+(z) &= & \sum_{\varphi \in \Delta_+} :
Q^i_\varphi(\gamma^\alpha(z)) \beta^\varphi(z) : + c_i \partial_z
\gamma^{\alpha_i}(z),
\end{eqnarray}
where the subscripts $\{\pm,c\}$ denote a Cartan decomposition of
the Lie algebra $\frak{sl}_N$ under which the $J(z)$ local
operators can be classified (as in $\S$2), the superscript
$\alpha_i$ denotes the free field that can be identified with the
$i^{th}$ positive root of $\frak{sl}_N$ where $i = 1, \dots,
N(N-1)/2$, $h^k$ is an element of the Cartan subalgebra of $\frak
{sl}_N$ where $k =1,\dots, N-1$, $\varphi(h^k)$ is the $k^{th}$
component of the root $\varphi$, the symbol $\Delta_+$ denotes the
set of positive roots of $\frak {sl}_N$, the $c_i$'s are complex
constants, and lastly, $P^i_\varphi, Q^i_\varphi$ are some
polynomials in the $\gamma^{\alpha}$ free fields.

Theorem 4.3 of \cite{frenkel lectures wakimoto} tells us that the
Laurent modes of the above set of local operators $\{J^i_{\pm},
J^k_c\}$ generate an affine $SL(N)$ algebra at the critical level
$k=-h^{\vee}$, i.e., the set $\{J^i_{\pm}, J^k_c\}$ will span an
affine $SL(N)$ OPE algebra at the critical level $k=-h^{\vee}$. In
fact, for the case of $SL(2)$, we have the identification
$J^i_{\pm} \leftrightarrow J_{\pm}$ and $J^k_c \leftrightarrow
J_3$, where $\{J_{+}, J_-, J_3\}$ is the set of local currents in
the holomorphic chiral algebra of the twisted sigma-model on
$SL(2)/B$ in (\ref{J_+}), (\ref{J_-}) and (\ref{J_3}) which
generates an affine $SL(2)$ OPE algebra at the critical level
$k=-2$. The fact that the currents $\{J^i_{\pm}, J^k_c\}$ are
composed purely out of free $\beta_i$ and $\gamma^i$ fields, and
the fact that there will always be classes in $H^0(X, {\cal
O}^{ch}_X)$ which correspond to operators that generate an affine
$SL(N)$ OPE algebra \cite{MSV}, will together mean that the set of
currents $\{J^i_{\pm}, J^k_c\}$ must correspond (up to $\overline
Q_+$-exact terms at worst) to classes in $H^0(X, {\cal
O}^{ch}_X)$, i.e., the global sections of the sheaf ${\cal
O}^{ch}_X$ of CDO's on $X=SL(N)/B$. Equivalently, this means that
the set of local current operators $\{J^i_{\pm}, J^k_c\}$ will be
$\overline Q_+$-closed and hence lie in the holomorphic chiral
algebra of the twisted sigma-model on $X = SL(N)/B$.

Proceeding as we did for the $SL(2)$ case discussed above, let us
now consider a modification $\{J^{i'}_{\pm}, J^{k'}_c\}$ of the
set of currents $\{J^i_{\pm}, J^k_c\}$, where
\begin{eqnarray}
\label{J^{i'}_-} J^{i'}_-(z) &= & \beta^{\alpha_i}(z) +
\sum_{\varphi \in \Delta_+}
: P^i_\varphi(\gamma^\alpha(z)) \beta^\varphi(z) :, \\
\label{J^{k'}_c} J^{k'}_c(z) &= & - \sum_{\varphi \in \Delta_+}
\varphi(h^k) :
\gamma^\varphi(z)\beta^\varphi(z): + b^i(z), \\
\label{J^{i'}_+} J^{i'}_+(z) &= & \sum_{\varphi \in \Delta_+} :
Q^i_\varphi(\gamma^\alpha(z)) \beta^\varphi(z) : + c_i \partial_z
\gamma^{\alpha_i}(z) + b^i(z) \gamma^{\alpha_i}(z),
\end{eqnarray}
and the $b^i(z)$'s are just classical $c$-number functions that
are holomorphic in $z$ and of conformal dimension one - it can be
Laurent expanded as $b^i(z) = \sum_{n \in \mathbb Z} b^i_n
z^{-n-1}$.\footnote{Note that the explicit expression of $b(z)$
cannot be arbitrary. It has to be chosen appropriately to ensure
that the Segal-Sugawara tensor and its higher spin analogs given
by the $S^{(s_i)}(z)$'s, can be identified with the space of
$^L\frak{sl}_N$-opers on the formal disc $D$ in $\Sigma$ as
necessitated by the isomorphism $\frak z(\widehat {\frak{sl}}_N)
\simeq {\cal W}_{\infty}(^L\widehat {\frak {sl}}_N)$ from the
duality of classical $\cal W$-algebras for $G=SL(N)$. For example,
the expression of $b(z)$ as ${1\over 2}c(z)$ in the $G=SL(2)$ case
ensures that $S'(z) = {1\over 4} c^2(z) - {1\over 2}\partial_z
c(z)$ can be identified with a projective connection on $D$ for
each choice of $c(z)$.} Since the $b^i(z)$'s are classical fields,
they will not participate as interacting quantum fields in any of
the OPE's amongst the quantum operators $\{J^{i'}_+, J^{i'}_-,
J^{k'}_3\}$. Rather, they will just act as a simple multiplication
on the $\gamma^{\alpha_i}(z)$ and $\beta^{\alpha_i}(z)$ fields, or
functions thereof. Moreover, this means that the $b^i(z)$'s, must
be trivial in the $\overline Q_+$-cohomology of the twisted
sigma-model on $SL(N)/B$ at the quantum level, i.e., it can be
expressed as a $\overline Q_+$-exact term $\{\overline Q_+, \dots
\}$ in the $\it{quantum}$ theory. Now, recall that we had the (non
quantum-corrected) geometrical gluing relation $\gamma^{\alpha_i}
= g^{\alpha_i}(\gamma^{\alpha})$, where each $\gamma^{\alpha_i}$
and $g^{\alpha_i}(\gamma^{\alpha})$ is defined in the open set
$U_1$ and $U_2$ respectively of the intersection $U_1 \cap U_2$ in
$X$. This expression means that the $\gamma^{\alpha_i}$'s define
global sections of the sheaf $\widehat{\cal O}^{ch}_X$. From our
$\overline Q_+$-Cech cohomology dictionary, this will mean that
each $\gamma^{\alpha_i}(z)$ must correspond to an operator in the
twisted sigma-model on $X$ that is annihilated by $\overline Q_+$
at the quantum level. This, together with the fact that $b^i(z)$'s
can be expressed as $\{\overline Q_+, \dots\}$, will mean that the
$b^i(z)\gamma^{\alpha_i}(z)$ term in $J^{i'}_+(z)$ of
(\ref{J^{i'}_+}) above, can be written as a $\overline Q_+$-exact
term $\{\overline Q_+, \dots \}$. Likewise, the $b^i(z)$ term in
$J^{k'}_c (z)$ of (\ref{J^{k'}_c}) can also be written as a
$\overline Q_+$-exact term $\{\overline Q_+, \dots \}$.
Consequently, since ${\overline Q}^2_+ =0$ even at the quantum
level, $\{J^{i'}_+, J^{i'}_-, J^{i'}_3\}$ continues to be a set of
quantum operators that are $\overline Q_+$-closed and
non-$\overline Q_+$-exact, which therefore correspond to classes
in $H^0(X, {\widehat{\cal O}^{ch}_X})$. Since the OPE's of
$\overline Q_+$-exact terms such as $b^i(z)\gamma^{\alpha_i}(z)$
and $b^i(z)$ with the other $\overline Q_+$-closed terms
 such as $(\sum_{\varphi \in \Delta_+} : Q^i_\varphi(\gamma^\alpha)
\beta^\varphi : + c_i \partial_z \gamma^{\alpha_i})$,
$(\beta^{\alpha_i} + \sum_{\varphi \in \Delta_+} :
P^i_\varphi(\gamma^\alpha) \beta^\varphi:)$, and $(- \sum_{\varphi
\in \Delta_+} \varphi(h^k) : \gamma^\varphi\beta^\varphi:)$ that
correspond respectively to the set of original operators $J^i_+$,
$J^i_-$, and $J^k_c$, must again result in $\overline Q_+$-exact
terms that are trivial in $\overline Q_+$-cohomology, they can be
discarded in the OPE's involving the set of operators $\{J^{i'}_+,
J^{i'}_-, J^{i'}_3\}$, i.e., despite being expressed differently
from the set of original operators $\{J^i_+, J^i_-, J^k_c\}$, the
set of operators $\{J^{i'}_+, J^{i'}_-, J^{i'}_c\}$ will persist
to generate an affine $SL(N)$ OPE algebra at the critical level
$k=-h^{\vee}$. In other words, via the set of modified operators
$\{J^{i'}_{\pm}, J^{k'}_c\}$ and their corresponding Laurent
modes, we have a different realisation of the affine $SL(N)$
algebra at the critical level $k=-h^{\vee}$. This is consistent
with Theorem 4.7 of \cite{frenkel lectures wakimoto}, which states
that the set $\{J^{i'}_{\pm}, J^{k'}_c\}$ of modified operators
will persist to generate an affine $SL(N)$ OPE algebra at the
critical level $k=-h^{\vee}$.

Obviously, from (\ref{J^{i'}_-})-(\ref{J^{i'}_+}), we see that the
above realisation depends on the choice of the $b^i(z)$'s. What
determines the $b^i(z)$'s then? To answer this, let us first
recall that the Segal-Sugawara tensor $S^{(2)'}(z)$ and its higher
spin analogs $S^{(s_i)'}(z)$ in the context of the modified
operators $\{J^{i'}_+, J^{i'}_-, J^{k'}_c\} \in \{J^{a'}\}$, can
be expressed as $S^{(s_i)'}(z) = {\tilde d}_{a_1 a_2 \dots
a_{s_i}}:J^{a_1'} J^{a_2'}\dots J^{a_{s_i}'}(z):$ in the quantum
theory. However, recall also that the original Segal-Sugawara
tensor and its higher spin analogs, expressed as $S^{(s_i)}(z) =
{\tilde d}_{a_1 a_2 \dots a_{s_i}}:J^{a_1} J^{a_2}\dots
J^{a_{s_i}}(z):$ in terms of the original operators $\{J^{i}_+,
J^{i}_-, J^{k}_c\} \in \{J^{a}\}$, act by zero in the quantum
theory. This means that the non-vanishing contributions to any of
the $S^{(s_i)}(z)$'s come only from terms that involve the
additional $b^i(z)$ fields. In fact, it is true that the
$S^{(s_i)'}(z)$'s also act by zero in the quantum theory at
$k=-h^{\vee}$, since they are also defined via a Sugawara-type
construction which results in their quantum definition being
$S^{(s_i)'}(z)= (k+h^{\vee})T^{(s_i)'}(z)$. In other words, the
$S^{(s_i)'}(z)$'s must be classical $c$-number fields of spin
$s_i$ that are holomorphic in $z$. This implies that the
$S^{(s_i)'}(z)$'s will be expressed solely in terms of the
$c$-number $b^i(z)$ fields. An explicit example of this general
statement has already been discussed earlier in the case of
$SL(2)$ - for the $SL(2)$ case, we have the identification
$J^{i'}_+ \leftrightarrow J'_+$, $J^{i'}_- \leftrightarrow J'_-$
$J^{k'}_c \leftrightarrow J'_3$, $S^{(2)'}(z) \leftrightarrow
S'(z)$, $b^i(z) \leftrightarrow {1\over 2}c(z)$ and $S^{(2)'}(z) =
{1\over 4} c^2(z) - {1\over 2} \partial_z c(z)$, whereby the
choice of $S^{(2)'}(z)$ determines $c(z)$. Consequently, a choice
of the set of $S^{(s_i)'}(z)$ fields will determine the $b^i(z)$
fields. Lastly, note that the $S^{(s_i)'}(z)$ fields lie in the
classical holomorphic chiral algebra of the purely bosonic sector
of the twisted sigma-model on $X=SL(N)/B$, and their Laurent modes
span the centre $\frak z(\widehat{\frak{sl}}_N)$ of the completed
universal enveloping algebra of $\widehat{\frak{sl}}_N$ at the
critical level $k=-h^{\vee}$. Hence, we effectively have a $\frak
z(\widehat{\frak {sl}}_N)$-dependent realisation of the affine
$SL(N)$ (OPE) algebra at the critical level as claimed.

\bigskip\noindent{\it A $\frak{z}(\widehat{\frak {sl}}_N)$-Dependent Parameterisation  of the
Sheaf of Coinvariants on $\textrm{Bun}_{SL(N)}$}

Now that we have seen how one can obtain a $\frak z(\widehat{\frak
{sl}}_N)$-dependent realisation of the affine $SL(N)$ (OPE)
algebra at the critical level, we can proceed to explain how,
within the context of the sigma-model on $X=SL(N)/B$, the sheaf of
coinvariants on $\textrm{Bun}_{SL(N)}$ can be parameterised by a
choice of the fields $S^{s_i}(z)$ for $i = 1,\dots,
\textrm{rank}({\frak{sl}_N})$.

To this end, notice that since the primary field operators
$\Phi^{\lambda}_s(z)$ are defined via the OPE's with the $J^a(z)$
currents of the $\widehat{\frak {sl}}_N$ algebra at the critical
level in (\ref{primary field OPE's}), a different realisation of
the $J^a(z)$ currents will also result in a different realisation
of the $\Phi^{\lambda}_s(z)$'s. Consequently, we will have a
$\frak z(\widehat{\frak {sl}}_N)$-dependent realisation of the
primary field operators $\Phi^{\lambda}_s(z)$. This amounts to a
$\frak z(\widehat{\frak {sl}}_N)$-dependent realisation of their
$n$-point correlation functions
$\left<\Phi^{\lambda_1}_s(z_1)\dots
\Phi^{\lambda_n}_s(z_n)\right>$. Since the correlation functions
can be associated to a (vector in the) space of coinvariants as
explained earlier, one will consequently have a $\frak
z(\widehat{\frak {sl}}_N)$-dependent realisation of the sheaf of
coinvariants on $\textrm{Bun}_{SL(N)}$ as well, i.e., the sheaf of
coinvariants will be parameterised by a choice of the fields
$S^{s_i}(z)$ for $i = 1,\dots, \textrm{rank}({\frak{sl}_N})$.

\bigskip\noindent{\it A Correspondence Between Hecke Eigensheaves on $\textrm{Bun}_{SL(N)}$ and Flat $^LSL(N)$-Bundles on $\Sigma$}

Finally, we shall now demonstrate that the above observation about
a $\frak z(\widehat{\frak {sl}}_N)$-dependent realisation of the
sheaf of coinvariants on $\textrm{Bun}_{SL(N)}$, and the duality
of classical $\cal W$-algebras for $G=SL(N)$ as an isomorphism of
Poisson algebras $\frak z(\widehat {\frak {sl}}_N) \simeq {\cal
W}_{\infty}(^L\widehat{\frak {sl}}_N)$, will result in a
correspondence between Hecke eigensheaves on
$\textrm{Bun}_{SL(N)}$ and flat holomorphic $^LSL(N)$-bundles on
the worldsheet $\Sigma$.

Firstly, note that the classsical $\cal W$-algebra ${\cal
W}_{\infty}(^L\widehat{\frak {sl}}_N)$ is isomorphic to
$\textrm{Fun}\ \textrm{Op}_{^L{\frak {sl}_N}}(D^{\times})$, the
algebra of functions on the space of $^L{\frak {sl}}_N$-opers on
the punctured disc $D^{\times}$ in $\Sigma$, where an
$\frak{sl}_N$-oper on $\Sigma$ is an $n^{th}$ order differential
operator acting from $\Omega^{-(n-1)/2}$ to $\Omega^{(n+1)/2}$
(where $\Omega$ is the canonical line bundle on $\Sigma$) whose
principal symbol is equal to 1 and subprincipal symbol is equal to
0 \cite{Frenkel}. Roughly speaking, it may be viewed as a (flat)
connection on an $^LSL(N)$-bundle on $\Sigma$. In turn,
$\textrm{Fun}\ \textrm{Op}_{^L{\frak {sl}_N}}(D^{\times})$ is
related to the algebra $\textrm{Fun}\ \textrm{Op}_{^L{\frak
{sl}_N}}(D)$ of functions on the space of $^L{\frak {sl}}_N$-opers
on the formal disc $D$ in $\Sigma$, via $\textrm{Fun}\
\textrm{Op}_{^L{\frak {sl}_N}}(D^{\times}) \simeq {\widetilde
U}(\textrm{Fun}\ \textrm{Op}_{^L{\frak {sl}_N}}(D))$, where
$\widetilde U$ is a functor from the category of vertex algebras
to the category of Poisson algebras \cite{frenkel lectures
wakimoto}. Since from the duality of classical $\cal W$-algebras
for $G=SL(N)$, we have an isomorphism of Poisson algebras $\frak
z(\widehat {\frak {sl}}_N) \simeq {\cal
W}_{\infty}(^L\widehat{\frak {sl}}_N)$, it will mean that the
$S^{(s_i)}(z)$'s will correspond to the components of the
(numeric) $^L{\frak {sl}_N}$-oper on the formal disc $D$ in
$\Sigma$ \cite{Frenkel}. Hence, a choice of the set of
$S^{(s_i)}(z)$ fields will amount to picking up an $^L{\frak
{sl}_N}$-oper on $D$. Since any $^L{\frak g}$-oper on $D$ can be
extended to a regular $^L{\frak g}$-oper that is defined globally
on $\Sigma$, it will mean that a choice of the set of
$S^{(s_i)}(z)$ fields will determine a unique $^LSL(N)$-bundle on
$\Sigma$ (that admits a structure of an oper $\chi$) with a
holomorphic connection.

Secondly, recall that we have a $\frak z(\widehat{\frak
{sl}}_N)$-dependent realisation of the sheaf of coinvariants on
$\textrm{Bun}_{SL(N)}$ which depends on the choice of the fields
$S^{s_i}(z)$ for $i = 1,\dots, \textrm{rank}({\frak{sl}_N})$.
Hence, from the discussion in the previous paragraph, we see that
we have a correspondence between a flat holomorphic
$^LSL(N)$-bundle on $\Sigma$ and a sheaf of coinvariants on
$\textrm{Bun}_{SL(N)}$.

Lastly, recall that $\Delta(V_{\chi})$ and therefore the sheaf of
of coinvariants on $\textrm{Bun}_{SL(N)}$ has a structure of a
$\it{twisted}$ $\cal D$-module on $\textrm{Bun}_{SL(N)}$. The
sought-after Hecke eigensheaf \cite{Frenkel} is an
$\it{untwisted}$ holonomic $\cal D$-module
$\Delta(V_{\chi})\otimes K^{-1/2}$ on $\textrm{Bun}_{SL(N)}$ with
eigenvalue $E_{\chi}$, where $K$ is the canonical line bundle on
$\textrm{Bun}_{SL(N)}$, and $E_{\chi}$ is the unique
$^LSL(N)$-bundle corresponding to a particular choice of the set
of $S^{(s_i)}(z)$ fields. In total, since tensoring with a
globally-defined $K$ on $\textrm{Bun}_{SL(N)}$ just maps, in a
one-to-one fashion, $\Delta(V_{\chi})$ to $\Delta(V_{\chi})\otimes
K^{-1/2}$, we find that we have a one-to-one correspondence
between a Hecke eigensheaf on $\textrm{Bun}_{SL(N)}$ and a flat
holomorphic $^LSL(N)$-bundle on $\Sigma$, where $^LSL(N) =
PSL(N)$, i.e., we have a geometric Langlands correspondence for
$G=SL(N)$.\footnote{Note that the above-mentioned flat holomorphic
$^LSL(N)$-bundles on $\Sigma$ are restricted to those that have a
structure of an $^L\frak g$-oper on $\Sigma$. The space of
connections of any such bundle only form a half-dimensional
subspace in the moduli stack $\textrm{Loc}_{^LG}$ of the space of
$\it{all}$ connections on a particular flat $^LG$-bundle, where
$G=SL(N)$. Thus, our construction establishes the geometric
Langlands correspondence only partially. However, it turns out
that our construction can be generalised to include all flat
$^LG$-bundles on $\Sigma$ by considering in the correlation
functions more general chiral operators that are labelled by
finite-dimensional representations of $\frak g$, which, in
mathematical terms, is equivalent to making manifest the singular
oper structure of any flat $^LG$-bundle on $\Sigma$
\cite{Frenkel}. }

\bigskip\noindent{\it Physical Interpretation of the Hecke Eigensheaves on $\textrm{Bun}_{SL(N)}$}

From all of our above results, we see that one can physically
interpret the Hecke eigensheaf as follows. A local section of the
fibre of the Hecke eigensheaf over a point $p$ in
$\textrm{Bun}_{SL(N)}$, will determine, for some holomorphic
$SL(N)$-bundle on $\Sigma$ that corresponds to the point $p$ in
the moduli space $\textrm{Bun}_{SL(N)}$ of all holomorphic
$SL(N)$-bundles on $\Sigma$, the value of any $n$-point
correlation function $\left < \Phi^{\lambda_1}_s (z_1) \dots
\Phi^{\lambda_n}_s (z_n) \right>$ of local bosonic operators in
the holomorphic chiral algebra of the twisted $(0,2)$ sigma-model
on $SL(N)/B$. And the geometric Langlands correspondence for our
case of $G=SL(N)$ just tells us that for every flat, holomorphic
$PSL(N)$-bundle that can be constructed over $\Sigma$, we have a
unique way of characterising how an $n$-point correlation function
of local bosonic primary operators in the holomorphic chiral
algebra of a $\textrm{\it quasi-topological}$ sigma-model with
$\it{no}$ boundaries like the twisted $(0,2)$ sigma-model on
$SL(N)/B$, will vary under the $\it{local}$
$SL(N)$-transformations generated by the affine $J^a(z)$ currents
on the worldsheet described earlier.

\newsubsection{Hecke Operators and the Correlation Functions of Local Operators}

Consider the quantum operator ${\cal J}(z) = {1 \over {(-n_1 -1)!
\dots (-n_m -1)!}} : \partial_z^{-n_1 -1}J^{a_1}(z) \dots
\partial_z^{-n_m -1}J^{a_m}(z):$. Note that since the $J^a(z)$'s
are $\overline Q_+$-closed and in the $\overline Q_+$-cohomology
or holomorphic chiral algebra of the sigma-model on $SL(N)/B$, so
will ${\cal J} (z)$ or polynomials $F({\cal J}(z))$ of arbitrary
positive degree in ${\cal J}(z)$ (modulo  polynomials of arbitrary
positive degree in the $S^{(s_i)}(z)$ operators which necessarily
act by zero and hence vanish in the quantum theory).\footnote{In
order to show this, first note that $\partial_z J^a(z) = [L_{-1},
J^a(z)] $, where $L_{-1} = \oint dz T_{zz}(z)$. Since $[\overline
Q_+, J^a(z)]=0$ even at the quantum level, it will mean that
$[\overline Q_+, \partial_zJ^a(z)] = [[\overline Q_+, L_{-1}],
J^a(z)] = \oint dz' [ [\overline Q_+, T_{zz}(z')], J^a(z)]= \oint
dz' [ \partial_{z'}(R_{i \bar j}\partial_{z'} \phi^i \psi^{\bar
j}), J^a(z)] = 0$. One can proceed to repeat this argument and
show that $[\overline Q_+, \partial^m_zJ^a(z)] =0$ for any $m\geq
1$ at the quantum level always.}

The set of local operators described by $F({\cal J}(z))$ can be
identified with the mathematically defined chiral vertex algebra
$V_{-h^{\vee}}(\frak g)$ associated to $\widehat{\frak g}$ at the
critical level $k=-h^{\vee}$, where $\frak g = \frak{sl}_N$ in our
present case. The action of the Hecke operator on a Hecke
eigensheaf as defined in the axiomatic CFT sense, is equivalent to
an insertion of an operator that lies in the chiral vertex algebra
given by $m$ copies of $V_{-h^{\vee}}(\frak g)$, i.e.,  $\oplus_m
V_{-h^{\vee}}(\frak g)$ \cite{Frenkel}. Such an operator is again
a polynomial operator of the form $F({\cal J}(z))$. In short, the
action of the Hecke operator is equivalent to inserting into the
correlation functions of local primary field operators of the
twisted $(0,2)$ sigma-model on $SL(N)/B$, other local operators
that also lie in the holomorphic chiral algebra of the twisted
$(0,2)$ sigma-model on $SL(N)/B$, which, as emphasised earlier, is
a $\textrm{\it quasi-topological}$ sigma-model with $\it{no}$
boundaries. This is to be contrasted with the description of the
Hecke operators (and Hecke eigensheaves) in the gauge-theoretic
approach to the geometric Langlands program, where they are
interpreted as 't Hooft line operators (and D-branes) in a
$\it{topological}$ sigma-model $\it{with}$ boundaries. Our results
therefore provide an alternative physical interpretation of these
abstract objects of the geometric Langlands correspondence for
$G=SL(N)$, to that furnished in the gauge-theoretic approach by
Kapustin and Witten in \cite{KW}.

\appendix

\renewcommand{\theequation}{\Alph{section}.\arabic{equation}}

\section{The Twisted $(0,2)$ Sigma-Model and Sheaves of CDO's}
\setcounter{equation}{0}

We shall review the relevant features of the twisted $(0,2)$
sigma-model considered by Witten in \cite{CDO} and its relation to
the theory of CDO's constructed by Malikov et al. in \cite{MSV}.
In our aim to keep this paper self-contained, we will present the
relevant details in as comprehensive and coherent a manner as
possible. The interested reader is encouraged to seek the original
references for other details not covered in this appendix.

\newsubsection{The Sigma-Model with $(0,2)$ Supersymmetry}

Let us first recall the two-dimensional non-linear sigma-model
with $(0,2)$ supersymmetry on a complex manifold $X$. It governs
maps $\Phi : \Sigma \to X$, with $\Sigma$ being the worldsheet
Riemann surface. By picking local coordinates $z$, $\bar z$ on
$\Sigma$, and $\phi^{i}$, $\phi^{\bar i}$ on $X$, the map $\Phi$
can then be described locally via the functions $\phi^{i}(z, \bar
z)$ and $\phi^{\bar i}(z, \bar z)$. Let ${\overline K}$ be the
anti-canonical bundle of $\Sigma$ (the bundle of one-forms of type
$(0,1)$), whereby the right-moving spinor bundle of $\Sigma$ is
given ${\overline K}^{1/2}$. Let $TX$ and $\overline {TX}$ be the
holomorphic and anti-holomorphic tangent bundle of $X$. The
right-moving fermi fields consist of $\psi^i$ and $\psi^{\bar i}$,
which are smooth sections of the bundles ${\overline K}^{1/2}
\otimes \Phi^*{TX}$ and ${\overline K}^{1/2} \otimes
\Phi^*{\overline {TX}}$ respectively. Here, $\psi^i$ and
$\psi^{\bar i}$ are superpartners of the scalar fields $\phi^i$
and $\phi^{\bar i}$. Let $g$ be the hermitian metric on $X$. The
action is then given by
\begin{eqnarray}
S& = & \int_{\Sigma} |d^2z| \  {1\over 2}g_{i{\bar j}} (\partial_z
\phi^i \partial_{\bar z}\phi^{\bar j} + \partial_{\bar z} \phi^i
\partial_z\phi^{\bar j} ) + g_{i{\bar j}} \psi^i D_z \psi^{\bar
j}, \label{action}
\end{eqnarray}
whereby $i, {\bar i} = 1 \dots, n={\textrm {dim}}_{\mathbb C}X$,
$|d^2 z| = i dz \wedge d{\bar z}$. In addition,  $D_z$ is the
$\partial$ operator on ${\overline K}^{1/2} \otimes
\phi^*{\overline {TX}}$ using the pull-back of the Levi-Civita
connection on $TX$. In formulas (using a local trivialisation of
${\overline K}^{1/2}$), we have\footnote{Note that we have used a
flat metric and hence vanishing spin connection on the Riemann
surface $\Sigma$ in writing these formulas.} \be D_z \psi^{\bar j}
= \partial_z \psi ^{\bar j} + \Gamma^{\bar j}_{\bar l \bar k}
\partial _z \phi^{\bar l} \psi^{\bar k}, \ee  where $\Gamma^{\bar j}_{\bar l \bar k}$ is
the affine connection of $X$.

The infinitesimal transformation  of the fields generated by the
supercharge $\overline Q_+$ under the first right-moving
supersymmetry,  is given by
\begin{eqnarray}
\label{tx1}
\delta \phi^{i} = 0, & \quad & \delta \phi^{\bar i} = {{\bar \epsilon}_-} \psi ^{\bar i}, \nonumber \\
\delta \psi ^{\bar i} = 0, & \quad & \delta \psi^i = - {{\bar
\epsilon}_-}\partial_{\bar z}\phi^i,
\end{eqnarray}
while the infinitesimal transformation  of the fields generated by
the supercharge $Q_+$ under the second right-moving supersymmetry,
is given by
\begin{eqnarray}
\label{tx2}
\delta \phi^{i} = {{\epsilon}_-}\psi ^{i}, & \quad & \delta \psi^{\bar i} = -{{\epsilon}_-} \partial_{\bar z} \phi^{\bar i}, \nonumber \\
\delta \psi ^{i} = 0, & \quad & \delta \phi^{\bar i} = 0.
\end{eqnarray}
where (${{\bar \epsilon}_-}$)${{\epsilon}_-}$ are
(anti-)holomorphic sections of ${\overline K}^{-1/2}$.

\newsubsection{Twisting the Model}

Classically, the action (\ref{action}) and therefore the model
that it describes, possesses a right-moving R-symmetry, giving
rise to a $U(1)_R$ global symmetry group. Denoting $q_R$ to be the
charge of the right-moving fermi fields under this symmetry group,
we find that $\psi^{\bar i}$ and $\psi^i$ will have charge $q_R =
\pm 1$ respectively. Quantum mechanically however, this symmetry
is anomalous because of non-perturbative worldsheet instantons;
the charge violations for the right-moving global symmetry is
given by $\Delta {q_R} = \int_{\Sigma} \Phi^* c_1(TX)$.

In order to define a twisted variant of the model, the spins of
the fermi fields need to be shifted by a multiple of their
corresponding right-moving charge $q_R$ under the global $U(1)_R$
symmetry group; by considering a shift in the spin $S$ via $S \to
S + {1\over 2} \left [(2{\bar s} -1)q_R \right]$ (where $\bar s$
is a real number), the fermi fields of the twisted model will
transform as smooth sections of the following bundles:
\begin{eqnarray}
\psi^i \in  \Gamma \left({\overline K}^{(1- \bar s)} \otimes
\Phi^*{TX} \right), & \qquad &  \psi^{\bar i} \in  \Gamma
\left({\overline K}^{\bar s} \otimes \Phi^*{\overline{TX}}\right).
\end{eqnarray} Notice that for $s = \bar s = {1\over 2}$, the
fermi fields transform as smooth sections of the same tensored
bundles defining the original $(0,2)$ sigma-model, i.e., we get
back the untwisted model.

To make contact with the theory of CDO's, we shall consider the
case where $\bar s = 0$. Then, the fermi fields of the twisted
model will transform as smooth sections of the following bundles:
\begin{eqnarray}
\psi^i_{\bar z} \in  \Gamma \left({\overline K}^1 \otimes
\Phi^*{TX} \right), & \qquad &  \psi^{\bar i} \in  \Gamma \left(
\Phi^*{\overline{TX}}\right).
\end{eqnarray}
Notice that we have included additional indices in the above
fields so as to reflect their new geometrical characteristics on
$\Sigma$; the fermi field without a $\bar z$ index transform as a
worldsheet scalar, while the  fermi field with a $\bar z$ index
transform as a $(0,1)$-form on the worldsheet. In addition, as
reflected by the $i$, and $\bar i$ indices, all fields continue to
be valued in the pull-back of the corresponding bundles on $X$.
Thus, the action of the twisted variant of the $(0,2)$ sigma-model
is given by
\begin{eqnarray}
S_{\mathrm twist}& = & \int_{\Sigma} |d^2z| \ {1\over 2}g_{i{\bar
j}} (\partial_z \phi^i \partial_{\bar z}\phi^{\bar j} +
\partial_{\bar z} \phi^i \partial_z\phi^{\bar j}) + g_{i{\bar j}} \psi_{\bar z}^i D_z \psi^{\bar j}.
\label{actiontwist}
\end{eqnarray}

A twisted theory is the same as an untwisted one when defined on a
$\Sigma$ which is flat. Hence, locally (where one has the liberty
to select a flat metric),  the twisting does nothing at all.
However, what happens non-locally may be non-trivial. In
particular, note that globally, the supersymmetry parameters
$\epsilon_-$ and ${\bar \epsilon}_-$ must now be interpreted as
sections of different line bundles; in the twisted model, the
transformation laws given by (\ref{tx1}) and (\ref{tx2}) are still
valid, and because of the shift in the spins of the various
fields, we find that for the laws to remain physically consistent,
${\bar \epsilon}_-$ must now be a function on $\Sigma$ while
$\epsilon_-$ must be a section of the non-trivial bundle
${\overline K}^{-1}$. One can therefore canonically pick ${\bar
\epsilon}_-$ to be a constant and $\epsilon_-$ to vanish, i.e.,
the twisted variant of the $(0,2)$ sigma-model has just $\it{one}$
canonical global fermionic symmetry generated by the supercharge
${\overline Q}_+$. Hence, the infinitesimal transformation of the
(twisted) fields under this single canonical symmetry must read
(after setting ${\bar \epsilon}_-$ to 1)
\begin{eqnarray}
\label{txtwist}
\delta \phi^{i} = 0, & \quad & \delta \phi^{\bar i} = \psi^{\bar i}, \nonumber \\
\delta \psi ^{\bar i} = 0, & \quad & \delta \psi_{\bar z}^i = -
\partial_{\bar z}\phi^i.
\end{eqnarray}
From the $(0,2)$ supersymmetry algebra, we have ${{\overline
Q}^2_+} = 0$. In addition, (after twisting) ${{\overline Q}_+} $
transforms as a scalar. Consequently, we find that the symmetry is
nilpotent i.e., $\delta^2 = 0$ (off-shell), and behaves as a
BRST-like symmetry.

Note at this point that the transformation laws of (\ref{txtwist})
can be expressed in terms of the BRST operator ${\overline Q}_+$,
whereby $\delta W = \{{\overline Q}_+, W\}$ for any field $W$. One
can then show that the action  (\ref{actiontwist}) can be written
as \be S_{\mathrm twist} = \int_{\Sigma} |d^2z| \{{\overline Q}_+,
V\} + S_{\mathrm top} \label{Stwist} \ee where \be V = - g_{i \bar
j}  \psi_{\bar z}^i \partial_z \phi^{\bar j}, \label{chi} \ee
while \be S_{\mathrm top} = {1\over 2}\int_{\Sigma} g_{i \bar j}
\left( \partial_z \phi^i
\partial_{\bar z} \phi^{\bar j} - \partial_{\bar z} \phi^i
\partial_z \phi^{\bar j} \right) \ee is $\int_{\Sigma} \Phi^*(K)$,
the integral of the pull-back to $\Sigma$ of the $(1,1)$-form $K=
{i \over 2} g_{i \bar j} d\phi^i \wedge d\phi^{\bar j}$.

Notice that since ${{\overline Q}^2_+} = 0$, the first term on the
RHS of (\ref{Stwist}) is invariant under the transformation
generated by ${\overline Q}_+$. In addition, as mentioned in the
introduction, we will be studying the twisted model in
$\it{perturbation}$ theory, where one does an expansion in the
inverse of the large-radius limit. Hence, only the degree-zero
maps of the term $\int_{\Sigma} \Phi^*(K)$ contribute to the path
integral factor $e^{-S_{\mathrm twist}}$. Therefore, in the
perturbative limit, one can set $\int_{\Sigma} \Phi^*(K) = 0$,
i.e., $S_{\mathrm{top}}$ can be set to zero. Thus, the action
given in (\ref{Stwist}) is invariant under the BRST symmetry as
required. Moreover, for the transformation laws of (\ref{txtwist})
to be physically consistent, ${\overline Q}_+$ must have charge
$q_R = 1$ under the global $U(1)_R$ gauge group. Since $V$ has a
corresponding charge of $q_R = -1$, $S_{\mathrm twist}$ in
(\ref{Stwist}) continues to be invariant under the $U(1)_R$
symmetry group at the classical level. In summary, the effective
action in perturbation theory reads
\begin{eqnarray}
S_{\mathrm pert}& = & \int_{\Sigma} |d^2z| \  g_{i{\bar j}}(
\partial_z \phi^{\bar j} \partial_{\bar z}\phi^i +
\psi_{\bar z}^i D_z \psi^{\bar j}), \label{actionpert}
\end{eqnarray}
where it can also written as \be S_{\mathrm pert} = \int_{\Sigma}
|d^2z| \{{\overline Q}_+, V\}. \label{Spert} \ee

Note that the original symmetries of the theory persist despite
limiting ourselves to perturbation theory; even though $S_{\mathrm
top} = 0$,  from (\ref{Spert}),  one finds that $S_{\mathrm pert}$
is invariant under the nilpotent BRST symmetry generated by
${\overline Q}_+$. It is also invariant under the $U(1)_R$ global
symmetry. $S_{\mathrm pert}$ shall henceforth be the action of
interest in all our subsequent discussions.

\newsubsection{Chiral Algebras from the Twisted Sigma-Model}

\bigskip\noindent{\it The Chiral Algebra}

Classically, the model is conformally invariant. The trace of the
stress tensor from $S_{\mathrm pert}$ vanishes, i.e., $T_{z \bar
z} = 0$. The other non-zero components of the stress tensor, at
the classical level, are given by \be T_{zz} = g_{i \bar j}
\partial_z \phi^i \partial_{z} \phi^{\bar j}, \label{Tzz} \ee and \be T_{\bar z \bar z} =g_{i \bar j}
\partial_{\bar z} \phi^i
\partial_{\bar z} \phi^{\bar j} + g_{i \bar j}  \psi_{\bar z}^i
\left ( \partial_{\bar z} \psi^{\bar j} + \Gamma^{\bar j}_{\bar l
\bar k}\partial_{\bar z} \phi^{\bar l} \psi^{\bar k} \right). \ee
Furthermore, one can go on to show that \be T_{\bar z \bar z} = \{
{\overline Q}_+ , - g_{i \bar j} \psi_{\bar z}^i \partial_{\bar z}
\phi^{\bar j} \}, \label{tZZ} \ee and
\begin{eqnarray}
\label{tzz} [{\overline Q}_+ , T_{zz} ] & = &   g_{i \bar
j}\partial_z \phi^i
D_z \psi^{\bar j} \nonumber \\
& = & 0 \hspace{0.2cm} (\textrm {on-shell}).
\end{eqnarray}
From (\ref{tzz}) and (\ref{tZZ}), we see that all components of
the stress tensor are ${\overline Q}_+$-invariant; $T_{zz}$ is an
operator in the ${\overline Q}_+$-cohomology while $T_{\bar z \bar
z}$ is ${\overline Q}_+$-exact and thus trivial in ${\overline
Q}_+$-cohomology. The fact that $T_{zz}$ is not ${\overline
Q}_+$-exact even at the classical level implies that the twisted
model is not a two-dimensional $\it{topological}$ field theory;
rather, it is a two-dimensional $\it{conformal}$ field theory.
This because the original model has $(0,2)$  and not $(2,2)$
supersymmetry. On the other hand, the fact that $T_{\bar z \bar
z}$ is ${\overline Q}_+$-exact has some non-trivial consequences
on the nature of the local operators in the ${\overline
Q}_+$-cohomology. Let us discuss this further.

We say that a local operator $\cal O$ inserted at the origin has
dimension $(n,m)$ if under a rescaling $z\to \lambda z$, $\bar
z\to \bar\lambda z$ (which is a conformal symmetry of the
classical theory), it transforms as $\partial^{n+m}/\partial
z^n\partial\bar z^m$, that is, as
$\lambda^{-n}\bar\lambda{}^{-m}$. Classical local operators have
dimensions $(n,m)$ where $n$ and $m$ are non-negative
integers.\footnote{Anomalous dimensions under RG flow may shift
the values of $n$ and $m$ quantum mechanically, but the spin given
by $(n-m)$, being an intrinsic property, remains unchanged.}
However, only local operators with $m = 0$ survive in ${\overline
Q}_+$-cohomology. The reason for the last statement is that the
rescaling of $\bar z$ is generated by $\bar L_0=\oint d\bar z\,
\bar z T_{\bar z\bar z}$.  As we noted in the previous paragraph,
$T_{\bar z\,\bar z}$ is of the form $\{{\overline Q}_+,\dots\}$,
so $\bar L_0=\{{\overline Q}_+,V_0\} $ for some $V_0$. If $\cal O$
is to be admissible as a local physical operator, it must at least
be true that $\{{\overline Q}_+, {\cal O}\}=0$. Consequently,
$[\bar L_0,{\cal O}]=\{{\overline Q}_+,\{V_0,{\cal O}\}\}$.  Since
the eigenvalue of $\bar L_0$ on $\cal O$ is $m$, we have $[\bar
L_0,{\cal O}]=m{\cal O}$. Therefore, if $m\not= 0$, it follows
that ${\cal O}$ is $\overline Q_+$-exact and thus trivial in
$\overline Q_+$-cohomology.

By a similar argument, we can show that $\cal O$,  as an element
of the $\overline Q_+$-cohomology, varies holomorphically with
$z$. Indeed, since the momentum operator (which acts on $\cal O$
as $\partial_{\bar z}$) is given by $\bar L_{-1}$, the term
$\partial_{\bar z} \cal O$ will be given by the commutator $[ \bar
L_{-1}, \cal O]$. Since $\bar L_{-1} = \oint d\bar z\,T_{\bar
z\bar z}$, we will  have $\bar L_{-1}=\{{\overline Q_+},V_{-1}\}$
for some $V_{-1}$. Hence, because $\cal O$ is physical such that
${\{\overline Q_+, \cal O\}} =  0$, it will be true that
$\partial_{\bar z}{\cal O}=\{{\overline Q_+},\{V_{-1},{\cal
O}\}\}$ and thus vanishes in $\overline Q_+$-cohomology.

The observations that we have so far are based solely on classical
grounds. The question that one might then ask is whether these
observations will continue to hold when we eventually consider the
quantum theory. The key point to note is that if it is true
classically that a cohomology vanishes, it should continue to do
so in perturbation theory, when quantum effects are small enough.
Since the above observations were made based on the classical fact
that $T_{\bar z \bar z}$ vanishes in $\overline Q_+$-cohomology,
they will continue to hold at the quantum level. Let us look at
the quantum theory more closely.

\vspace{0.4cm}{\noindent{\it The Quantum Theory}}

Quantum mechanically, the conformal structure of the theory is
violated by a non-zero one-loop $\beta$-function; renormalisation
adds to the classical action $S_{\mathrm pert}$ a term of the
form: \be  \Delta_{\textrm 1-loop}\ = \ c_{1} \ R_{ i \bar j}
\partial_{z}\phi^{\bar j}\psi_{\bar z}^{i}, \label{1-loop}
\ee for some divergent constants $c_{1}$, where $R_{ i \bar j}$ is
the Ricci tensor of $X$. In the Calabi-Yau case, one can choose a
Ricci-flat metric such that $\Delta_{\textrm 1-loop}$ vanishes and
the original action is restored. In this case, the classical
observations made above continue to hold true. On the other hand,
in the ``massive models'' where $c_1(X) \neq 0$, there is no way
to set $\Delta_{\textrm 1-loop}$ to zero. Conformal invariance is
necessarily lost, and there {\it is} nontrivial RG running.
However, one can continue to express $T_{\bar z \bar z}$ as
$\{\overline Q_+, \dots \}$, i.e., it remains $\overline
Q_+$-exact, and thus continues to vanish in $\overline
Q_+$-cohomology. Hence, the above observations about the
holomorphic nature of the local operators having dimension $(n,0)$
$\it{continue}$ to hold in the quantum theory.

We would also like to bring to the reader's attention another
important feature of the $\overline Q_+$-cohomology at the quantum
level. Recall that classically, we had $[\overline Q_+, T_{zz}] =
0$ via the classical equations of motion. Notice that the
classical expression for $T_{zz}$ is not modified at the quantum
level (at least up to one-loop), since even in the non-Calabi-Yau
case, the additional term of $\Delta_{\mathrm 1-loop}$ in the
quantum action does not contribute to $T_{zz}$. However, due to
one-loop corrections to the action of $\overline Q_+$, we have, at
the quantum level \be [\overline Q_+, T_{zz}] = \partial_z (R_{ i
\bar j}\partial_z \phi^i    \psi^{\bar j}).\label{tzzanomaly} \ee
Note that the term on the RHS of (\ref{tzzanomaly}) $\it{cannot}$
be eliminated through the equations of motion in the quantum
theory. Neither can we modify $T_{zz}$ (by subtracting a total
derivative term) such that it continues to be $\overline
Q_+$-invariant. This implies that in a `massive'  model, operators
do not remain in the $\overline Q_+$-cohomology after general
holomorphic coordinate transformations on the worldsheet, i.e.,
the model is $\it{not}$ conformal at the level of the $\overline
Q_+$-cohomology.\footnote{In $\S$2 and $\S$3, we will examine more
closely, from a different point of view, the one-loop correction
to the action of $\overline Q_+$ associated with the
beta-function, where (\ref{tzzanomaly}) will appear in a different
guise.} However, $T_{zz}$ continues to be holomorphic in $z$ up to
$\overline Q_+$-trivial terms; from the conservation of the stress
tensor, we have $\partial_{\bar z}T_{zz} = - \partial_z T_{z \bar
z}$, and $T_{z \bar z}$, while no longer zero, is now given by
$T_{z \bar z} = \{\overline Q_+, G_{z \bar z}\}$ for some $G_{z
\bar z}$, i.e., $\partial_z T_{z \bar z}$ continues to be
$\overline Q_+$-exact,  and  $\partial_{\bar z}T_{zz} \sim 0$ in
$\overline Q_+$-cohomology. The holomorphy of $T_{zz}$, together
with the relation (\ref{tzzanomaly}), has further implications for
the $\overline Q_+$-cohomology of local operators; by a Laurent
expansion of $T_{zz}$,\footnote{Since we are working modulo
$\overline Q_+$-trivial operators, it suffices for $T_{zz}$ to be
holomorphic up to $\overline Q_+$-trivial terms before an
expansion in terms Laurent coefficients is permitted.} one can use
(\ref{tzzanomaly}) to show that $[\overline Q_+, L_{-1} ] = 0$.
This means that operators remain in the $\overline Q_+$-cohomology
after global translations on the worldsheet. In addition, recall
that $\overline Q_+$ is a scalar with spin zero in the twisted
model. As shown few paragraphs before, we have the condition $\bar
L_0 = 0$. Let the spin be $S$, where $S= L_0 - \bar L_0$.
Therefore, $[\overline Q_+, S]= 0$ implies that $[\overline Q_+,
L_0 ] = 0$. In other words, operators remain in the $\overline
Q_+$-cohomology after global dilatations of the worldsheet
coordinates.

One can also make the following observations about the correlation
functions of these local operators. Firstly, note that $\left < \{
\overline Q_+, W \} \right> = 0$ for any $W$, and recall that for
any local physical operator ${\cal O_{\alpha}}$, we have
${\{\overline Q_+, {\cal O_{\alpha}}\}} = 0$. Since the
$\partial_{\bar z}$ operator on $\Sigma$ is given by ${\bar
L}_{-1} = \oint d{\bar z}\ T_{\bar z \bar z}$, where  $T_{\bar z
\bar z} = \{ \overline Q_+, \dots \}$, we find that
${\partial_{\bar z}\left <{\cal O}_1(z_1) {\cal O}_2(z_2) \dots
{\cal O}_s(z_s)  \right >}$ is given by $ \oint d{\bar z} \left
<\{ \overline Q_+, \dots \} \ {\cal O}_1(z_1) {\cal O}_2(z_2)
\dots {\cal O}_s(z_s) \right > = \oint d{\bar z} \left <
\{\overline Q_+, \dots \prod_{i} {{\cal O}_i(z_i)} \} \right> =
0$. Thus, the correlation functions are always holomorphic in $z$.
Secondly, $T_{z \bar z} = \{\overline Q_+, G_{z \bar z}\}$ for
some $G_{z \bar z}$ in the `massive' models. Hence, the variation
of the correlation functions due to a change in the scale of
$\Sigma$ will be given by $\left <{{\cal O}_1(z_1)} {{\cal
O}_2(z_2)} \dots {{\cal O}_s(z_s)} {\{\overline Q_+, G_{z \bar z}
\}} \right >= \left < \{\overline Q_+, \prod_{i} {{\cal O}_i(z_i)}
\cdot G_{z \bar z} \} \right> = 0$.   In other words, the
correlation functions of local physical operators will continue to
be invariant under arbitrary scalings of $\Sigma$. Thus, the
correlation functions are always independent of the K\"ahler
structure on $\Sigma$ and depend only on its complex structure.

\vspace{0.4cm}{\noindent{\it  A Holomorphic Chiral Algebra $\cal
A$}}

Let ${\cal O} (z)$ and $\widetilde {\cal O} (z')$ be two
$\overline Q_+$-closed operators such that their product is
$\overline Q_+$-closed as well. Now, consider their operator
product expansion or OPE: \be {\cal O}(z) {\widetilde {\cal
O}}(z') \sim \sum_k f_k (z-z') {\cal O}_k (z'), \label{OPE} \ee in
which the explicit form of the coefficients $f_k$ must be such
that the scaling dimensions and $U(1)_R$ charges of the operators
agree on both sides of the OPE. In general, $f_k$ is not
holomorphic in $z$. However, if we work modulo $\overline
Q_+$-exact operators in passing to the $\overline Q_+$-cohomology,
the $f_k$'s which are non-holomorphic and are thus not annihilated
by $\partial / \partial {\bar z}$, drop out from the OPE because
they multiply operators ${\cal O}_k$ which are $\overline
Q_+$-exact. This is true because $\partial / \partial{\bar z}$
acts on the LHS of (\ref{OPE}) to give terms which are
cohomologically trivial.\footnote{Since $\{\overline Q_+,{\cal
O}\}=0$, we have $\partial_{\bar z}{\cal O}=\{\overline Q_+,
V(z)\}$ for some $V(z)$, as argued before. Hence $\partial_{\bar
z}{\cal O}(z)\cdot {\widetilde {\cal O}}(z')=\{\overline
Q_+,V(z){\widetilde {\cal O}}(z')\}$.} In other words, we can take
the $f_k$'s to be holomorphic coefficients in studying the
$\overline Q_+$-cohomology. Thus, the OPE of (\ref{OPE}) has a
holomorphic structure.

In summary, we have established that the $\overline
Q_+$-cohomology of holomorphic local operators has a natural
structure of a holomorphic chiral algebra (as defined in the
mathematical literature) which we shall henceforth call $\cal A$;
it is always preserved under global translations and dilatations,
though (unlike the usual physical notion of a chiral algebra) it
may not be preserved under general holomorphic coordinate
transformations on the Riemann surface $\Sigma$ depending on
whether $c_1(X)$ vanishes or not. Likewise, the OPEs of the chiral
algebra of local operators obey the usual relations of holomorphy,
associativity, and invariance under translations and scalings of
$z$, but not necessarily invariance under arbitrary holomorphic
reparameterisations of $z$. The local operators are of dimension
(n,0) for $n \geq 0$, and the chiral algebra of such operators
requires a flat metric up to scaling on $\Sigma$ to be defined.
Therefore, the chiral algebra that we have obtained can either be
globally-defined on a $\Sigma$ of genus-one, or locally-defined on
an arbitrary but curved $\Sigma$. The sigma-model is also plagued
by anomalies of the form ${1\over 2} c_1(\Sigma)c_1(X)$ and
${1\over 2}p_1(X)$, where $p_1(X)$ is the first Pontryagin class
of $TX$. However, for the flag manifolds $X$ considered in this
paper, $p_1(X)$ vanishes. In addition, since the chiral algebra
that we will be analysing depends only on a local coordinate $z$
on $\Sigma$, i.e., we will only be working locally on $\Sigma$,
the first anomaly will also be irrelevant in our context.
Therefore, we shall henceforth have nothing more to say about
these anomalies. Last but not least, as is familiar for chiral
algebras, the correlation functions of these operators depend on
$\Sigma$ only via its complex structure. The correlation functions
are holomorphic in the parameters of the theory and are therefore
protected from perturbative corrections.

\newsubsection{Local Operators as Perturbative Observables}

\bigskip\noindent{\it Local Operators}

In general, a local operator is an operator $\cal F$ that is a
function of the $\it{physical}$ fields $\phi^i$, $\phi^{\bar i}$,
$\psi^i_{\bar z}$, $\psi^{\bar i}$, and their derivatives with
respect to $z$ and $\bar z$.\footnote{Note here that since we are
interested in local operators which define a holomorphic chiral
algebra on the Riemann surface $\Sigma$, we will work locally on
$\Sigma$ with local parameter $z$. Hence, we need not include in
our operators the dependence on the scalar curvature of $\Sigma$.}
However, as we saw in $\S$A.3, the $\overline Q_+$-cohomology
vanishes for operators of dimension $(n,m)$ with $m \neq 0$. Since
$\psi^i_{\bar z}$ and the derivative $\partial_{\bar z}$ both have
$m=1$ (and recall from $\S$A.3 that a physical operator cannot
have negative $m$ or $n$), $\overline Q_+$-cohomology classes can
be constructed from just $\phi^i$, $\phi^{\bar i}$, $\psi^{\bar
i}$ and their derivatives with respect to $z$. Note that the
equation of motion for $\psi^{\bar i}$ is $D_z \psi^{\bar i}= 0$.
Thus, we can ignore the $z$-derivatives of $\psi^{\bar i}$, since
it can be expressed in terms of the other fields and their
corresponding derivatives. Therefore, a chiral (i.e. $\overline
Q_+$-invariant) operator which represents a $\overline
Q_+$-cohomology class is given by $ {\cal
F}(\phi^i,\partial_z\phi^i,\partial_z^2\phi^i,\dots; \phi^{\bar
i},\partial_z\phi^{\bar i},\partial_z^2\phi^{\bar i},\dots;
\psi^{\bar i})$, where we have tried to indicate that $\cal F$
might depend on $z$ derivatives of $\phi^{i}$ and $\phi^{\bar i}$
of arbitrarily high order, though not on derivatives of
$\psi^{\bar i}$. If the scaling dimension of $\cal F$ is bounded,
it will mean that $\cal F$ depends only on the derivatives of
fields up to some finite order or is a polynomial of bounded
degree in those. Notice that $\cal F$ will always be a polynomial
of finite degree in $\psi ^{\bar i}$, simply because $\psi^{\bar
i}$ is fermionic and can only have a finite number of components
before they vanish due to their anticommutativity. However, the
dependence of $\cal F$ on $\phi^i$, $\phi^{\bar i}$ (as opposed to
their derivatives) need not have any simple form. Nevertheless, we
can make the following observation - from the $U(1)_R$ charges of
the fields listed in $\S$A.2, we see that if $\cal F$ is
homogeneous of degree $k$ in $\psi^{\bar i}$, then it has
$U(1)_R$-charge $q_R =k$.

A general $q_R= k$ operator ${\cal F}
(\phi^i,\partial_z\phi^i,\dots; \phi^{\bar i},\partial_z\phi^{\bar
i},\dots; \psi^{\bar i})$ can be interpreted as a $(0,k)$-form on
$X$ with values in a certain tensor product bundle. In order to
illustrate the general idea behind this interpretation, we will
make things explicit for operators of dimension $(0,0)$ and
$(1,0)$. Similar arguments will likewise apply for operators of
higher dimension. For dimension $(0,0)$, the most general operator
takes the form ${\cal F}(\phi^i,\phi^{\bar i}; \psi^{\bar j})=
f_{\bar j_1,\dots,\bar j_k}(\phi^i, \phi^{\bar i}) \psi^{\bar
j_i}\dots \psi^{\bar j_k}$; thus, $\cal F$ may depend on $\phi^i$,
and $\phi^{\bar i}$, but not on their derivatives, and is $k^{th}$
order in $\psi^{\bar j}$. Mapping $\psi^{\bar j}$ to $d\phi^{\bar
j}$ (which one can do so as both $\psi^{\bar j}$ and $d\phi^{\bar
j}$ are anticommuting quantities), such an operator corresponds to
an ordinary $(0,k)$-form $f_{\bar j_1,\dots,\bar j_k}(\phi^i,
\phi^{\bar i})d\phi^{\bar j_1}\dots d\phi^{\bar j_k}$ on $X$. For
dimension $(1,0)$, there are two general cases. In the first case,
we have an operator ${\cal F}(\phi^l,\partial_z\phi^i; \phi^{\bar
l}; \psi^{\bar j})=f_{i,\bar j_1,\dots,\bar j_k}(\phi^l,\phi^{\bar
l}) \partial_z\phi^i \psi^{\bar j_1}\dots\psi^{\bar j_k}$ that is
linear in $\partial_z\phi^i$ and does not depend on any other
derivatives. It is a $(0,k)$-form on $X$ with values in the bundle
$T^*X$; alternatively, it is a $(1,k)$-form on $X$. Similarly, in
the second case, we have an operator ${\cal F}(\phi^l; \phi^{\bar
l}, \partial_z \phi^{\bar s}; \psi^{\bar j})=f^i{}_{\bar
j_1,\dots,\bar j_k}(\phi^l, \phi^{\bar l}) g_{i \bar s}\partial_z
\phi^{\bar s}\psi^{\bar j_i}\dots \psi^{\bar j_k}$ that is linear
in $\partial_z \phi^{\bar s}$ and does not depend on any other
derivatives. It is a $(0,k)$-form on $X$ with values in the bundle
$TX$. In a similar fashion, for any integer $n>0$, the operators
of dimension $(n,0)$ and charge $q_R = k$ can be interpreted as
$(0,k$)-forms with values in a certain bundle over $X$. This
structure persists in quantum perturbation theory, but there may
be perturbative corrections to the complex structure of the
bundle.

\bigskip\noindent{\it The Quantum Action of $\overline Q_+$}

The action of $\overline Q_+$ on such operators can be easily
described at the classical level. If we interpret $\psi^{\bar i}$
as $d\phi^{\bar i}$, then $\overline Q_+$ acts on functions of
$\phi^i$ and $\phi^{\bar i}$, and  is simply the $\bar\partial $
operator on $X$. This follows from the transformation laws
$\delta\phi^{\bar i}=\psi^{\bar i}$, ${\delta\phi^i} = 0$,
${\delta \psi^{\bar i}}=0$. The interpretation of $\overline Q_+$
as the $\bar\partial$ operator will remain valid when $\overline
Q_+$ acts on a more general operator ${\cal
F}(\phi^i,\partial_z\phi^i,\dots;\phi^{\bar i},\partial_z
\phi^{\bar i},\dots; \psi^{\bar i})$ that does depend on the
derivatives of $\phi^i$ and $\phi^{\bar i}$. The reason for this
is because we have the equation of motion $D_z \psi^{\bar i}=0$.
This means that one can neglect the action of $\overline Q_+$ on
derivatives $\partial_z^m\phi^{\bar i}$ with $m>0$.

Perturbatively however, there will be corrections to the action of
$\overline Q_+$. In fact, as briefly mentioned in $\S$A.3 earlier,
(\ref{tzzanomaly}) provides such an example - the holomorphic
stress tensor $T_{zz}$, though not corrected at 1-loop, is no
longer $\overline Q_+$-closed because the action of $\overline
Q_+$ has received perturbative corrections. The fact that
$\overline Q_+$ does not always act as the $\bar
\partial$ operator at the quantum level suggests
that one needs a more general framework than just ordinary
Dolbeault or $\bar \partial$-cohomology to describe the $\overline
Q_+$-cohomology of the twisted $(0,2)$ sigma-model. Indeed, as we
will show shortly, the appropriate description of the $\overline
Q_+$-cohomology of local operators spanning the chiral algebra
will be given in terms of the more abstract notion of Cech
cohomology.

\newsubsection{A Sheaf of Chiral Algebras}

We shall now explain the idea of a ``sheaf of chiral algebras'' on
$X$. To this end, note that both the $\overline Q_+$-cohomology of
local operators (i.e., operators that are local on the Riemann
surface $\Sigma$), and the fermionic symmetry generator $\overline
Q_+$, can be described locally on $X$. Hence, one is free to
restrict the local operators to be well-defined not throughout
$X$, but only on a given open set $U \subset X$. Since in
perturbation theory, we are considering trivial maps $\Phi :\Sigma
\to X$ with no multiplicities, an operator defined in an open set
$U$ will have a sensible operator product expansion with another
operator defined  in $U$. From here, one can naturally proceed to
restrict the definition of the operators to smaller open sets,
such that  a global definition of the operators can be obtained by
gluing together the open sets on their unions and intersections.
From this description, in which one associates a chiral algebra,
its OPEs, and chiral ring to every open set $U \subset X$, we get
what is known mathematically as a ``sheaf of chiral algebras''. We
shall call this sheaf $\widehat {\cal A}$.

\bigskip \noindent{\it Description of $\cal A$ via Cech Cohomology}

In perturbation theory, one can also describe    the  $\overline
Q_+$-cohomology   classes by a form of Cech cohomology. This
alternative description will take us to the mathematical point of
view on the subject \cite{MSV, GMS1}. In essence, we will show
that the chiral algebra $\cal A$ of the $\overline Q_+$-cohomology
classses of the twisted $(0,2)$ sigma-model on $X$, can be
represented, in perturbation theory, by the classes of the Cech
cohomology of the sheaf $\widehat {\cal A}$ of locally-defined
chiral operators. To this end, we shall demonstrate an isomorphism
between the $\overline Q_+$-cohomology classes and the classes of
the Cech cohomology of $\widehat {\cal A}$.

Let us start by considering an open set $U \subset X$ that is
isomorphic to a contractible space such as an open ball in
$\mathbb C^n$, where $n = {\textrm {dim}}_{\mathbb C} (X)$.
Because $U$ is a contractible space, any bundle over $U$ will be
trivial. In the absence of perturbative corrections at the
classical level, any operator ${\cal F}$ in the $\overline
Q_+$-cohomology will be classes of $H^{0,k}_{\bar \partial}(U,
\widehat {F})$ on $U$ as explained earlier. Since $\widehat F$
will be a trivial bundle over $U$, which means that $\widehat {F}$
will always possess a global section, i.e., it corresponds to a
soft sheaf, and because the higher Cech cohomologies of a soft
sheaf are trivial \cite{Wells}, we will have $
{H_{\textrm{Cech}}^{k} (U, {\widehat {F}} )} = 0$ for $k > 0$.
Mapping this back to Dolbeault cohomology via the Cech-Dolbeault
isomorphism, we find that $H^{0,k}_{\bar
\partial}(U, \widehat {F}) = 0$ for $k > 0$.  Note that small
quantum corrections in the perturbative limit can only annihilate
cohomology classes and not create them. Hence, in perturbation
theory, it follows that the local operators ${\cal F}$  with
positive values of $q_R$,  must vanish in    $\overline
Q_+$-cohomology on $U$.

Now consider a good cover of  $X$ by open sets $\{U_a \}$.  Since
the intersection of open sets $\{U_a \}$ also give open sets
(isomorphic to open balls in $\mathbb C^n$),  $\{U_a \}$ and all
of their intersections have the same property as $U$    described
above: $\bar\partial$-cohomology and hence $\overline
Q_+$-cohomology vanishes for positive values of $q_R$ on $\{U_a
\}$ and their intersections.

Let the operator ${\cal F}_1$ on $X$ be a $\overline
Q_+$-cohomology class with $q_R = 1$. It is here that we shall
import the usual arguments relating a $\bar\partial$ and Cech
cohomology, to demonstrate an isomorphism between the $\overline
Q_+$-cohomology and a Cech cohomology.  When restricted to an open
set        $U_a$, the operator ${\cal F}_1$ must be trivial in
$\overline Q_+$-cohomology, i.e., ${\cal F}_1 =\{\overline
Q_+,{\cal C}_a\}$, where ${\cal C}_a$ is an operator of $q_R=0$
that is well-defined in $U_a$.

Now, since $\overline Q_+$-cohomology classes such as ${\cal F}_1$
can be globally-defined on $X$, we have ${\cal F}_1 =\{\overline
Q_+ ,{\cal C}_a\}=\{\overline Q_+,{{\cal C}_b}\}$ over the
intersection  $U_a\cap U_b$, so $\{\overline Q_+,{{\cal C}_a}-
{{\cal C}_b}\}=0$. Let ${\cal C}_{ab}= {{\cal C}_a}- {{\cal
C}_b}$.  For each $a$ and $b$, ${\cal C}_{ab}$ is defined in
$U_a\cap U_b$. Therefore, for all $a,b,c$, we have \be {\cal
C}_{ab}=     -{\cal C}_{ba}, \quad {{\cal C}_{ab}}+ {{\cal
C}_{bc}} + {{\cal C}_{ca}} =0. \label{cab} \ee Moreover, for ($q_R
=0$) operators ${\cal K}_a$ and ${\cal K}_b$, whereby $\{
\overline Q_+, {{\cal K}_a} \} = \{ \overline Q_+, {{\cal K}_b}
\}= 0$, we have an equivalence relation \be {\cal C}_{ab} \sim
{{\cal C'}_{ab} = {{\cal C}_{ab} + {\cal K}_a - {\cal K}_b}}.
\label{cab1} \ee Note that the collection $\{C_{ab} \}$ are
operators in the $\overline Q_+$-cohomology with well-defined
operator product expansions.

Since the local operators with positive values of $q_R$ vanish in
$\overline Q_+$-cohomology on an arbitrary open set $U$, the sheaf
$\widehat {\cal A}$ of the chiral algebra of operators has for its
local sections  the $\psi^{\bar i}$-independent (i.e. $q_R =0$)
operators  ${\widehat {\cal F}} (\phi^i,\partial_z\phi^i,\dots;
\phi^{\bar i},\partial_z\phi^{\bar i},\dots)$ that are annihilated
by $\overline Q_+$. Each $C_{ab}$ with $q_R =0$ is thus a section
of $\widehat{\cal A}$ over the intersection $U_a\cap U_b$. From
(\ref{cab}) and (\ref{cab1}), we find that the collection
$\{C_{ab} \}$ defines the elements  of the first Cech cohomology
group $H_{{\rm Cech}}^1(X, \widehat{\cal A})$.

Next, note that the $\overline Q_+$-cohomology classes are defined
as those operators which are $\overline Q_+$-closed, modulo those
which can be globally written as $\{ \overline Q_+, \dots \}$ on
$X$. In other words, ${\cal F}_1$ vanishes in $\overline
Q_+$-cohomology if we can write it as ${\cal F}_1 = \{\overline
Q_+ ,{\cal C}_a\}=\{\overline Q_+,{{\cal C}_b}\} =  \{\overline
Q_+ ,{\cal C}\}$, i.e., ${\cal C}_a = {\cal C}_b$ and hence ${\cal
C}_{ab} = 0$. Therefore, a vanishing $\overline Q_+$-cohomology
with $q_R =1$ corresponds to a vanishing first Cech cohomology.
Thus, we have obtained a map between the $\overline
Q_+$-cohomology with $q_R =1$ and a first Cech cohomology. Similar
to the case of relating a $\bar \partial$ and Cech cohomology, one
can also run everything backwards and construct an inverse of this
map \cite{CDO}. Since there is nothing unique about the $q_R =1$
case, we can repeat the above procedure for operators with $q_R >
1$. In doing so, we find that  the $\overline Q_+$-cohomology
coincides with the Cech cohomology of $\widehat {\cal A}$ for all
$q_R$. Hence, the chiral algebra $\cal A$ of the twisted $(0,2)$
sigma-model will be given by  $\bigoplus_{q_R} H^{q_R}_{\textrm
{Cech}} (X, {\widehat{\cal A}})$ as a vector space. As there will
be no ambiguity, we shall henceforth omit the label ``Cech'' when
referring to the cohomology of $\widehat {\cal A}$.

Note that in the mathematical literature, the sheaf $\widehat
{\cal A}$, also known as a sheaf of vertex algebras, is studied
purely from the Cech viewpoint; the field $\psi^{\bar i}$ is
omitted and locally on $X$, one considers operators constructed
only from $\phi^i$, $\phi^{\bar i}$ and their $z$-derivatives. The
chiral algebra $\cal A$ of $\overline Q_+$-cohomology classes with
positive $q_R$ are correspondingly constructed as Cech
$q_R$-cocycles. However, in the physical description via a
Lagrangian and $\overline Q_+$ operator, the sheaf  $\widehat
{\cal A}$ and its cohomology are given a $\bar
\partial$-like description, where Cech $q_R$-cycles are
represented by operators that are $q^{th}_R$ order in the field
$\psi^{\bar i}$. Notice that the mathematical description does not
involve any form of perturbation theory at all. Instead, it
utilises the abstraction of Cech cohomology to define the spectrum
of operators in the quantum sigma-model. It is in this sense that
the study of the sigma-model is given a rigorous foundation in the
mathematical literature.

\newsubsection{Relation to a Free $\beta\gamma$ System}

Now, we shall express in a physical language a few key points that
are made in the mathematical literature \cite{GMS1} starting from
a Cech viewpoint.  Let us start by providing a convenient
description of the local structure of the sheaf $\widehat {\cal
A}$. To this end, we will describe in a new way the $\overline
Q_+$-cohomology of operators that are regular in a small open set
$U \subset X$. We assume that $U$ is isomorphic to an open ball in
$\mathbb C^n$ and is thus contractible.

Notice from $S_{\mathrm pert}$ in (\ref{Spert}) and $V$ in
(\ref{chi}), that the hermitian metric on $X$ only appears inside
a term of the form $\{\overline Q_+, \dots \}$  in the action.
Thus, any shift in the metrics will also appear inside $\overline
Q_+$-exact (i.e. $\overline Q_+$-trivial) terms. Consequently, for
our present purposes, we can arbitrarily redefine the value of the
hermitian metric on $X$, since it does not affect the analysis of
the $\overline Q_+$-cohomology. Therefore, to describe the local
structure, we can pick a hermitian metric that is flat when
restricted to $U$. Thus, the local action (derived from the flat
hermitian metric) of the twisted $(0,2)$ sigma-model on $U$ is \be
I = {1 \over 2 \pi} \int_{\Sigma} |d^2 z| \sum_{i, \bar j}
\delta_{i \bar j} \left ( \partial_z \phi^{\bar j} \partial_{\bar
z}\phi^i + \psi ^i_{\bar z}
\partial_z \psi^{\bar j} \right ).\label{Su}
\ee

Now let us describe the $\overline Q_+$-cohomology classes of
operators regular in $U$.  As explained earlier, these are
operators of dimension $(n,0)$ that are independent of $\psi^{\bar
i}$. In general, such operators  are of the form ${\widehat {\cal
F}} (\phi^i,\partial_z\phi^i,\dots; \phi^{\bar i},\partial_z
\phi^{\bar i},\dots)$. Recall that $\overline Q_+$ will act  as
the $\bar \partial$ operator at the classical level. Because
perturbative corrections to the action of $\overline Q_+$ can be
ignored on a flat open set $U$ \cite{CDO}, on the classes of
operators in $U$, $\overline Q_+$ will continue to act as $\bar
\partial =\psi^{\bar i}\partial/\partial\phi^{\bar i}$, and the
condition that $\widehat {\cal F}$ is annihilated by $\overline
Q_+$ is precisely that, as a function of $\phi^i$, $\phi^{\bar
i}$, and their $z$-derivatives, it is independent of $\phi^{\bar
i}$ (as opposed to its derivatives), and depends only on the other
variables, namely $\phi^i$, and the derivatives of $\phi^i$ and
$\phi^{\bar i}$.\footnote{We can again ignore the action of
$\overline Q_+$ on $z$-derivatives of $\phi^{\bar i}$ because of
the equation of motion $\partial_z\psi^{\bar i}=0$ and the
symmetry transformation law $\delta \phi^{\bar i} = \psi^{\bar
i}$.} Hence, the $\overline Q_+$-invariant operators are of the
form ${\widehat {\cal
F}}(\phi^i,\partial_z\phi^i,\dots;\partial_z\phi^{\bar
i},\partial_z^2\phi^{\bar i},\dots)$. In other words, the
operators, in their dependence on the center of mass coordinate of
the string whose worldsheet theory is the twisted $(0,2)$
sigma-model, is holomorphic. The local sections of $\widehat {\cal
A}$ are just given by the operators in the $\overline
Q_+$-cohomology of the local, twisted $(0,2)$ sigma-model with
action (\ref{Su}).

Let us set $\beta_i  =  \delta_{i \bar j} \partial_z \phi^{\bar
j}$ and $\gamma^i = \phi^i$, whereby $\beta_i$ and $\gamma^i$ are
bosonic operators of dimension $(1,0)$ and $(0,0)$ respectively.
Then, the $\overline Q_+$-cohomology of operators regular in $U$
can be represented by arbitrary local functions of $\beta$ and
$\gamma$ of the form ${\widehat {\cal F}} (\gamma,
\partial_z \gamma,
\partial_z^2 \gamma, \dots, \beta, \partial_z \beta, \partial_z^2
\beta, \dots)$. The operators $\beta$ and $\gamma$ have the
operator products of a standard $\beta\gamma$ system.  The
products $\beta\cdot\beta$ and $\gamma\cdot\gamma$ are
non-singular, while \be \beta_i(z)\gamma^j(z')=-{\delta_{ij}\over
z-z'}+{\rm regular}. \ee  These statements can be deduced from the
flat action (\ref{Su}) by standard field-theoretic methods. We can
write down an action for the fields $\beta$ and $\gamma$, regarded
as free elementary fields, which reproduces these OPE's. It is
simply the following action of a $\beta\gamma$ system: \be I_{
\beta\gamma}= {1\over 2\pi} \int |d^2z| \ \sum_i\beta_i
\partial_{\bar z}\gamma^i. \label{bcaction} \ee Hence, we find
that the local $\beta\gamma$ system above reproduces the
$\overline Q_+$-cohomology of $\psi^{\bar i}$-independent
operators of the sigma-model on $U$, i.e., the local sections of
the sheaf $\widehat{\cal A}$.

At this juncture, one can make another important observation
concerning the relationship between the local twisted $(0,2)$
sigma-model with action (\ref{Su}) and the local version of the
$\beta\gamma$ system of (\ref{bcaction}). To begin with, note that
the holomorphic stress tensor ${\widehat T}(z) = -2 \pi T_{zz}$ of
the local sigma-model is given by \be {\widehat T}(z) = -
\delta_{i \bar j} \partial_z \phi^{\bar j}\partial_z \phi^i.
\label{T(z)} \ee (Here and below, normal ordering is understood
for ${\widehat T}(z)$). Via the respective identification of the
fields $\beta$ and $\gamma$ with $\partial_z \phi$ and $\phi$, we
find that ${\widehat T}(z)$ can be written in terms of the $\beta$
and $\gamma$ fields as \be {\widehat T}(z) = - \beta_i
\partial_z \gamma^i. \label{Tz} \ee
${\widehat T}(z)$, as given by (\ref{Tz}), coincides with the
holomorphic stress tensor of the local $\beta \gamma$ system.
Simply put, the twisted $(0,2)$ sigma-model and the $\beta\gamma$
system have the same $\it{local}$ holomorphic stress tensor. This
means that locally on $X$, the sigma-model and the $\beta\gamma$
system have the same generators of general holomorphic coordinate
transformations on the worldsheet.


One may now ask the following question: does the $\beta\gamma$
system reproduce the $\overline Q_+$-cohomology of $\psi^{\bar
i}$-independent operators globally on $X$, or only in a small open
set $U$? Well, the $\beta\gamma$ system will certainly reproduce
the $\overline Q_+$-cohomology of $\psi^{\bar i}$-independent
operators globally on $X$ if there is no obstruction to defining
the system globally on $X$, i.e., one finds, after making global
sense of the action (\ref{bcaction}), that the corresponding
theory remains anomaly-free. Let's look at this more closely.

First and foremost, the classical action (\ref{bcaction}) makes
sense globally if we interpret the bosonic fields $\beta$,
$\gamma$ correctly.  $\gamma$ defines a map $\gamma:\Sigma\to X$,
and $\beta$ is a $(1,0)$-form on $\Sigma$ with values in the
pull-back $\gamma^*(T^*X)$. With this interpretation,
(\ref{bcaction}) becomes the action of what one might call a
non-linear $\beta\gamma$ system. However, by choosing $\gamma^i$
to be local coordinates on a small open set $U\subset X$, one can
make the action linear. In other words, a local version of
(\ref{bcaction}) represents the action of a linear $\beta \gamma$
system.

Now that we have made global sense of the action of the
$\beta\gamma$ system at the classical level, we move on to discuss
what happens at the quantum level. The anomalies that enter in the
twisted $(0,2)$ sigma model also appear in the nonlinear
$\beta\gamma$ system as follows. Expand around a classical
solution of the nonlinear $\beta\gamma$ system, represented by a
holomorphic map $\gamma_0:\Sigma \to X$. Setting ${\gamma}
=\gamma_0 +\gamma'$, the action, expanded to quadratic order about
this solution, is $(1/2\pi) \ (\beta , {\overline D \gamma'})$.
$\gamma'$, being a deformation of the coordinate $\gamma_0$ on
$X$, is a section of the pull-back $\gamma_0^* (TX)$. Thus, the
kinetic operator of the $\beta$ and $\gamma$ fields is the
$\overline D$ operator on sections of $ \gamma_0^*(TX)$; it is the
complex conjugate of the $D$ operator of the fermion kinetic term
of the twisted sigma-model action $S_{\textrm{pert}}$ that results
in its observed anomalies. Complex conjugation reverses the sign
of the anomalies, but here the fields are bosonic, while in the
twisted sigma-model, they are fermionic; this gives a second sign
change.\footnote{Notice that the $D$ operator in
$S_{\textrm{pert}}$ acts on sections of the pull-back of the
anti-holomorphic bundle $\overline {TX}$ instead of the
holomorphic bundle $TX$. However, this difference is irrelevant
with regard to anomalies since $p_1(\overline {TX}) = p_1(TX)$.}
Hence, the non-linear $\beta\gamma$ system has exactly the same
anomalies as the underlying twisted $(0,2)$ sigma-model. And if
the anomalies vanish, the $\beta\gamma$ system will reproduce the
$\overline Q_+$-cohomology of $\psi^{\bar i}$-independent
operators globally on $X$. In other words, one can find a global
section of $\widehat {\cal A}$ in such a case.

However, note that the $\beta\gamma$ system lacks the presence of
right-moving fermions and thus the $U(1)_R$ charge $q_R$ carried
by the fields $\psi^i_{\bar z}$ and $\psi^{\bar i}$ of the
underlying twisted $(0,2)$ sigma-model. Locally, the $\overline
Q_+$-cohomology of the sigma model is non-vanishing only for $q_R
=0$. Globally however, there can generically be cohomology in
higher degrees. Since the chiral algebra of operators furnished by
the linear $\beta\gamma$ system gives the correct description of
the $\overline Q_+$-cohomology of $\psi^{\bar i}$-independent
operators on $U$, one can then expect the globally-defined chiral
algebra of operators furnished by the non-linear $\beta\gamma$
system to correctly describe the $\overline Q_+$-cohomology
classes of zero degree (i.e. $q_R =0$) on $X$. How then can one
use the non-linear $\beta\gamma$ system to describe the higher
cohomology? The answer lies in the analysis carried out in the
$\S$A.5. In the $\beta\gamma$ description, we do not have a close
analog of $\bar \partial$ cohomology at our convenience.
Nevertheless, we can use the more abstract notion  of Cech
cohomology. As before, we begin with a good cover of $X$ by small
open sets $\{U_a \}$, and, as explained in $\S$A.5, we can then
describe the $\overline Q_+$-cohomology classes of positive degree
(i.e. $q_R > 0$) by Cech $q_R$-cocycles, i.e., they can be
described by the $q^{th}_R$ Cech cohomology of the sheaf $\widehat
{\cal A}$ of the chiral algebra of the linear $\beta\gamma$ system
with action being a linearised version of (\ref{bcaction}).
Although unusual from a physicist's perspective, this Cech
cohomology approach has been taken as a starting point for the
present subject in the mathematical literature \cite{MSV, GMS1}.

\newsubsection{Local Symmetries, Gluing the Free $\beta\gamma$ Systems, and the Sheaf of CDO's}

\noindent{\it Conserved Currents and Local Symmetries}

So far, we have obtained an understanding of the local structure
of the $\overline Q_+$-cohomology. We shall now proceed towards
our real objective of obtaining an understanding of its global
structure, since after all, the sigma-model is defined on all of
$X$, and not just some open set $U$. In order to do, we will need
to glue the local descriptions that we have studied above
together, so that we will appropriately have a globally-defined
$\beta\gamma$ system and its chiral algebra at our disposal.

To this end, we must first cover $X$ by small open sets $\{U_a\}$.
Recall here that in each $U_a$, the $\overline Q_+$-cohomology is
described by the chiral algebra of local operators of a free
$\beta\gamma$ system on $U_a$. Next, we will need to glue these
local descriptions together over the intersections $\{ U_a \cap
U_b \}$, so as to describe the global structure of the $\overline
Q_+$-cohomology  in terms of a globally-defined sheaf of chiral
algebras over the entire manifold $X$.

Note that the gluing has to be carried out using the automorphisms
of the free $\beta\gamma$ system. Thus, one must first ascertain
the underlying symmetries of the system, which are in turn divided
into  geometrical and non-geometrical symmetries. The geometrical
symmetries are used in gluing together the local sets $\{ U_a \}$
into the entire manifold $X$. The non-geometrical symmetries on
the other hand, are used in gluing the local descriptions at the
algebraic level.

As usual, the generators of these symmetries will be given by the
charges of the conserved currents of the free $\beta\gamma$
system. Since the conserved charges must also be
conformally-invariant, it will mean that they must be given by an
integral of a dimension one current, modulo total derivatives. The
dimension one currents of the free $\beta \gamma$ system can be
constructed as follows.

Let us describe the currents which are associated with the
geometrical symmetries first. Firstly, if we have a holomorphic
vector field $V$ on $X$ where $V = V^i (\gamma) {\partial \over
{\partial \gamma^i}}$, we can construct a dimension one current
$J_V=-V^i \beta_i$. The corresponding conserved charge is then
given by $K_V=\oint J_V dz $. A computation of the operator
product expansion with the elementary fields $\gamma$ gives \be
J_V(z)\gamma^k(z')\sim {V^k(z')\over z-z'}. \label{jv} \ee Under
the symmetry transformation generated by $K_V$, we have $\delta
\gamma^k = i \epsilon [ K_V, \gamma^k ]$, where $\epsilon$ is a
infinitesinal transformation parameter. Thus, we see from
(\ref{jv}) that $K_V$ generates the infinitesimal diffeomorphism
$\delta\gamma^k=i \epsilon V^k$ of $U$. In other words, $K_V$
generates the holomorphic diffeomorphisms of the target space $X$.
For finite diffeomorphisms, we will have a coordinate
transformation ${\tilde \gamma}^k = g^k (\gamma)$, where each $g^k
(\gamma)$ is a holomorphic function in the $\gamma^k$s. Since we
are using the symmetries of the $\beta \gamma$ system to glue the
local descriptions over the intersections $\{U_a \cap U_b\}$, on
an arbitrary intersection $U_a \cap U_b$, $\gamma^k$ and ${\tilde
\gamma}^k$ must be defined in $U_a$ and $U_b$ respectively.

We shall now determine the current associated with the
non-geometrical symmetries. The charge of the current should not
generate any transformations on the $\gamma^i$'s at all since
these fields have a geometrical interpretation as the coordinates
on $X$. In other words, the current must be constructed out of the
$\gamma^i$'s and their derivatives only. Thus, a suitable
dimension one current would be given by $J_B=B_i
(\gamma)\partial_z\gamma^i$, where the $B_i(\gamma)$'s are just
holomorphic functions in the $\gamma^i$'s. The conserved charge is
then given by $K_B = \oint J_B dz$. As explained in \cite{CDO},
the $B_i(\gamma)$'s must be the components of an arbitrary
holomorphic $(1,0)$-form $B = \sum_i B_i (\gamma) d{\gamma^i}$ on
$X$ that is non-exact, i.e., for every non-vanishing $K_B$, there
is a $(2,0)$-form $C = \partial B$, that is $\partial$-closed
(since $\partial^2 =0$). Thus, $C$ corresponds to a sheaf
$\Omega^{2,cl}_X$ of $\partial$-closed $(2,0)$-forms on $X$, which
is related via the first Cech cohomology group
$H^1(X,\Omega^{2,cl}_X)$ to the moduli of the chiral algebra of
the sigma-model \cite{CDO,MSV}. This point will be important in
our forthcoming paper, where we will investigate the physical
interpretation of a ``quantum'' geometric Langlands correspondence
in a similar context, albeit with fluxes that correspond to the
moduli of the chiral algebra turned on.

\bigskip\noindent{\it Local Field Transformations and Gluing the
Free $\beta\gamma$ Systems}

Let us now describe how the different fields of the  free $\beta
\gamma$ system on any $U$ will transform under the geometrical and
non-geometrical symmetries generated by $K_V$ and $K_B$
respectively. Via a computation of the relevant OPEs, we have
\begin{eqnarray}
\label{autoCDOgamma}
{\tilde \gamma}^i & = & g^i (\gamma) ,\\
\label{autoCDObeta} {\tilde \beta}_i  & = &  \beta_k   D^k{}_i +
\partial_z \gamma^j E_{i j},
\end{eqnarray}
where $i,j,k = 1, 2, \dots, N={\textrm{dim}_{\mathbb C} X}$. Here,
$D$ and $E$ are $N \times N$ matrices such that $[D]^T = [\partial
g]^{-1}$ and $[E] = [\partial B]$, that is, $[(D^T)^{-1}]_i{}^k =
\partial_i g^k$ and $[E]_{ij} =
\partial_i B_j$.

Note that in order to consistently glue a pair of free
$\beta\gamma$ systems in any overlap region $U_a \cap U_b$, one
will need to use the relations in
(\ref{autoCDOgamma})-(\ref{autoCDObeta}) to glue their free fields
together. As required, (\ref{autoCDOgamma})-(\ref{autoCDObeta})
defines an automorphism of the free $\beta\gamma$ system - the
$\tilde \gamma^i$ and $\tilde \beta^i$ fields produce the correct
OPE's amongst themselves.

\smallskip\noindent{\it A Sheaf of CDO's}

Last but not least, note that
(\ref{autoCDOgamma})-(\ref{autoCDObeta}) actually define the
automorphism relations of a sheaf $\widehat{\cal O}^{ch}_X$ of
Chiral Differential Operators or CDO's on $X$ \cite{MSV}. In other
words, $\widehat A \simeq \widehat{\cal O}^{ch}_X$. Hence, the
$\overline Q_+$-cohomology and therefore the holomorphic chiral
algebra $\cal A$ of the twisted $(0,2)$ sigma-model will be given
by $\bigoplus_{q_R} H^{q_R}(X, {\widehat{\cal O}^{ch}_X})$ - the
sum of all Cech cohomology groups of the sheaf of CDO's on $X$, as
a vector space.

\vspace{1.0cm}
\hspace{-1.0cm}{\large \bf Acknowledgements:}\\
\vspace{-0.5cm}

I'm especially indebted to E. Frenkel for explaining many things
in connection with the geometric Langlands program to me over our
often tedious email correspondences. I'm also thankful to I.
Bakas, D. Ben-Zvi, S. Gukov, F. Malikov and V. Pestun for
illuminating exchanges. Last but not least, I would like to thank
H. Sati for encouragement, E. Witten for many constructive
criticisms, and the JHEP referee for the suggested clarifications.
This work is supported by the Institute for Advanced Study and the
NUS - Overseas Postdoctoral Fellowship.

\end{document}